\documentclass[]{aa}  

\usepackage{graphicx}
\usepackage{txfonts}
\begin{document} 

   \title{Continuum sources from the THOR survey between 1 and 2\,GHz\thanks{Table \ref{table_catalog_example} is available in electronic form at the CDS via anonymous ftp to cdsarc.u-strasbg.fr (130.79.128.5)}}

   \author{
                                S. Bihr\inst{1}
                \and
                        K.G. Johnston\inst{2}
                \and
                        H. Beuther\inst{1}
            \and
                L.D. Anderson\inst{3}            
            \and
                J. Ott\inst{4}
            \and
                M. Rugel\inst{1}
\and
F. Bigiel\inst{5}
\and
A. Brunthaler\inst{6}
\and
S.C.O. Glover\inst{5}
\and
T. Henning\inst{1}
\and
M.H. Heyer\inst{7}
\and
R.S. Klessen\inst{5}
\and
H. Linz\inst{1}
\and
S.N. Longmore\inst{10}
\and
N.M. McClure-Griffiths\inst{11}
\and
K.M. Menten\inst{6}
\and
R. Plume\inst{12}
\and
T. Schierhuber\inst{1}
\and
R. Shanahan\inst{12}
\and
J.M. Stil\inst{12}
\and
J.S. Urquhart\inst{6,13}
\and
A.J. Walsh\inst{14}
}
   \institute{
                                Max Planck Institute for Astronomy, K\"onigstuhl 17, 69117 Heidelberg, Germany\\
                                                \email{name@mpia.de}
         \and
                School of Physics and Astronomy, University of Leeds, Leeds LS2 9JT, UK
         \and
               Department of Physics and Astronomy, West Virginia University, Morgantown, WV 26506, USA              
          \and
               National Radio Astronomy Observatory, P.O. Box O, 1003 Lopezville Road, Socorro, NM 87801, USA
           \and     
               Universit\"at Heidelberg, Zentrum f\"ur Astronomie, Institut f\"ur Theoretische Astrophysik, Albert-Ueberle-Str. 2, D-69120 Heidelberg, Germany
                        \and               
               Max-Planck-Institut f\"ur Radioastronomie, Auf dem H\"ugel 69, D-53121 Bonn, Germany
            \and
               Department of Astronomy, University of Massachusetts, Amherst, MA 01003-9305, USA
            \and
              Kavli Institute for Particle Astrophysics and Cosmology, Stanford University, SLAC National Accelerator Laboratory, Menlo Park, CA 94025, USA
             \and
              Department of Astronomy and Astrophysics, University of California, 1156 High Street, Santa Cruz, CA 95064, USA 
                        \and             
              Astrophysics Research Institute, Liverpool John Moores University, 146 Brownlow Hill, Liverpool L3 5RF, UK
             \and
             Research School of Astronomy and Astrophysics, The Australian National University, Canberra, ACT, Australia 
             \and
             Department of Physics and Astronomy, University of Calgary, 2500 University Drive NW, Calgary, AB T2N 1N4, Canada.
             \and
              Centre for Astrophysics and Planetary Science, University of Kent, Canterbury CT2 7NH, UK
             \and
              International Centre for Radio Astronomy Research, Curtin University, GPO Box U1987, Perth WA 6845, Australia         
               }

   \date{Received September xx, xxxx; accepted March xx, xxxx}

  \abstract{
We carried out a large program with the Karl G. Jansky Very Large Array (VLA): `THOR: The \ion{H}{i}, OH, Recombination line survey of the Milky Way'. We observed a significant portion ($\sim$100\,deg$^2$) of the Galactic plane in the first quadrant of the Milky Way in the 21\,cm \ion{H}{i} line, 4 OH transitions, 19 radio recombination lines, and continuum from 1 to 2\,GHz. In this paper we present a catalog of the continuum sources in the first half of the survey ($l = 14.0-37.9\degr$ and $l = 47.1-51.2$\degr, $|b| \leq 1.1$\degr) at a spatial resolution of $10-25\arcsec$, depending on the frequency and sky position with a spatially varying noise level of $\sim$0.$3-1$\,mJy\,beam$^{-1}$. The catalog contains $\sim$4400 sources. Around 1200 of these are spatially resolved, and $\sim$1000 are possible artifacts, given their low signal-to-noise ratios. Since the spatial distribution of the unresolved objects is evenly distributed and not confined to the Galactic plane, most of them are extragalactic. Thanks to the broad bandwidth of the observations from 1 to 2\,GHz, we are able to determine a reliable spectral index for $\sim$1800 sources. The spectral index distribution reveals a double-peaked profile with maxima at spectral indices of $\alpha \approx -1$ and $\alpha \approx 0$, corresponding to steep declining and flat spectra, respectively. This allows us to distinguish between thermal and non-thermal emission, which can be used to determine the nature of each source. We examine the spectral index of $\sim$300 known \ion{H}{ii} regions, for which we find thermal emission with spectral indices around $\alpha \approx 0$. In contrast, supernova remnants (SNR) show non-thermal emission with $\alpha \approx -0.5$ and extragalactic objects generally have a steeper spectral index of $\alpha \approx -1$. Using the spectral index information of the THOR survey, we investigate potential SNR candidates. We classify the radiation of four SNR candidates as non-thermal, and for the first time, we provide strong evidence for the SNR origin of these candidates.}


   \maketitle

\section{Introduction}

At present, high resolution (<20\arcsec) Galactic plane surveys are available for studying different questions concerning star formation and the interstellar medium (ISM). These surveys cover a large fraction of the spectral range, from the near- \citep[UKIDSS,][]{Lucas2008}, mid- \citep[GLIMPSE,][]{Churchwell2009} and far-infrared \citep[MIPSGAL, HIGAL,][]{Carey2009, Molinari2010}, to the submm \citep[ATLASGAL, BOLOCAM,][]{Schuller2009, Rosolowsky2010, Aguirre2011, Csengeri2014}, to longer radio wavelengths studying the continuum as well as molecular lines \citep[e.g., GRS, MAGPIS, CORNISH, HOPS, MALT90, MALT-45, ][]{Jackson2006, Helfand2006, Hoare2012, Purcell2013, Walsh2011, Purcell2012, Jackson2013}. As hydrogen is the most common element in our universe, observations of this element are a crucial ingredient to complete the picture of our Galaxy. Molecular hydrogen is difficult to observe directly as its rotational energy levels are not readily excited in the cold ISM. However, the 21\,cm \ion{H}{i} line provides a direct measurement of the atomic hydrogen. To date, the Galactic plane surveys of the 21\,cm \ion{H}{i} line have a spatial resolution of >1$\arcmin$ \citep[CGPS, SGPS, VGPS,][]{Taylor2003, McClure-Griffiths2005, Stil2006}, which is not sufficient in comparison to the other Galactic plane surveys. This was the motivation to initiate a Galactic plane survey using the Karl G. Jansky Very Large Array (VLA) in C-configuration, achieving a spatial resolution of $\sim$20\arcsec: `THOR - The \ion{H}{i}, OH, Recombination line survey of the Milky Way'. The angular resolution of 20$\arcsec$ gives us a linear resolution of $\sim$0.1 to 1.5\,pc at typical Galactic distances of 1 to 15\,kpc. Since the new WIDAR correlator at the VLA offers a broad bandwidth, including high resolution sub-bands, we are able to observe the 21\,cm \ion{H}{i} line, 4 OH lines, 19 H$\alpha$ radio recombination lines (RRL) and the continuum from 1-2\,GHz simultaneously. Starting in 2012 with a pilot study around the giant molecular cloud (GMC) associated with the W43 star formation complex \citep{Bihr2015b, Walsh2016}, we observed a large fraction of the Galactic plane in the first quadrant of the Milky Way ($l = 14-65\degr$, $|b| \leq 1.1$\degr) in consecutive semesters. In this paper, we present the results of the continuum observations of the first half of the survey ($l = 14.0-37.9\degr$ and $l = 47.1-51.2$\degr, $|b| \leq 1.1$\degr), covering $\sim$56\,deg$^2$, including a catalog of the extracted sources. The full survey will be presented in a forthcoming paper by Beuther et al. (in prep.).\\
\begin{figure*}
  \resizebox{\hsize}{!}{
  \includegraphics[width=9cm]{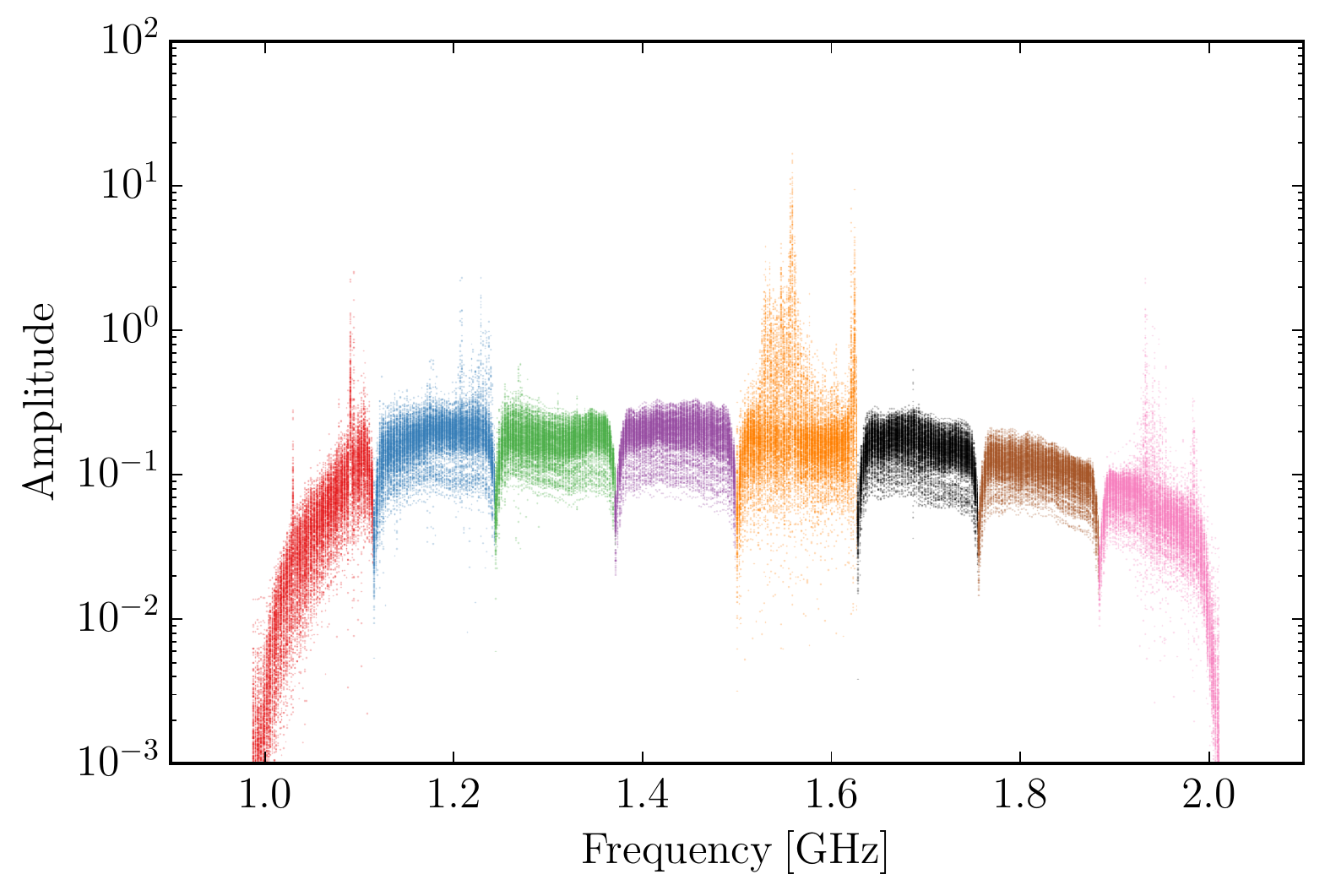}
  \includegraphics[width=9cm]{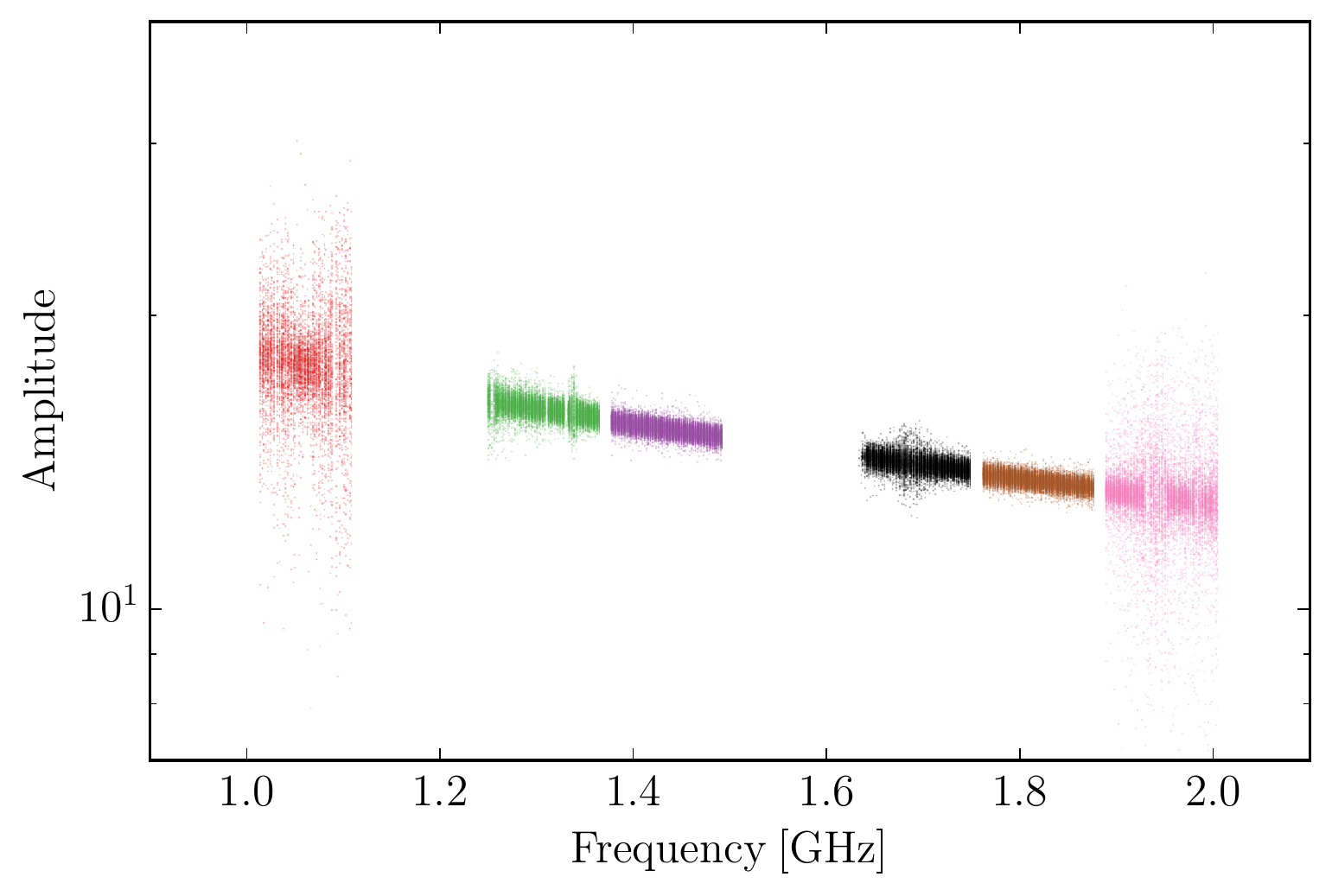}
  }
  \caption{Observed amplitude of the bandpass/flux calibrator 3C286 as a function of frequency prior to the calibration on the left side and after the calibration on the right side. The color coding represents the eight spectral windows. The spectral windows at 1.2 and 1.6\,GHz were flagged because of RFI (see Sect. \ref{sec_rflag}).}
  \label{Fig_spw_before_and_after_calibration}
\end{figure*}
The discrete continuum sources between 1 and 2\,GHz are dominated by two distinct emission classifications: thermal and non-thermal emission \citep{Wilson2010}. The thermal emission is mostly due to free-free emission from electrons, whereas the non-thermal emission is due to the synchrotron emission of relativistic electrons in magnetic fields. These different emission mechanisms can be distinguished by the spectral index $\alpha$, which is defined as $I(\nu) \varpropto \nu^{\alpha}$, where $I(\nu)$ is the frequency dependent intensity. The thermal free-free emission shows a flat or positive spectral index, depending on the optical depth. The values can vary between 2 and -0.1 for the optically thick and thin regime, respectively \citep[e.g.,][]{Mezger1967, Keto2003, Wilson2010}. In contrast to this, synchrotron emission shows a negative spectral index depending on the particle energy distribution. One usually finds spectral indices below -0.5 \citep[e.g.,][]{Rybicki1979, Meisenheimer1999}. Supernova remnants (SNR) show a spatially varying spectral index around $\alpha = -0.5$ \citep[e.g.,][]{Bhatnagar2011, Green2014, Reynoso2015, Dubner2015}. The broad bandpass of our VLA observations allows us to determine the spectral index for bright sources and therefore distinguish between the two radiation mechanisms. However, knowing the kind of radiation does not directly disclose the source type. Thermal free-free emission can emerge from \ion{H}{ii} regions or planetary nebulae. Non-thermal synchrotron radiation can be produced by extragalactic jets powered by an active galactic nucleus (AGN) or from Galactic SNR. Thermal radiation from extragalactic sources is possible, but might be too weak to be detected in our observations. As a result, thermal emission is most likely of Galactic origin, and the non-thermally emitting sources could be extragalactic AGN or Galactic SNR. The ability to characterize continuum sources and thus distinguish between Galactic and extragalactic emission is crucial for prospective THOR \ion{H}{i} and OH absorption studies.

\section{Observations and data reduction}
\subsection{VLA observations}
\begin{figure*}
   \resizebox{\hsize}{!}
            {\includegraphics[width=18cm]{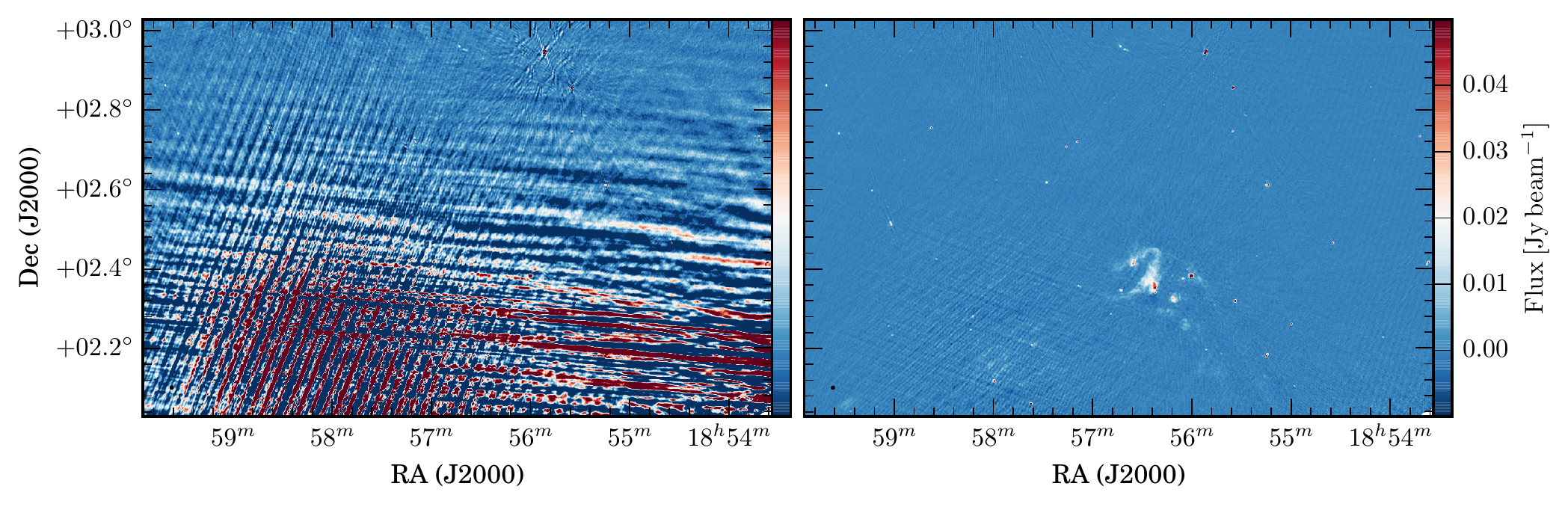}}
      \caption{Left panel: the imaged data for one spectral window around 1.3\,GHz after the calibration, without automated flagging. Strong RFI features are present. Right panel: the same region after applying the automated flagging method RFlag.}
         \label{Fig_RFI_image}
\end{figure*}
\begin{table}
\caption{Summary of spectral windows}            
\label{table_spectral_windows}      
\centering                          
\begin{tabular}{c c c}       
\hline\hline                 
Frequency & Lowest & Highest   \\  
$\rm{[MHz]}$ & resolution & resolution\\
\hline 
$989-1117$ & 24.4$\arcsec$ $\times$ 15.1$\arcsec$ & 16.5$\arcsec$ $\times$ 15.7$\arcsec$\\
$1244-1372$ & 19.7$\arcsec$ $\times$ 12.5$\arcsec$
  & 13.1$\arcsec$ $\times$ 12.3$\arcsec$
\\
$1372-1500$ & 18.1$\arcsec$ $\times$ 11.1$\arcsec$
  & 12.6$\arcsec$ $\times$ 11.9$\arcsec$
\\
$1628-1756$ & 15.4$\arcsec$ $\times$ 9.1$\arcsec$
 & 10.5$\arcsec$ $\times$ 9.9$\arcsec$
\\
$1756-1884$ & 14.5$\arcsec$ $\times$ 8.9$\arcsec$
 & 10.0$\arcsec$ $\times$ 9.7$\arcsec$
\\
$1884-2012$ & 13.1$\arcsec$ $\times$ 8.1$\arcsec$
 & 9.0$\arcsec$ $\times$ 8.3$\arcsec$
\\
\hline 
\end{tabular}
\tablefoot{Owing to the varying declination of different observing blocks, we obtain different resolution elements.}
\end{table}

\begin{figure}
   \centering
   \includegraphics[width=\hsize]{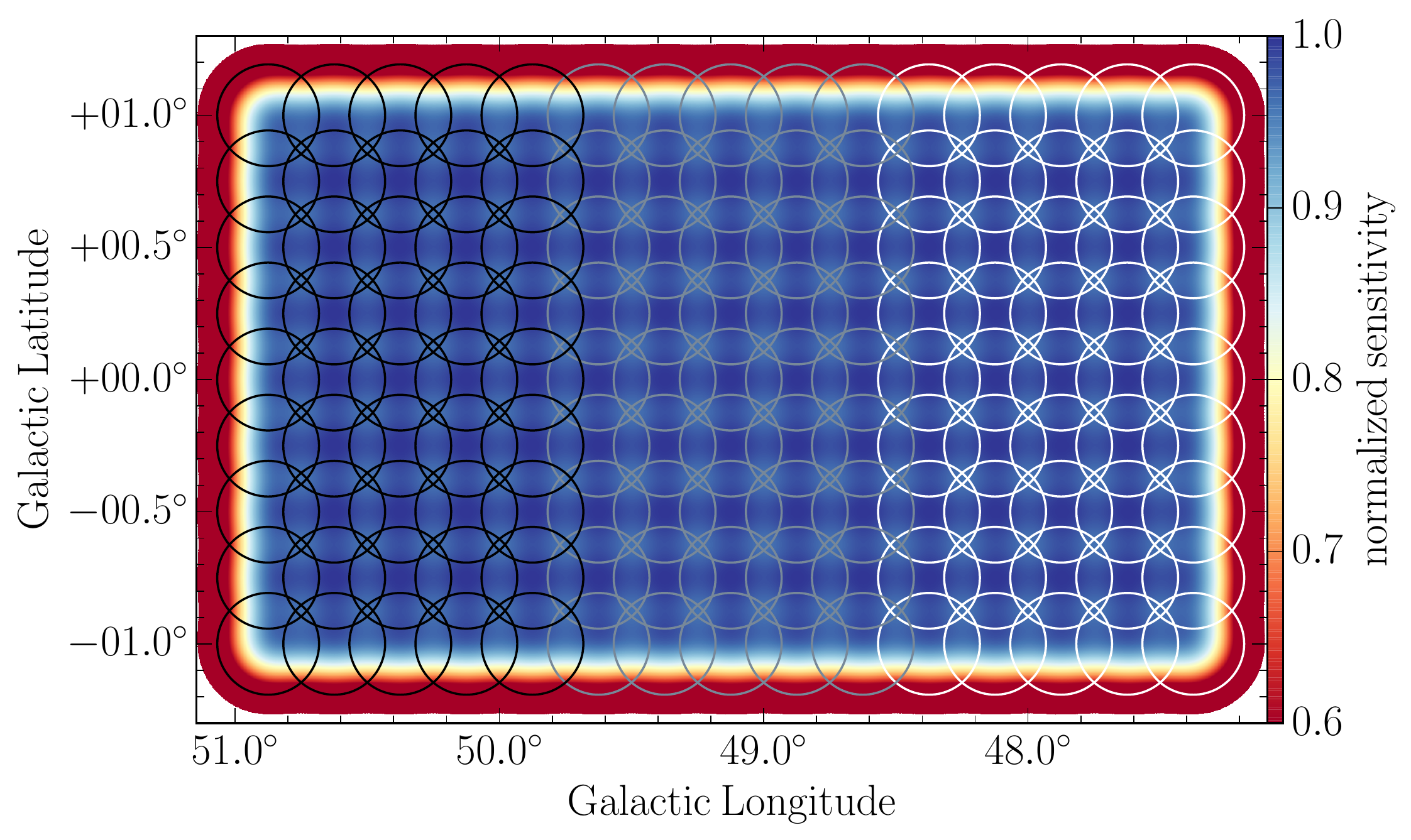}
      \caption{Normalized sensitivity pattern in color of the observed mosaic for the spectral window with the highest frequency at 1.95\,GHz. The sensitivity drops toward the edge, and the variations within the mosaic are smaller than 4\%. The circles represent the primary beam at this frequency, which is $\sim23\arcmin$. The different colors of the circles represent three different observing blocks.}
         \label{Fig_primary_beam_coverage}
\end{figure}

We used the VLA in New Mexico in C configuration to map the continuum in the L band from 1 to 2\,GHz simultaneously with the \ion{H}{i} 21\,cm line, 4 OH lines, and 19 H$\alpha$ recombination lines. For the VLA in C-configuration, the baselines range from 35 to 3400\,m. The corresponding primary beam changes with frequency from $\sim$45$\arcmin$ at 1\,GHz to $\sim$23$\arcmin$ at 2\,GHz and therefore the actual size of the mosaics changes as well. The data presented in this paper were observed in two campaigns. The first campaign was the THOR pilot observations ($l = 29.2-31.5$\degr, $|b| \leq 1.1$\degr) during the 2012A semester \citep[Project 12A-161, see also][]{Bihr2015b}. We used a hexagonal geometry for the mosaic for this $2\degr\times2\degr$ field at 17.9\arcmin spacing, which results in 59 pointings. Each pointing was observed 4 $\times$ 2\,min, which results in an overall integration time of ten hours for the pilot field, including around two hours overhead for flux, bandpass, and complex gain calibration. The second campaign covered a large section of the first quadrant of the Milky Way ($l = 14.0-29.2\degr$ and $l = 31.5-37.9\degr$ and $l = 47.1-51.2$\degr, $|b| \leq 1.1$\degr) and was observed during the 2013A semester (Project 13A-120). In contrast to the pilot field, we used a rectangular grid for the mosaic (see Fig.\,\ref{Fig_primary_beam_coverage}) with a smaller spacing of 15\arcmin. The close spacing meant that the sensitivity variations are at most 4\% for the spectral window around 1.95\,GHz and less for smaller frequencies. The second campaign was split into 20 observing blocks, each covering a field of $\Delta l = 1.25\degr$ and $|b| \leq 1.1\degr$ with 45 pointings each. Each pointing was observed 3 $\times$ $\sim$2\,min, which results in a total integration time of five hours for each observing block, including $\sim$50\,min overhead for flux, bandpass, and complex gain calibration. We chose the quasar 3C286 as a flux and bandpass calibrator for all fields. As complex gain calibrator, we used the quasar  J1822-0938 for all observing blocks between $l = 14.0-37.9\degr$ (including the pilot field) and the quasar J1925+2106 for all observing blocks between $l = 47.1-51.2$\degr. The achieved resolution depends on the frequency and the sky position and varies between 10 and 25$\arcsec$ (see Table \ref{table_spectral_windows} for further details). By the date of publication of this paper, the other half of the survey will have been observed. However, since the calibration and imaging is an enormous computing and person power effort, the data reduction of that second half is still going on. The full survey will be presented in a future article.\\We used the new WIDAR correlator and observed the continuum between 1 and 2\,GHz using eight sub-bands, so-called spectral windows, each with a bandwidth of 128\,MHz. Owing to strong contamination of radio frequency interference (RFI), we could not use two spectral windows. The frequencies of the six remaining spectral windows are given in Table \ref{table_spectral_windows}. We split each spectral window further into 64 channels with a channel width of 2\,MHz. This setup allows us to flag individual channels that might be contaminated by, for instance, RFI without significantly losing sensitivity.

\subsection{Calibration}
We used the CASA package (version 4.1.0) in combination with a modified VLA pipeline\footnote{https://science.nrao.edu/facilities/vla/data-processing/pipeline} (version 1.2.0) to edit and calibrate the data. Prior to the calibration, we manually flagged strong RFI and bad antennas. The pipeline uses automated flagging algorithms such as RFlag on the calibrator observations to improve the calibration solutions, but does not flag the target fields. Subsequently, the pipeline applies the bandpass, flux, and gain calibration. At this point, we neither used Hanning smoothing nor recalculated the data weights (CASA command `statwt'), since this could influence very bright continuum sources. We implemented some modifications to the pipeline to improve the quality checking and performed further flagging on the target fields with automated flagging algorithms (see Sect. \ref{sec_rflag} for further details) and by hand after the pipeline run. A detailed description of our calibration procedure will be given in the THOR survey overview paper (Beuther et al., in prep). \\

\subsection{Automated flagging algorithm RFlag}
\label{sec_rflag}
\begin{figure}
   \centering
   \includegraphics[width=\hsize]{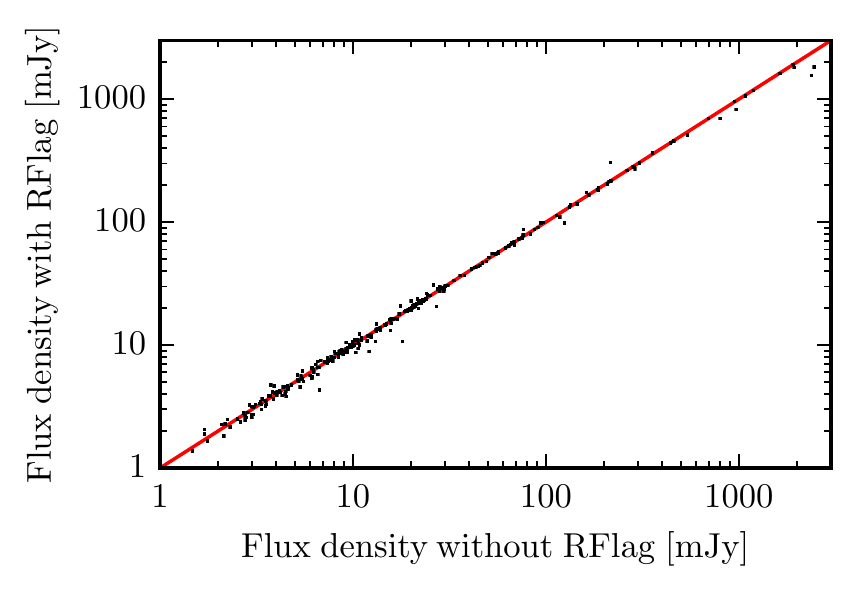}
      \caption{Extracted flux density of sources with the RFlag method applied as a function of the extracted flux density for the same sources without the RFlag method applied. For this comparison we used the spectral window around 1.4\,GHz and the field around $l = 22$\degr. The red line represents a one-to-one relation. Over a wide range of flux densities, the RFlag method does not influence the actual source flux densities.}
         \label{Fig_rflag_flux_comparison}
\end{figure}

As shown in the lefthand panel of Fig.\,\ref{Fig_spw_before_and_after_calibration}, some spectral windows in our data are affected by RFI. The spectral windows around 1.2 and 1.6\,GHz have the strongest contamination, and we cannot use them. The spectral window around 1.6\,GHz is severely affected by the GPS satellites, which can be seen as outliers from the normal bandpass  shape in the lefthand panel of Fig.\,\ref{Fig_spw_before_and_after_calibration}, and we are not even able to calibrate the data. The spectral window around 1.2\,GHz can be calibrated. However, the images show a consistently strong RFI contamination, which cannot be removed by the automated flagging algorithm discussed below. Within the other spectral windows, we found RFI contamination varying in frequency, sky position, and time. Therefore it is very difficult and time-consuming to flag all data manually, so we explored the possibility of automated flagging algorithms. CASA provides the so-called RFlag algorithm, which was introduced previously to AIPS by E. Greisen in 2011. The RFlag algorithm iterates the data in chunks of time and performs a time analysis for each channel, as well as a spectral analysis, for each time step and flags outliers (see the CASA manual\footnote{available on the CASA webpage: http://casa.nrao.edu} for further details). Using the standard threshold greatly improved the results as shown in Fig.\,\ref{Fig_RFI_image}. The RFlag algorithm flags almost all RFI reliably. However, a useful automated flagging algorithm must not only flag the RFI reliably, but also keep the actual scientific signal unchanged. We therefore tested the effects of the RFlag algorithm on the thermal noise in our data, as well as the flux densities of our sources. For these two tests, we investigated the spectral window at $\sim$1.4\,GHz of the field around $l = 22$\degr. The frequency range around the \ion{H}{i} 21\,cm line is a protected band and is indeed almost free of terrestrial RFI. Applying the RFlag algorithm on this spectral window should not affect the thermal noise and the flux densities of our sources. We calibrated and cleaned the data exactly the same way, but on one data set, we applied the RFlag algorithm before cleaning, whereas we cleaned the other data set without automated flagging and used this as a reference.\\
As a first test we compared the noise between the two data sets. Because the spectral window around 1.4\,GHz is mostly free of RFI, we did not find different noise levels for the two data sets. This shows that the RFlag algorithm does not flag good data, which would increase the noise level. For both data sets we extracted the continuum sources using the method described in Sect. \ref{sec_blobcat}, cross-matched the two data sets, and compared the flux densities for each source. Figure \ref{Fig_rflag_flux_comparison} shows the result of this comparison of the flux densities with and without the RFlag algorithm applied. Over the full range of flux density values, we see no significant deviation for unresolved and small sources (smaller than $\sim$100\arcsec). However, more extended sources might be affected by the RFlag algorithm, and this effect will be discussed in Sect.\,\ref{sec_extended_sources}.\\ 
In summary, the RFlag algorithm provides a reliable tool for removing RFI from the continuum data. While the noise level and unresolved and small sources are not affected significantly by the RFlag algorithm, large extended sources have to be treated more carefully. We discuss this in more detail in Sect. \ref{sec_extended_sources}.

\subsection{Imaging and deconvolution}
\label{sec_imaging_and_deconvolution}
For the imaging and deconvolution, we used the task \textit{clean} in the CASA package. Since we cover a large area on the sky, we created mosaics consisting of several pointings. This is an algorithmic, as well as a computational challenge, and we extensively tested different versions of the mosaicking algorithm in the CASA package, including versions 4.2.2, 4.3, 4.4, and a test version of 4.5. Our main focus was to obtain consistent flux density measurements, so we compared flux density and intensity values of point sources in mosaics created with the above-mentioned CASA versions with their corresponding values in individually cleaned pointings. The clean and deconvolution algorithms for single pointings are simpler and well tested and were therefore used as reference. In collaboration with the CASA developer team, we could identify several problems in the mosaic algorithm of the CASA versions 4.3, 4.4, and a test version of 4.5. Therefore we decided to use version 4.2.2 because this version provides flux density values in the final mosaics that are within ten percent of the flux density values measured in single pointings.\\
Since we cover a wide range of frequency from 1 to 2\,GHz, we cannot clean all spectral windows together without considering the frequency dependence of the sources, as well as primary beam effects. While the CASA package is able to clean wide-band images for single pointings (using the parameter \textit{nterm}), to date (up to version 4.4) this is not available for mosaics. We could clean each observed channel separately, but this would reduce the signal-to-noise significantly and requires immense computational resources. As a compromise, we cleaned each 128\,MHz-wide spectral window separately, thus neglecting the frequency dependence inside each spectral window. Thereafter, we compared the peak intensity between the spectral windows to determine the spectral index (see Sect. \ref{sec_spectral_index_determination} for further details). To suppress the sidelobes and increase the resolution, we chose $robust=0$ as a weighting parameter, which is a compromise between uniform and natural weighting. As a pixel size, we chose 2.5\arcsec, which is sufficient to sample the smallest possible resolution element (synthesized beam width) of $\sim$8\arcsec.\\
To achieve a uniform noise between the separate observing blocks, we included the neighboring observing blocks in the clean process. Because the clean command in CASA works with equatorial coordinates, we have to choose a large image size of 4600$\times$4600\,pixels to cover one field, consisting of three observing blocks. We applied primary beam corrections to obtain reliable flux densities. Because the continuum emission covers a wide range of spatial scales, we used the multiscale clean in CASA to recover the large scale structure. In this cleaning method we selected four different scales: besides the point source, also 1, 3, and 6$\times$ of the resolution element. We stopped the cleaning process at a threshold of 5\,mJy\,beam$^{-1}$ or $10^{5}$ iterations, whichever was reached first. As the noise in our data is dominated by the sidelobe noise, the cleaning threshold is higher than the thermal noise level. The final resolution depends on the frequency of each spectral window and the declination of the observed field. Table \ref{table_spectral_windows} provides an overview of the highest and lowest resolution for each spectral window. The noise level of the images are discussed in Sec.\ref{sec_completeness}.

\subsection{Extended sources}
\label{sec_extended_sources}
\begin{figure}
   \centering
   \includegraphics[width=\hsize]{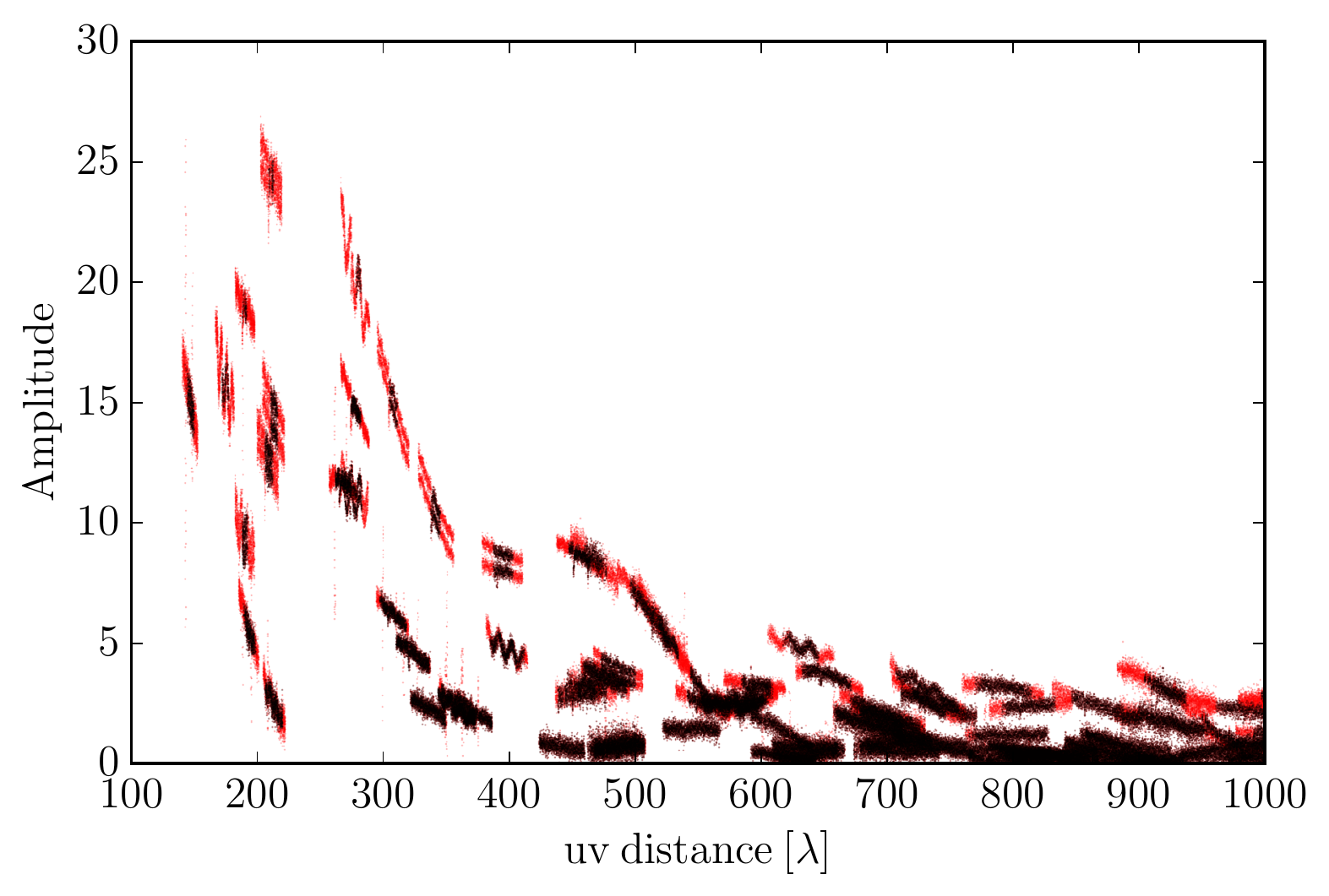}
      \caption{Amplitude as a function of uv distance for a single pointing close to the spatially very extended SNR G021.8-00.6 (pointing center at J2000 18:32:52.2, $-10$:03:32.3). The black points represent the data we used for the imaging whereas the red points represent the data points flagged by the RFlag method.}
         \label{Fig_uv_plot_rflag}
\end{figure}

Extended sources suffer from different filtering effects owing to both the interferometer and the applied RFlag method. The first effect is due to the incomplete sampling in the uv plane. Each interferometer suffers from this effect, and it depends to first order on the shortest available baseline. Theoretically the VLA in C-configuration can observe all spatial scales up to 970\arcsec for the L band (VLA-manual). However, this value is for a 12-hour observation near the zenith and snapshot observations may recover scales diminished by a factor of two. However, this rule-of-thumb estimate might be too optimistic, and more realistic observations are not able to reach this value. To examine the insensitivity of the large spatial scales of the interferometer in a more realistic environment, we performed simulated observations of artificial sources with the THOR observation setup. We tested sources with a Gaussian intensity profile and varying sizes. These tests showed that we are able to recover sources with sizes up to $\sim$120\arcsec reasonably well (80\% flux recovery) for all frequencies between 1 and 2\,GHz. To achieve this result, the use of the multiscale clean was crucial. However, Galactic sources do not show simple 2D Gaussian profiles so that to quantify the filtering effect in detail is difficult. Nevertheless, these simulations show that sources up to $\sim$120\arcsec are not severely affected by the insensitivity to large spatial scales of the interferometer. Since this insensitivity depends on the frequency, this can artificially change the spectral index of extended sources. However, our simulated observations revealed that this only affects sources larger than $\sim$120\arcsec.\\
The second filtering effect for extended sources is due to the applied RFlag algorithm. Extended sources show high amplitudes for short uv distances (see Fig. \ref{Fig_uv_plot_rflag}). As the RFlag algorithm searches for outliers in a frequency and time domain, it recognizes some of these high values as outliers and flags them accordingly. Quantifying this effect is complicated because the flagging depends on the source size, its intensity, and the internal intensity structure. However, similar to our tests in Sect. \ref{sec_rflag}, we used the spectral window around $\sim$1.4\,GHz with and without applied RFlag to examine the effects of the algorithm on the large scale structure. Figure \ref{Fig_uv_plot_rflag} shows the amplitude as a function of uv distance for one pointing close to the extended SNR G021.8-00.6. The red data points show the data points flagged by the RFlag algorithm. For a uv distance smaller than 300$\rm{\lambda}$ the RFlag algorithm flags significantly more data ($\sim$70\%) in comparison to larger uv distances ($\sim$25\%). In the simple approximation of $\theta = \lambda / D$, where $\theta$, $\lambda$, and $D$ are the angular scale, the wavelength, and the diameter or baseline length of the telescope, respectively, the uv distance of 300$\rm{\lambda}$ describes an angular scale of $\sim$600$\arcsec$. The flagging of the data points for short uv distances removes part of the large scale structure of the source. However, only large and bright sources are affected by this filtering. The SNR G021.8-00.6 has a spatial extent of $\sim$1200$\times$400$\arcsec$ and the RFlag algorithm flags $\sim$40\% of the flux density. Smaller sources on the order of $\sim$100 to 300$\arcsec$ show lower values of $\sim$5 to 10\% flux density removal. Owing to these two filtering effects for largely extended (>400\arcsec) and bright (>1\,Jy) sources, we refrained from analyzing these sources in detail and the corresponding flux values have to be treated cautiously. For the spectral index determination (see Sect. \ref{sec_spectral_index_determination}), we use the peak intensity rather than the integrated flux density since the former is less affected by the explained filtering effects.

\section{Source extraction}
In this section, we explain the source extraction method using the BLOBCAT software as well as the method of determining the spectral index. To achieve a higher signal-to-noise ratio for the source extraction, we use the average of two spectral windows to detect the sources. Thereafter we extract the peak intensity for each source in each spectral window separately to subsequently fit the spectral index.

\subsection{Averaging spectral windows}
\label{sec_averaging_spectral_window}
The extraction algorithm and method can influence the resulting catalog, and several different methods are common \citep[e.g.,][]{Williams1994, Hancock2012, Berry2015}. For our data we must solve several challenges: we want to achieve the best signal-to-noise ratio, but avoid picking up artifacts in the images caused by RFI or sidelobes. To get the best signal-to-noise ratio, mosaicking the entire spectral range from 1 to 2\,GHz would be preferable; however, CASA is currently (up to version 4.4) not able to perform wide-band mosaics (see Sect. \ref{sec_imaging_and_deconvolution}), and several spectral windows are severely affected by RFI (see Sect. \ref{sec_rflag}). We therefore cleaned each spectral window separately. To achieve a higher signal-to-noise ratio, we averaged the two spectral windows around 1.4 and 1.8\,GHz, because these spectral windows contain no significant RFI. Prior to the averaging process, we smoothed the spectral window around 1.8\,GHz to the lower resolution of the spectral window around 1.4\,GHz. Averaging over more than the two mentioned spectral windows does not increase the detection of sources significantly, but increases the detection of artifacts due to RFI contamination in the other spectral windows.\\

\subsection{Noise estimate}
\label{sec_noise_estimate}
\begin{figure*}
   \resizebox{\hsize}{!}
            {\includegraphics[width=18cm]{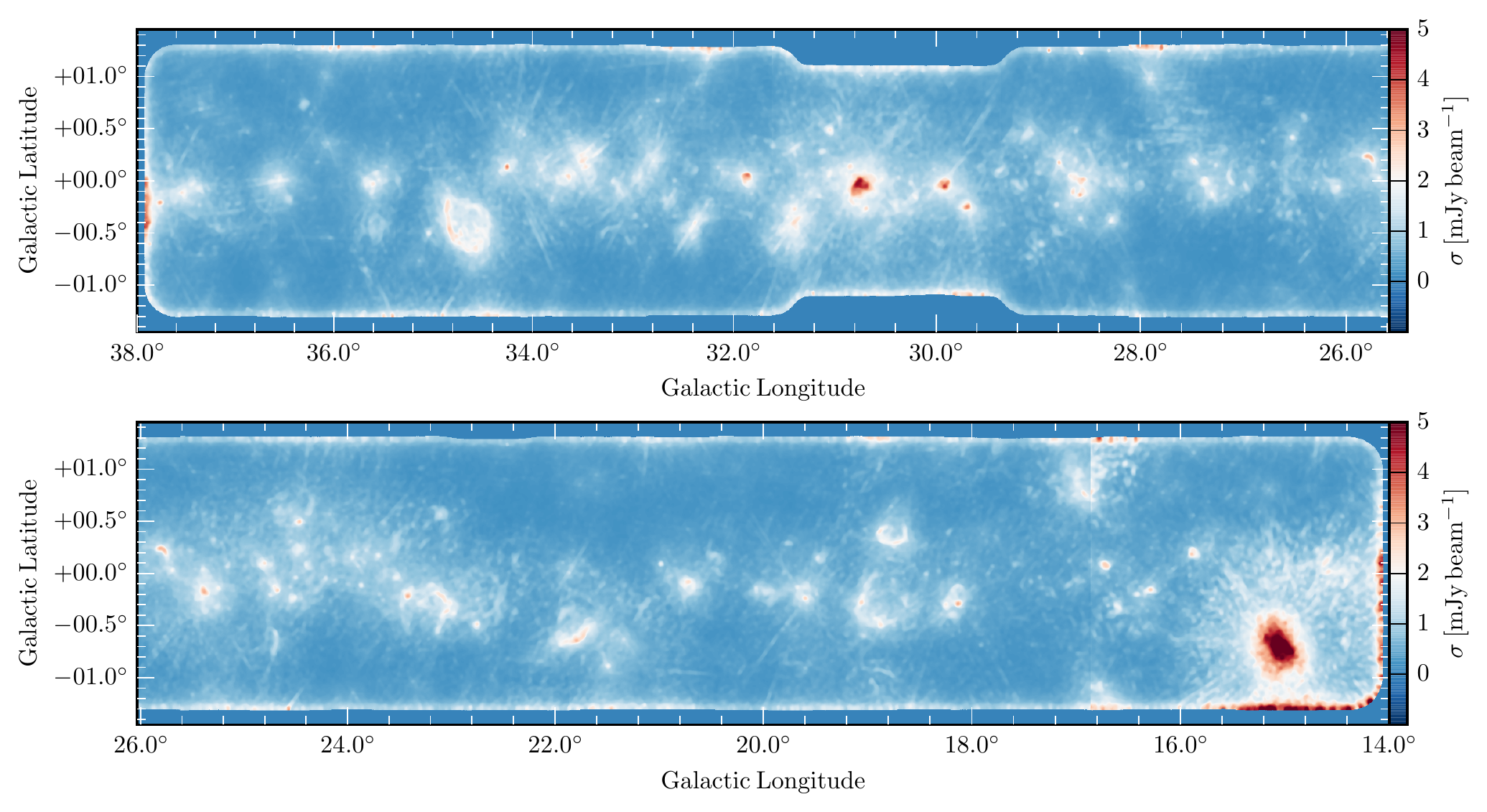}}
      \caption{Noise map of the first part of the THOR survey using the average of two spectral windows 1.4 and 1.8\,GHz (see Sect. \ref{sec_averaging_spectral_window}).}
         \label{Fig_noise_map_large}
\end{figure*}
\begin{figure}
   \centering
   \includegraphics[width=\hsize]{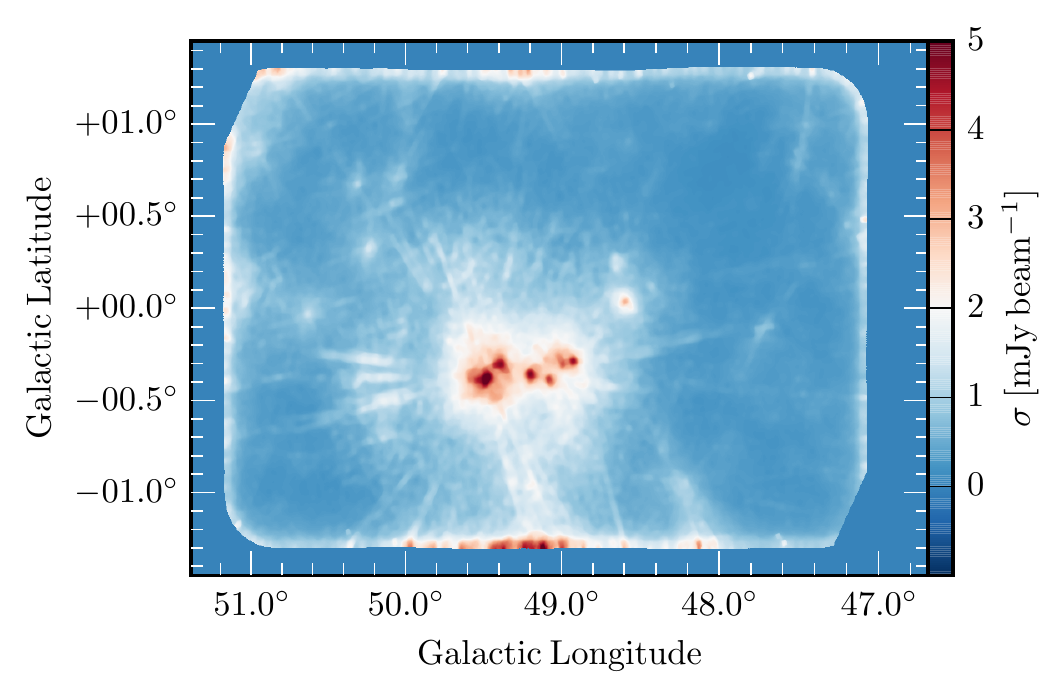}
      \caption{Noise map of the region around l=49$\degr$ of the THOR survey using the average of two spectral windows 1.4 and 1.8\,GHz (see Sect. \ref{sec_averaging_spectral_window}).}
         \label{Fig_noise_map_49deg}
\end{figure}
Since our data are limited by the sidelobe noise, we have to consider strongly varying noise within our observed region. Close to strong emission sources, the noise is dominated by the sidelobes and can be an order of magnitude higher than in emission free regions (see Figs. \ref{Fig_noise_map_large} and \ref{Fig_noise_map_49deg}). As a result, the main challenge is to consider this varying noise during the process of source extraction. To create a reliable noise map, we followed the instructions given in \citet{Hales2012}. The described method determines the rms value for each pixel by determining the median in a specified area (50$\times$50\,px) around the pixel in the residual image from the clean process. Prior to the median estimate, the algorithm clips all peak values in the specified area until all values are within $\pm$3$\sigma$, where $\sigma$ is the median in the specified area. This method ensures that most real emission, which might still be present in the residual image is removed from the noise image and the determined noise map consists of the thermal and sidelobe noise. The noise maps are given in Figs. \ref{Fig_noise_map_large} and \ref{Fig_noise_map_49deg}.

\subsection{BLOBCAT}
\label{sec_blobcat}
We used the BLOBCAT software \citep{Hales2012} to extract the sources from the averaged continuum images. This software is a flood-fill algorithm that considers locally varying noise. BLOBCAT creates a signal-to-noise ratio map by dividing the actual input image by the given noise map. This dimensionless map is used for the source extraction by searching for all pixels above a given detection threshold, which we set to 5$\sigma$. Thereafter, BLOBCAT identifies all neighboring pixels around the peak pixel\ down to a given flooding threshold, which we set to the standard value of 2.6$\sigma$ \citep{Hales2012}. These "islands" of pixels are labeled and written to a table. BLOBCAT also performs several corrections for pixellation errors, peak, and integrated surface brightness biases (see \citet{Hales2012} for further details). Using BLOBCAT, we extracted in total 4772 sources, however, this includes artifacts that we subsequently removed by hand (see Sect. \ref{sec_visual_inspection}).

\subsection{Visual inspection}
\label{sec_visual_inspection}
\begin{figure}
   \centering
   \includegraphics[width=\hsize]{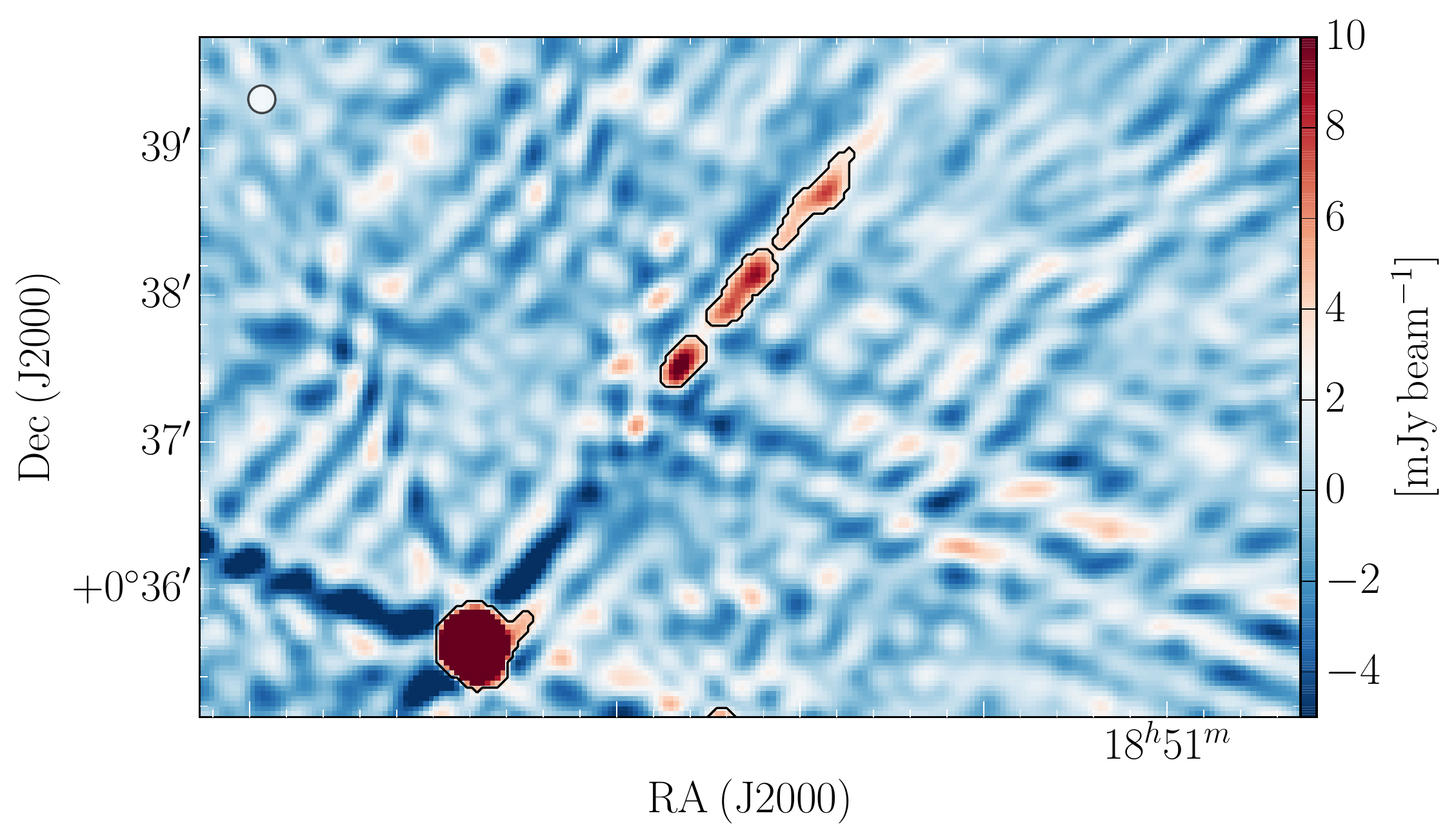}
      \caption{Example of an obvious artifact. A strong emission source, which is located in the bottom left part, creates sidelobes that were identified by the extraction software as actual sources. The black contours show the area of extracted sources identified by the BLOBCAT software.}
         \label{Fig_visual_inspection_artifact}
\end{figure}

Even though we have considered the spatially varying noise during the source extraction process, strong artifacts, especially sidelobes, can be picked up by the BLOBCAT extraction software. Especially problematic is sidelobe contamination from strong sources located just outside our survey boundaries, which cannot be removed by the algorithm. We therefore inspected each source visually and removed obvious artifacts by hand. Figure \ref{Fig_visual_inspection_artifact} shows an example of an obvious sidelobe, which was picked up by the extraction software. We identified 349 sources as obvious artifacts and removed them from the catalog. This leaves 4422 sources in the catalog. Besides the obvious artifacts, it can be difficult to distinguish between artifacts and actual sources for certain extracted sources. We classified these sources as "possible artifacts" and labeled them accordingly in the catalog. Besides visually inspected possible artifacts, we classified and labeled all sources with a signal-to-noise ratio lower than 7$\sigma$ as "possible artifacts". Out of the 4772 extracted sources, we classified 1057 as "possible artifacts", 349 as artifacts, and therefore 3366 sources remain as reliable detections. The following analysis is based on the reliable detections; however, in the catalog, we also present the "possible artifacts".

\subsection{Completeness}
\label{sec_completeness}
\begin{figure}
   \centering
   \includegraphics[width=\hsize]{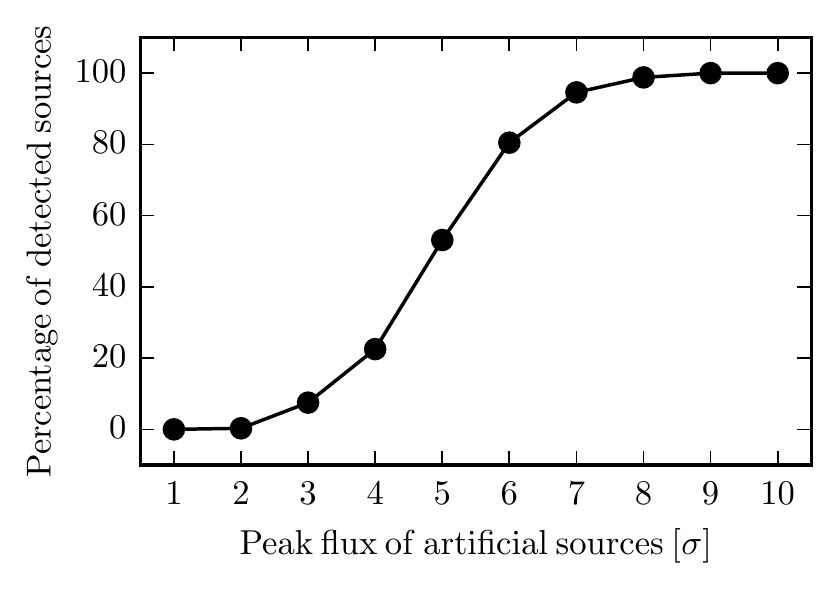}
      \caption{Percentage of detected sources as a function of peak intensity of the added artificial sources in units of the noise level $\sigma$.}
         \label{Fig_fraction_of_detected_sources}
\end{figure}

\begin{figure}
   \centering
   \includegraphics[width=\hsize]{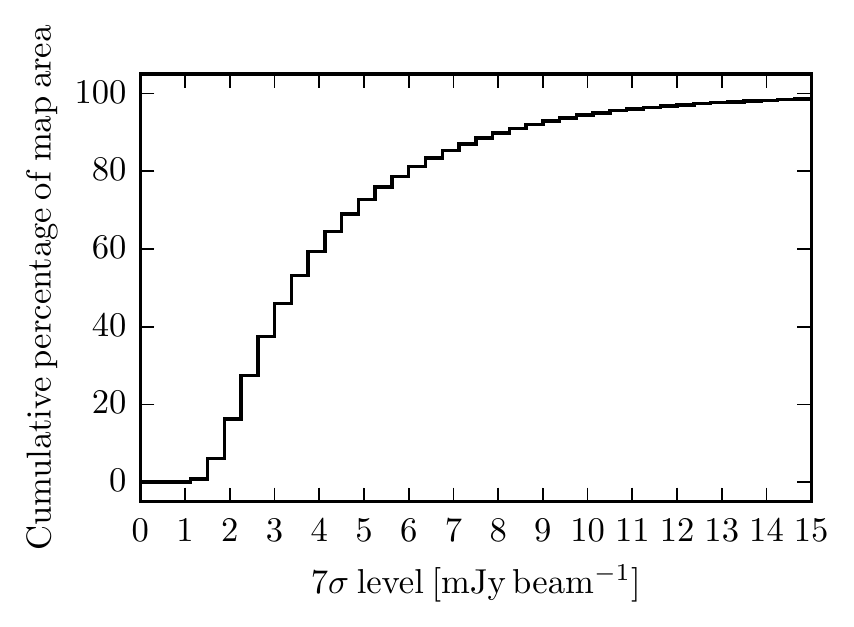}
      \caption{Cumulative percentage of the map area as a function of the corresponding noise level 7$\sigma$ in mJy\,beam$^{-1}$. 50\% of the survey area has a noise level of $7\sigma \sim3$\,mJy\,beam$^{-1}$ or better.}
         \label{Fig_histogram_sensitivity_area}
\end{figure}

As our noise is spatially varying, it is difficult to estimate the completeness of our catalog. In the vicinity of strong extended Galactic sources, it is not possible to detect weak extragalactic sources. Our survey is therefore incomplete in these regions. However, we performed several tests to verify our source extraction method. We chose a region of 0.5\degr$\times$0.5$\degr$ with a constant noise level and added artificial 2D Gaussian sources that have the size of the resolution element and different peak intensities. Using the source extraction method described in Sect. \ref{sec_blobcat}, we extracted these artificial sources and estimated the completeness. The result is shown in Fig. \ref{Fig_fraction_of_detected_sources}. Above the chosen threshold of 7$\sigma$ for reliable sources, we detected 95\% of all sources. Furthermore, we determined the fraction of the area that covers a certain noise level, which is shown in Fig. \ref{Fig_histogram_sensitivity_area}. The lowest noise level in our survey of $7\sigma = 1-2$\,mJy\,beam$^{-1}$, which is dominated by the thermal noise, is achieved in only a small fraction ($\sim$10\%) of the survey area. About half the survey area has a noise level of $7\sigma <3$\,mJy\,beam$^{-1}$, and only 10\% of the survey area shows a noise level of $7\sigma >8$\,mJy\,beam$^{-1}$. Using this information, we can create completeness maps for different sources intensities, which are shown in the appendix in Figs. \ref{Completeness_map_large_2mJy} to \ref{Completeness_map_49deg_10mJy}.

\subsection{Resolved and unresolved sources}
\label{sec_resolved_and_unresolved_sources}
\begin{figure}
   \centering
   \includegraphics[width=\hsize]{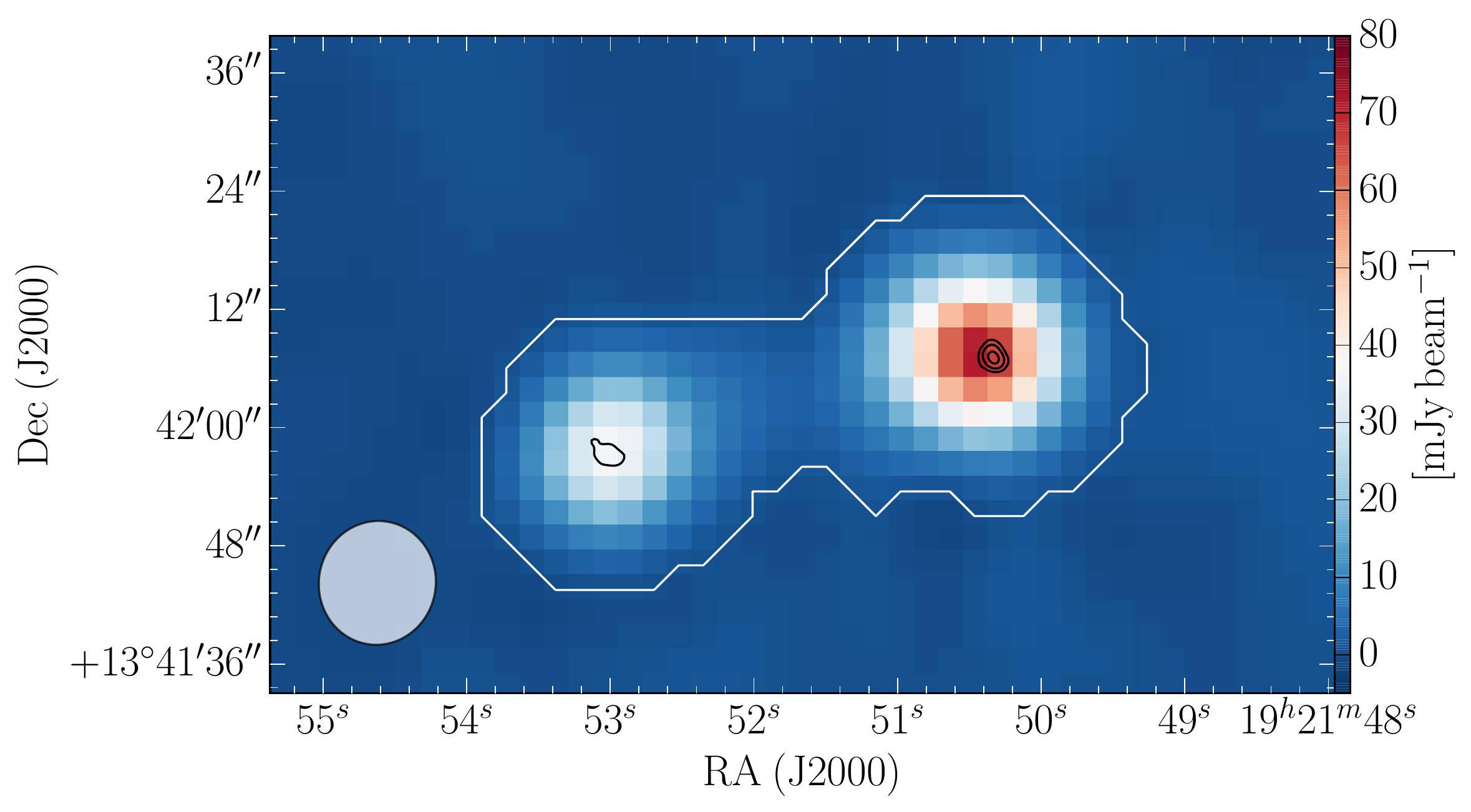}
      \caption{Example of an hourglass-shaped source (G48.561-0.364) that consists of two unresolved sources close together. The white contours represent the area of the source extracted by BLOBCAT. The black contours show observations from CORNISH (see Sect. \ref{sec_comparison_CORNISH}) at 5\,GHz with a resolution of 1.5\arcsec at levels of 2, 5, and 10\,mJy\,beam$^{-1}$. }
         \label{Fig_extra_galactic_radio_lobe}
\end{figure}

As a first classification of the sources, we divide them in two groups: resolved and unresolved sources. The BLOBCAT software provides the number of pixels as an output, but this is not a good measurement to distinguish between resolved and unresolved sources. Because the BLOBCAT software uses a fixed threshold of 2.6$\sigma$ to flood-fill the neighboring pixels around the peak pixel, the number of pixels of a source depends on the corresponding peak intensity. If we use a simple cut based on the number of pixels, we would, on the one hand, misclassify strong unresolved sources as resolved, and on the other, we would misclassify weak but closely resolved sources as unresolved. Therefore we use a comparison of the peak intensity and the flux density to distinguish between resolved and unresolved sources. The peak intensity and flux density have the same value for unresolved sources, whereas resolved sources show a higher flux density value in comparison to the peak intensity value. However, we have to consider the uncertainties in the peak intensity as well as in the flux density, so we use a less strict condition and classify all sources as unresolved sources that have $S_{\nu} < 1.2 \times I_{\nu}$, where $S_{\nu}$ and $I_{\nu}$ are the flux density in Jy and peak intensity in Jy\,beam$^{-1}$, respectively. In our full catalog we classify 3184 sources in total as unresolved and 1238 sources (28\%) as resolved. For the sources with the flag "possible artifacts" (see Sect. \ref{sec_visual_inspection}), the ratio of unresolved and resolved sources is similar with 76\% of the sources being unresolved. \\
This classification scheme classifies two overlapping, but unresolved sources as resolved. For unresolved sources that are randomly distributed in the sky, this arrangement is unlikely, however for extragalactic radio lobes, this overlap can occur frequently. Figure \ref{Fig_extra_galactic_radio_lobe} shows an example of two radio lobes that are close together. Even though each radio lobe is unresolved, we extracted them as one source and hence the flux density is larger than the peak intensity and we classify them as resolved. This affects the classification of extragalactic and Galactic sources. However, in many cases (e.g., Fig \ref{Fig_extra_galactic_radio_lobe}), the spectral index helps to resolve this problem.\\
The flux density of unresolved sources can be affected in several ways and therefore has to be treated cautiously. We find for the ratio of $S_{\nu} / I_{\nu}$ values less than one, which means that the flux density is lower than the peak intensity. For unresolved sources, this ratio should be one. We could identify three reasons for this low ratio. First, the source extraction software BLOBCAT does not fit enough pixels for weak sources, which lowers the flux density. In extreme cases, the fitted area of BLOBCAT can be smaller than the resolution element. Second, unresolved sources can be situated in slightly negative sidelobes from nearby strong extended sources, which affects the flux density, as well as the peak intensity, and this can change the ratio. Third, weak sources ($I_{\nu} \lesssim 5$\,mJy\,beam$^{-1}$) are not cleaned properly because these sources are below our cleaning threshold, which lowers the measured flux density and changes the ratio of $S_{\nu} / I_{\nu}$ to values below one. We therefore suggest using peak intensities for unresolved sources for further analysis, and we indicate the corresponding flux densities within our catalog with brackets.

\subsection{Spectral index determination}
\label{sec_spectral_index_determination}
\begin{figure}
   \centering
   \includegraphics[width=\hsize]{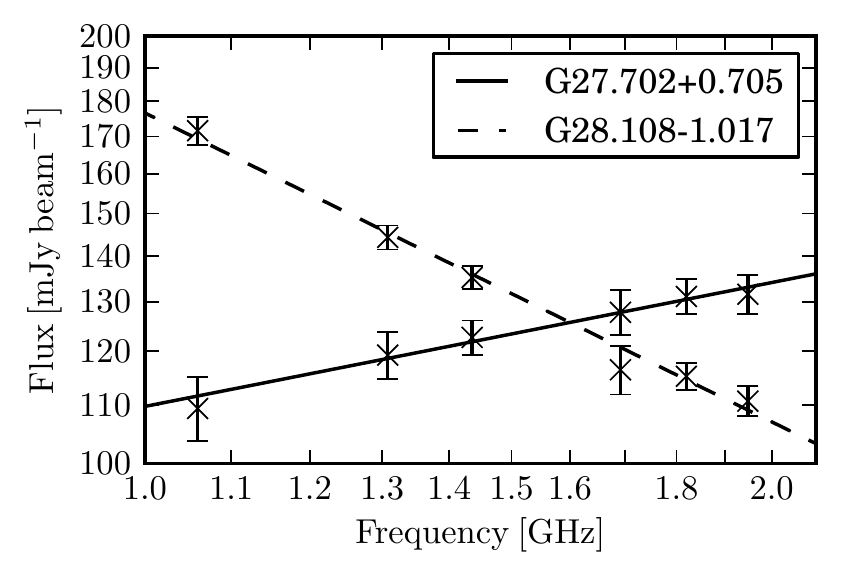}
      \caption{Example of the extracted peak intensity as a function of frequency. Each spectral window is represented by one data point including the 3$\sigma$ uncertainty. G27.702+0.705 is represented by the solid line and has a spectral index of $\alpha = 0.29 \pm 0.03,$ whereas G28.108-1.017 is represented by the dashed line and has a spectral index of $\alpha = -0.72 \pm 0.02$. }
         \label{Fig_spectral_index_good_sources}
\end{figure}

\begin{figure}
   \centering
   \includegraphics[width=\hsize]{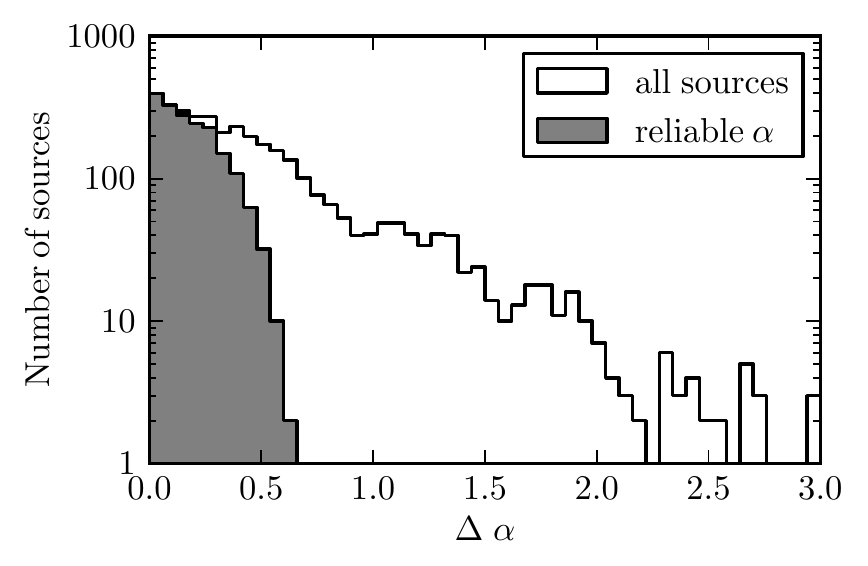}
      \caption{Histogram of the uncertainty of the determined spectral index. The black line includes all sources for which we are able to determine a spectral index, whereas the gray shaded area represents the sources for which we have an intensity measurement in all six spectral windows and therefore a reliable spectral index measurement.}
         \label{Fig_spectral_index_uncertainty_histogram}
\end{figure}

As our observations cover a wide bandwidth from 1 to 2\,GHz, we are able to determine spectral indices by extracting the peak intensity of each source within each spectral window and perform a fit of the spectral index $\alpha$ with the form $I(\nu) \varpropto \nu^{\alpha}$. As explained in Sect. \ref{sec_extended_sources}, we use the peak intensity instead of the integrated flux density to determine the spectral index, since the peak intensity is less affected by filtering effects for extended sources. For unresolved sources, both quantities reveal the same result. To overcome problems due to different resolutions, we smooth all spectral windows to a common resolution of $25\arcsec$ prior to extracting the peak intensity. Furthermore, we use the same technique as described in Sect. \ref{sec_noise_estimate} to determine the spatially varying noise and to estimate the noise within each spectral window. Because we smooth two spectral windows to perform the source extraction, the signal-to-noise ratio is higher for the source extraction in comparison to the intensity extraction within each spectral window separately. We therefore use a less rigid threshold for the intensity extraction of 3 sigma in comparison to 5 sigma for the source extraction. The extracted peak intensities for each spectral window are given in the catalog presented in this paper. Figure \ref{Fig_spectral_index_good_sources} shows an example for the extracted intensities, including the fit of the spectral index for two different sources. In the appendix in Figs. \ref{Fig_source_example_1} to \ref{Fig_source_example_3}, we present three example sources showing the images of all spectral windows that include the spectral index fit. We use the scipy function "curve\_fit" to fit the data points and use the uncertainty of the fit as the uncertainty for the spectral index. With this method, we can determine a spectral index for 3625 sources.\\
For some sources, we are not able to extract the peak intensity for all six spectral windows, owing to higher noise or contamination by RFI, for example. In such cases, we determine the spectral slope from the remaining data points. Naturally, this leads to larger uncertainties. As a result, we introduce the label "reliable spectral index" for all sources that have a reliable intensity for all six spectral windows, hence a reliable spectral index fit. The catalog contains 1840 sources that fulfill this criterion, which is about 50\% of the sources where it is possible to determine a spectral index. Figure \ref{Fig_spectral_index_uncertainty_histogram} shows the distribution of the uncertainty of the determined spectral index for all sources and for the sources with the label "reliable spectral index". The labeled sources show a significantly smaller uncertainty with a mean of $\Delta \alpha = 0.18$, whereas all sources show a mean uncertainty of the spectral index of $\Delta \alpha = 0.62$. In the following, we concentrate our analysis of the spectral index on the sources with reliable spectral indices.\\

\section{Catalog}
\begin{figure}
   \centering
   \includegraphics[width=\hsize]{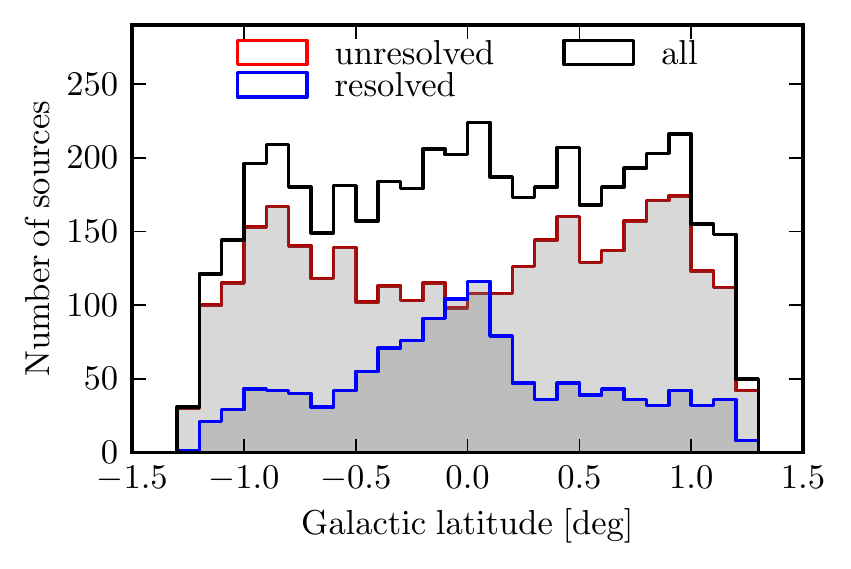}
      \caption{Histogram for the number of sources as a function of Galactic latitude.}
         \label{Fig_gal_lat_histogram}
\end{figure}

The published catalog contains 27 entries for each source. Table \ref{table_catalog_entries_description} describes each entry in detail. As explained, we use a smaller beam for the source extraction than for the intensity extraction. This makes the published values for the corresponding peak intensities different. Table \ref{table_catalog_example} shows an example, and Figs. \ref{Fig_source_example_1} - \ref{Fig_source_example_3} present three example sources showing all the different data we used, including the spectral index fit.\\
\begin{table*}
\caption{Description of the catalog entries.}             
\label{table_catalog_entries_description}      
\centering          
\begin{tabular}{c l l l  }
\hline\hline       
Col. Num. & Name & Unit  & Description\\
\hline
1 & Gal. ID & & Name of the source the form G`Gal. long'$\pm$`Gal. latitude'\tablefootmark{a}.\\
2 & RA & deg & RA in J2000 of the peak position.\\
3 & Dec & deg &Dec in J2000 of the peak position.\\
4 & S\_p \tablefootmark{b}  & Jy beam$^{-1}$ & Peak intensity of aver. image used for source extraction (see Sect. \ref{sec_averaging_spectral_window}).\\
5 & SNR & & Signal-to-noise ratio in the averaged image.\\
6 & S\_int &  Jy & Integrated flux density of the averaged image (see Sect. \ref{sec_blobcat}).\\
7 & BMAJ & arcsec & Major axis of the resolution element used for the source extraction.\\
8 & BMIN & arcsec & Minor axis of the resolution element used for the source extraction.\\
9 & BPA & deg & Rotation angle of the resolution element used for the source extraction.\\
10 & n\_pix & & Number of pixels flooded by BLOBCAT (see Sect. \ref{sec_blobcat}).\\ 
11 & resolved\_source & & Resolved source label (see Sec \ref{sec_resolved_and_unresolved_sources}). 1 = True, 0 = False.\\
12 & possible\_artifact & & Label for possible artifacts and/or SNR<7. 1 = True, 0 = False.\\ 
13 & S\_p(spw-1060) \tablefootmark{c}  &  Jy beam$^{-1}$   & Peak intensity around 1.06\,GHz used for spectral index (see Sect. \ref{sec_spectral_index_determination}).\\
14 & delta\_S\_p(spw-1060) \tablefootmark{c}  &  Jy beam$^{-1}$  & Uncertainty of peak intensity around 1.06\,GHz.\\ 
15 & S\_p(spw-1310)  \tablefootmark{c} &  Jy beam$^{-1}$  & Peak intensity around 1.31\,GHz used for spectral index (see Sect. \ref{sec_spectral_index_determination}).\\
16 & delta\_S\_p(spw-1310) \tablefootmark{c}  &  Jy beam$^{-1}$  & Uncertainty of peak intensity around 1.31\,GHz.\\
17 & S\_p(spw-1440) \tablefootmark{c}  &  Jy beam$^{-1}$  & Peak intensity around 1.44\,GHz used for spectral index (see Sect. \ref{sec_spectral_index_determination}).\\
18 & delta\_S\_p(spw-1440) \tablefootmark{c} &  Jy beam$^{-1}$  & Uncertainty of peak intensity around 1.44\,GHz.\\
19 & S\_p(spw-1690) \tablefootmark{c}  &  Jy beam$^{-1}$  & Peak intensity around 1.69\,GHz used for spectral index (see Sect. \ref{sec_spectral_index_determination}).\\
20 & delta\_S\_p(spw-1690) \tablefootmark{c}  &  Jy beam$^{-1}$  & Uncertainty of peak intensity around 1.69\,GHz.\\
21 & S\_p(spw-1820) \tablefootmark{c}  &  Jy beam$^{-1}$  & Peak intensity around 1.82\,GHz used for spectral index (see Sect. \ref{sec_spectral_index_determination}).\\
22 & delta\_S\_p(spw-1820) \tablefootmark{c}  &  Jy beam$^{-1}$  & Uncertainty of peak intensity around 1.82\,GHz.\\
23 & S\_p(spw-1950) \tablefootmark{c}  &  Jy beam$^{-1}$  & Peak intensity around 1.95\,GHz used for spectral index (see Sect. \ref{sec_spectral_index_determination}).\\
24 & delta\_S\_p(spw-1950) \tablefootmark{c}  &  Jy beam$^{-1}$  & Uncertainty of peak intensity around 1.95\,GHz.\\
25 & alpha & & Spectral index of source using all available data points (see Sect.\ref{sec_spectral_index_determination}).\\
26 & delta\_alpha & & Uncertainty of spectral index.\\
27 & reliable\_alpha & & Label for reliable spectral index (see Sect. \ref{sec_spectral_index_determination}). 1 = True, 0 = False.\\
\hline        
\end{tabular}
\tablefoot{
\tablefoottext{a}{Indicating the peak position.}\\
\tablefoottext{b}{Synthesized beam is different for different fields and is given in rows 7-9.}\\
\tablefoottext{c}{Synthesized beam is smoothed to 25\arcsec$\times$25\arcsec.}
}          
\end{table*}
Table \ref{table_simple_statistics} summarizes the number of extracted sources, including the introduced labels. The exact numbers have to be treated cautiously. Compact sources superimposed on large regions of extended emission are missed in the catalog. In contrast to this, large, extended sources, such as SNRs, can be split up in different sources and therefore create multiple entries in our catalog, even though the emission occurs most likely from the same object. The majority (72\%) of the extracted sources are not resolved. Most of them might be extragalactic in origin since their spectral indices are negative (see Sect. \ref{sec_spectral_index}). About 28\% of the extracted sources are classified as resolved, but as explained in Sect. \ref{sec_resolved_and_unresolved_sources}, some of them might be two closely separated sources. The distribution of resolved and unresolved sources as a function of Galactic latitude is shown in Fig. \ref{Fig_gal_lat_histogram}. This reveals an over-density of resolved sources close to the Galactic midplane, whereas unresolved sources are equally distributed. The distribution drops for $|b|>1\degr$ as the noise increases at the survey edges. The distribution of the unresolved sources also indicates a slight drop toward the Galactic plane ($b\sim 0$\degr) because we miss weak extragalactic sources in the close vicinity of strong Galactic sources, which are mostly located along the Galactic plane. Similar results can be found in \citet{Helfand2006}. This shows that a large number of the sources in our catalog are not confined to the Galactic plane and therefore have an extragalactic origin.
\begin{table}
\caption{Statistics of the catalog}            
\label{table_simple_statistics}      
\centering                          
\begin{tabular}{l c c}       
\hline\hline                 
Description & Number  & Percentage  \\    
\hline                       
All & 4422 & 100\%\\
Unresolved sources & 3184 & 72\%\\
Resolved sources & 1238 & 28\%\\
Possible artifacts & 1057 & 24\%\\
Reliable alpha & 1840 & 41\%\\
\hline 
\end{tabular}
\end{table}

\section{Discussion}

\subsection{Comparison with other surveys}
Since the THOR survey is not the first cm-continuum survey in the Galactic plane, we compare our results to previous surveys to check for consistency in the flux density, intensity, and position. We focus our comparison on three major surveys: the Multi-Array Galactic Plane Imaging Survey \citep[MAGPIS,][]{Helfand2006}, The NRAO VLA Sky Survey \citep[NVSS,][]{Condon1998}, and the Co-Ordinated Radio `N' Infrared Survey for High-mass star formation \citep[CORNISH,][]{Hoare2012,Purcell2013}. 
\subsection{MAGPIS}
\label{sec_comparison_MAGPIS}
\begin{figure}
   \centering
   \includegraphics[width=\hsize]{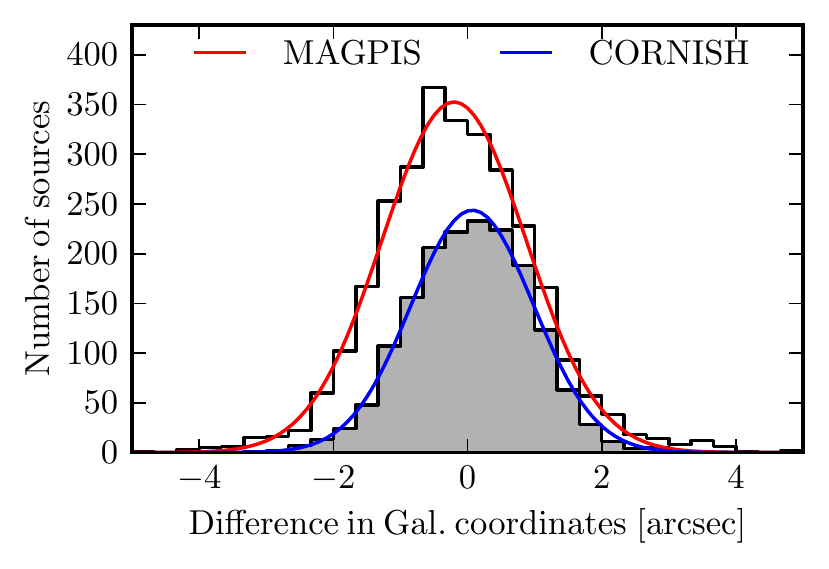}
      \caption{Histogram for the number of sources as a function of the difference in Galactic coordinates of the peak position for the matched sources with the MAGPIS and CORNISH catalogs in red and blue, respectively. The black histogram represents the actual data, whereas the colored lines show the corresponding fits.}
         \label{Fig_gal_coordinates_comparison}
\end{figure}

\begin{figure}
   \centering
   \includegraphics[width=\hsize]{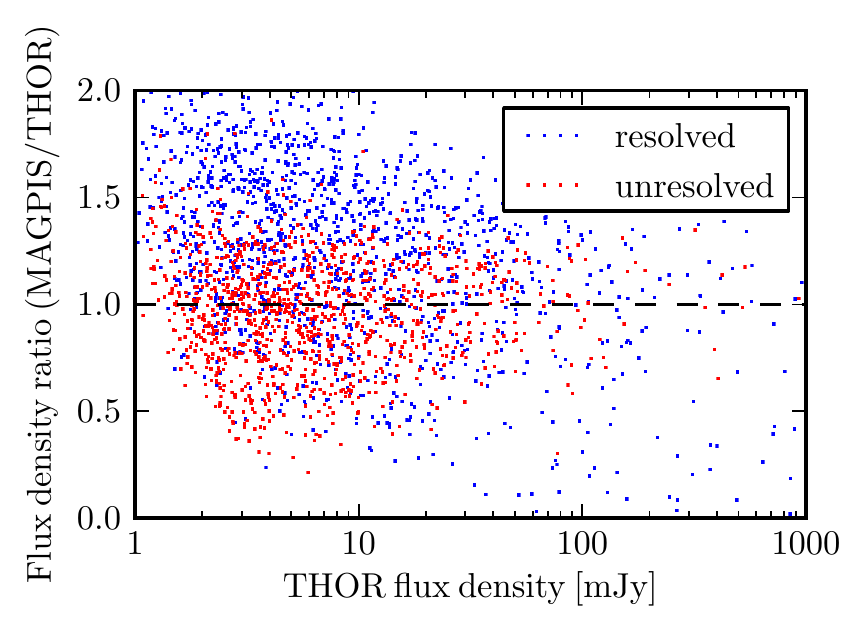}
      \caption{Ratio of the MAGPIS and THOR flux density as a function of the THOR flux density. The red and blue points represent the unresolved and resolved sources, respectively, as defined in Sect. \ref{sec_resolved_and_unresolved_sources}. The dashed black line represents a one-to-one relation.}
         \label{Fig_flux_comparison_magpis}
\end{figure}

The MAGPIS survey \citep{Helfand2006} used the VLA in D, C, and B configurations to map the Galactic plane in the region $5\degr < l < 48\degr$ and $|b|<0.8$\degr with two continuum bands at 1365 and 1435\,MHz, achieving a resolution of $\sim$6$\arcsec$ and a sensitivity limit of 1-2\,mJy, depending on neighboring bright extended emission. They cataloged 3000 discrete sources in the region $5\degr < l < 32\degr$ with diameters less than 30$\arcsec$ and 400 diffuse sources. Within the overlap region of the THOR survey ($14.2\degr < l < 32\degr$, $|b|<0.8$\degr), the MAGPIS catalog contains 2256 discrete and 290 extended sources. The THOR continuum catalog contains 1848 sources in the same area, including possible artifacts and therefore fewer sources than the MAGPIS catalog. Using a best match method and a circular matching threshold of 20\arcsec, we match 1568 sources in total. Choosing a smaller matching threshold of 5$\arcsec$ does not change the result significantly. Owing to different spatial filtering of the THOR and MAGPIS data, the determined area for extended sources is different within the two surveys. This effect accounts for the majority of the non-matches. Merely matching the point sources of the THOR survey reveals a matching rate of $\sim$92\%, including possible artifacts. If we do not consider the possible artifacts, the matching rate is even higher with $\sim$97\% and the matching rate considering only the possible artifacts is $\sim$78\%. This shows that almost all reliable sources within the THOR catalog have a counterpart in the MAGPIS catalog, and therefore the number of false positives due to artifacts or sidelobes is low within our catalog. Since the matching rate for possible artifacts is still high, the majority of these sources will also be real detections.\\
Because the matching with the MAGPIS survey worked well, we used the matched sources to verify the positions, as well as the flux density. For these comparisons we employed the MAGPIS discrete source catalog and neglected the diffuse sources because they suffer from different spatial filtering, which makes the comparison inaccurate. Figure \ref{Fig_gal_coordinates_comparison} shows the histogram of the difference in Galactic coordinates for the peak position, along with the corresponding fits. We used a Gaussian function to fit the distribution and find a shift of -0.2$\arcsec$ and a FWHM for the distributions of 2.5$\arcsec$, which is the size of one pixel. The comparison of the flux density is shown in Fig. \ref{Fig_flux_comparison_magpis}. Similar to the NVSS sources, the unresolved sources show a tight correlation. In contrast to this, the resolved sources show higher flux density values in the MAGPIS data, owing to less filtering. These tests show that our observation, calibration, and imaging processes work well, and our work is consistent with previous observations. 


\subsection{NVSS}
\begin{figure}
   \centering
   \includegraphics[width=\hsize]{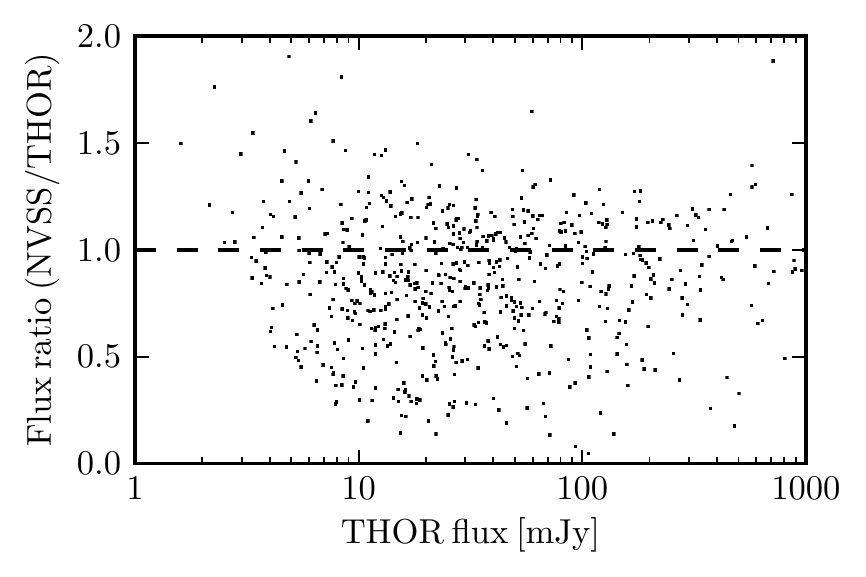}
      \caption{Flux density comparison with the NVSS data. The ratio of NVSS  and THOR flux density is shown as a function of the THOR flux density for all matched sources with a matching threshold of 5\arcsec for the peak positions. The black dashed line represents a one-to-one relation.}
         \label{Fig_flux_comparison_nvss}
\end{figure}

The NVSS \citep{Condon1998} is a continuum survey at 1.4\,GHz with the VLA in D and DnC configuration covering the northern sky for J2000 $\delta>-40$\degr. The catalog contains $\sim 2\times10^6$ sources with a sensitivity limit of $\sim$2.5\,mJy and a resolution of 45\arcsec. Within the region of the THOR continuum catalog, the NVSS catalog contains 7587 sources and therefore almost twice as many sources as our catalog. We find a match of 1351 sources for a circular matching threshold of 20$\arcsec$ and only 657 for a circular matching threshold of 5\arcsec for the peak position. Further analysis of the NVSS images showed that the NVSS catalog is severely contaminated with obvious false detections due to strong sidelobes from sources close to the Galactic plane or due to ghost artifacts \citep{Grobler2014}. Therefore the matching process is not reliable for large matching radius as we match THOR sources with false positives in the NVSS catalog. To overcome this problem, we only compare the measured flux densities for all matched sources with a tight matching radius of 5\arcsec. The result is shown in Fig. \ref{Fig_flux_comparison_nvss}. For these tightly matched sources, the flux density comparison shows a good correlation over three orders of magnitude, with a slight bias. As shown in Fig. \ref{Fig_flux_comparison_nvss}, the THOR flux density values are slightly higher than the NVSS flux density values. This bias is visible for resolved and unresolved sources so is not a filtering effect. We do not have a good explanation for this bias. However, the THOR and MAGPIS flux densities are consistent (see Sect.\,\ref{sec_comparison_MAGPIS}), and we report a slight inconsistency with the NVSS flux densities. Since the matching with the NVSS catalog is difficult due to artifacts in the NVSS images, we refrain from comparing the peak positions of the sources, but we perform this comparison with MAGPIS and CORNISH.

\subsection{CORNISH}
\label{sec_comparison_CORNISH}
CORNISH \citep{Hoare2012,Purcell2013} is a Galactic plane survey from $10\degr < l < 65\degr$ and $|b|<1\degr$ using the VLA in B and BnA configuration at a frequency of 5\,GHz. Therefore, the resulting resolution of 1.5$\arcsec$ is higher than the THOR observations, but objects larger than $\sim$14$\arcsec$ are filtered out. The mean noise level is $\sim$0.4\,mJy\,beam$^{-1}$ and 3062 sources are detected above a 7$\sigma$ limit. Within the THOR region, CORNISH includes 1367 reliable sources. We used all THOR sources, which we classify as unresolved to match the CORNISH sources, and we find a best match of 834 sources using a circular matching threshold of 20$\arcsec$ for the peak position. As the frequency and filtering is different, we refrained from comparing the flux densities, but we verified the peak positions. Figure \ref{Fig_gal_coordinates_comparison} shows the difference in Galactic coordinates for the matched sources. Similar to the comparison with the MAGPIS survey (see Sect. \ref{sec_comparison_MAGPIS}), we do not detect a significant shift or offset in the distribution, and the corresponding Gaussian fit has a shift of 0.07$\arcsec$ and a FWHM of 2.3$\arcsec$. The small position offset between the sources in the MAGPIS and CORNISH surveys with the THOR survey show that our data do not suffer significantly from systematical uncertainties for the position and the uncertainty of the position depends on the synthesized beam and the signal-to-noise ratio for each source and is better than 2$\arcsec$. 

\subsection{Spectral index}
\label{sec_spectral_index}
\begin{figure}
   \centering
   \includegraphics[width=\hsize]{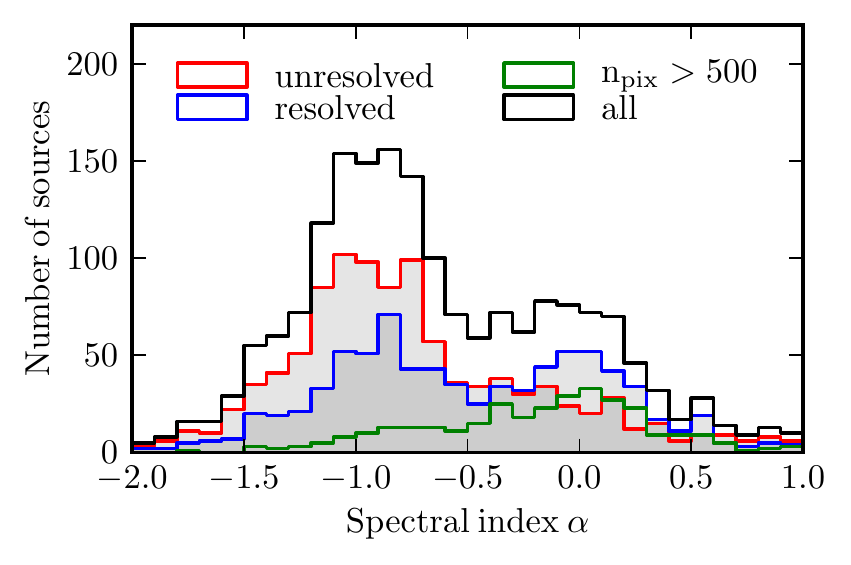}
      \caption{Histogram of the spectral index for all sources with a reliable spectral index measurement ($\sim$1800 sources). The black histogram represents all sources, whereas the red and blue histogram represents unresolved and resolved sources. The green histogram shows the spectral index of all sources that are larger than 500\,pixels.}
         \label{Fig_histogram_spectral_index}
\end{figure}

As outlined in the introduction, the spectral index allows us to distinguish various physical processes. In Sect. \ref{sec_spectral_index_determination}, we determine a reliable spectral index for $\sim$1800 sources. This information helps to distinguish between thermal and non-thermal radiation, showing positive and negative spectral indices, respectively. Figure \ref{Fig_histogram_spectral_index} shows the distribution for the spectral index. Considering all sources (black line in Fig. \ref{Fig_histogram_spectral_index}), we find a prominent peak around $\alpha \sim -1$ and a second weaker peak around $\alpha \sim 0$. Considering only the unresolved sources (red line in Fig. \ref{Fig_histogram_spectral_index}), we recover the strong peak around $\alpha \sim -1$, whereas the second peak around $\alpha \sim 0$ decreases. As a result, most of the unresolved sources show a negative spectral index that indicates non-thermal synchrotron radiation. Therefore, we classify the unresolved sources with a negative spectral index of $\alpha < -0.2$ as extragalactic sources. In contrast to the unresolved sources, the resolved sources (blue line in Fig. \ref{Fig_histogram_spectral_index}) show two peaks at $\alpha \sim -1$ and $\alpha \sim 0$. Most of the resolved sources with a flat spectral index can be matched with Galactic \ion{H}{ii} regions (see Sect. \ref{sec_HII_region}). The resolved sources with negative spectral index are mainly radio lobes from extragalactic jets that were classified as resolved sources, but might be two overlapping unresolved sources (see Sect. \ref{sec_resolved_and_unresolved_sources}). If we consider only the largest sources in our sample with more than 500 pixels, which corresponds to an effective radius of $\sim32\arcsec$ (green line in Fig. \ref{Fig_histogram_spectral_index}), we find a broad distribution from $\alpha \sim -1$ to $0.5$. The sources with flat spectral index can be classified as \ion{H}{ii} region, and the sources with negative spectral index can be SNR. We explore the \ion{H}{ii} regions and SNR in more detail in Sects. \ref{sec_HII_region} and \ref{sec_SNR}, respectively.

\subsection{\ion{H}{ii} regions}
\label{sec_HII_region}

\begin{figure}
   \centering
   \includegraphics[width=\hsize]{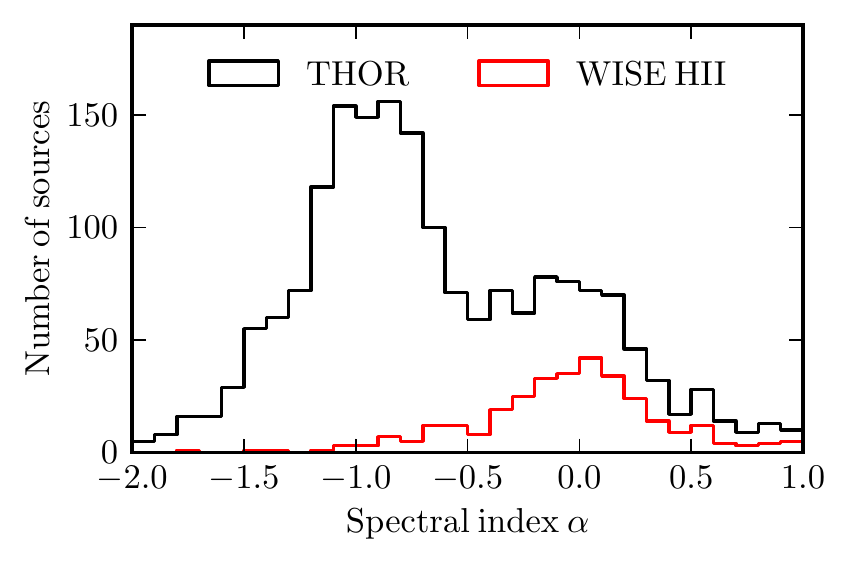}
      \caption{Histogram of the spectral index determined with the THOR data as explained in Sect.\ref{sec_spectral_index_determination}. The red and black lines represent the matched WISE \ion{H}{ii} sources and all THOR continuum sources that reveal a reliable spectral index, respectively.}
         \label{Fig_histogram_spectral_index_HII_regions}
\end{figure}

\begin{figure}
   \centering
   \includegraphics[width=\hsize]{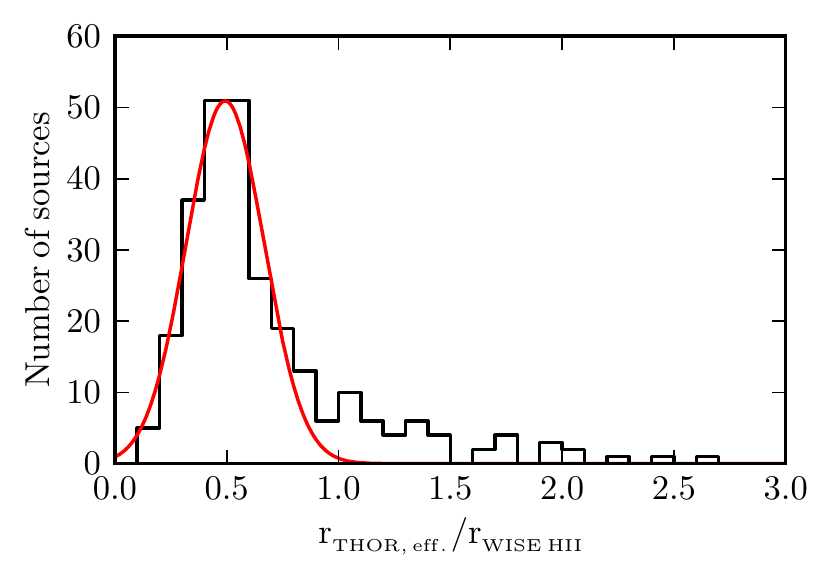}
      \caption{Radius of the \ion{H}{ii} regions measured by the THOR data compared to the mid-infrared WISE data given by \citet{Anderson2014}. The red line represents a Gaussian fit.}
         \label{Fig_histogram_size_comparison_HII_regions}
\end{figure}

Since \ion{H}{ii} regions are formed by OB stars, they are ideal objects to locate high-mass star formation. \citet{Anderson2014} present the most complete catalog of these objects, using mid-infrared observations from the Wide-Field Infrared Survey Explorer (WISE) satellite \citep{Wright2010}. They detected $\sim$8000 sources within the Milky Way.  Approximately 2000 of these sources are \ion{H}{ii} regions with measured velocities from ionized gas spectroscopy, whereas the remaining 6000 are \ion{H}{ii} region candidates that either show ($\sim$2000 objects) or do not show ($\sim$4000 objects) radio continuum emission. The mid-infrared sizes range from 10$\arcsec$ to more than 20$\arcmin$ with a mean of $\sim$100$\arcsec$ for all previously known \ion{H}{ii} regions. The wide range of sizes for the \ion{H}{ii} regions makes it challenging to match them with our continuum catalog. Within a single large \ion{H}{ii} region, we usually detect several extragalactic background sources, which are not related to the \ion{H}{ii} region. A visual inspection of all sources is very time consuming. However, we visually inspected 6\,deg$^2$ ($\sim$10\%) of the THOR region and used this result to test several automated matching methods. For the automated matching, the best result was achieved if we exclude large \ion{H}{ii} regions from WISE with $r>150\arcsec$ and use only the reliable THOR continuum sources. As a matching threshold, we used the size of the \ion{H}{ii} region. This method could recover over 90\% of the visually inspected sources with less than 10\% false detections. Within the THOR region, the WISE \ion{H}{ii} region catalog \citep{Anderson2014} contains 791 sources that show radio emission and are smaller than $r<150\arcsec$, including known \ion{H}{ii} regions, as well as \ion{H}{ii} region candidates. Using the described matching threshold, we match 388 sources. \\
As \ion{H}{ii} regions exhibit thermal radio emission, we expect a flat or positive spectral index, depending on the optical depth. Out of the 388 matched sources, 326 show a reliable spectral index. Figure \ref{Fig_histogram_spectral_index_HII_regions} presents the distribution of the reliable spectral index for all matched sources (red histogram) in comparison to the entire THOR continuum catalog (black histogram). As expected, we find a single peak around zero, which confirms the thermal origin of the radiation for these sources. About 80\% of the matched WISE \ion{H}{ii} regions are resolved, which is a significantly larger percentage than for the entire set of THOR continuum sources ($\sim$30\%). For the resolved sources, we can compare the sizes of the \ion{H}{ii} regions measured in mid-infrared emission by the size measured in the radio emission. For the THOR sources we can estimate an effective radius for the area determined with the BLOBCAT extraction algorithm. The mid-infrared emission at 12\,$\rm{\mu}$m traces the photo-dissociation region at the outer edge of the \ion{H}{ii} regions, whereas the radio emission traces the enclosed ionized gas in the interior of the \ion{H}{ii} regions. This can be seen for several \ion{H}{ii} regions presented by \citet{Watson2008}. We therefore expect that the ratio of the THOR radius divided by the WISE radius is less than one. The result of this comparison is shown in Fig. \ref{Fig_histogram_size_comparison_HII_regions}. We find a close correlation around 0.5. This close correlation has to be treated cautiously, as the comparison suffers from systematic uncertainties due to the different methods of the size determination. The radius for the WISE \ion{H}{ii} regions is determined visually with circles, whereas the radius of the THOR sources is an effective radius of the extracted source area.

\subsection{Supernova remnants }
\label{sec_SNR}
\begin{figure}
   \centering
   \includegraphics[width=\hsize]{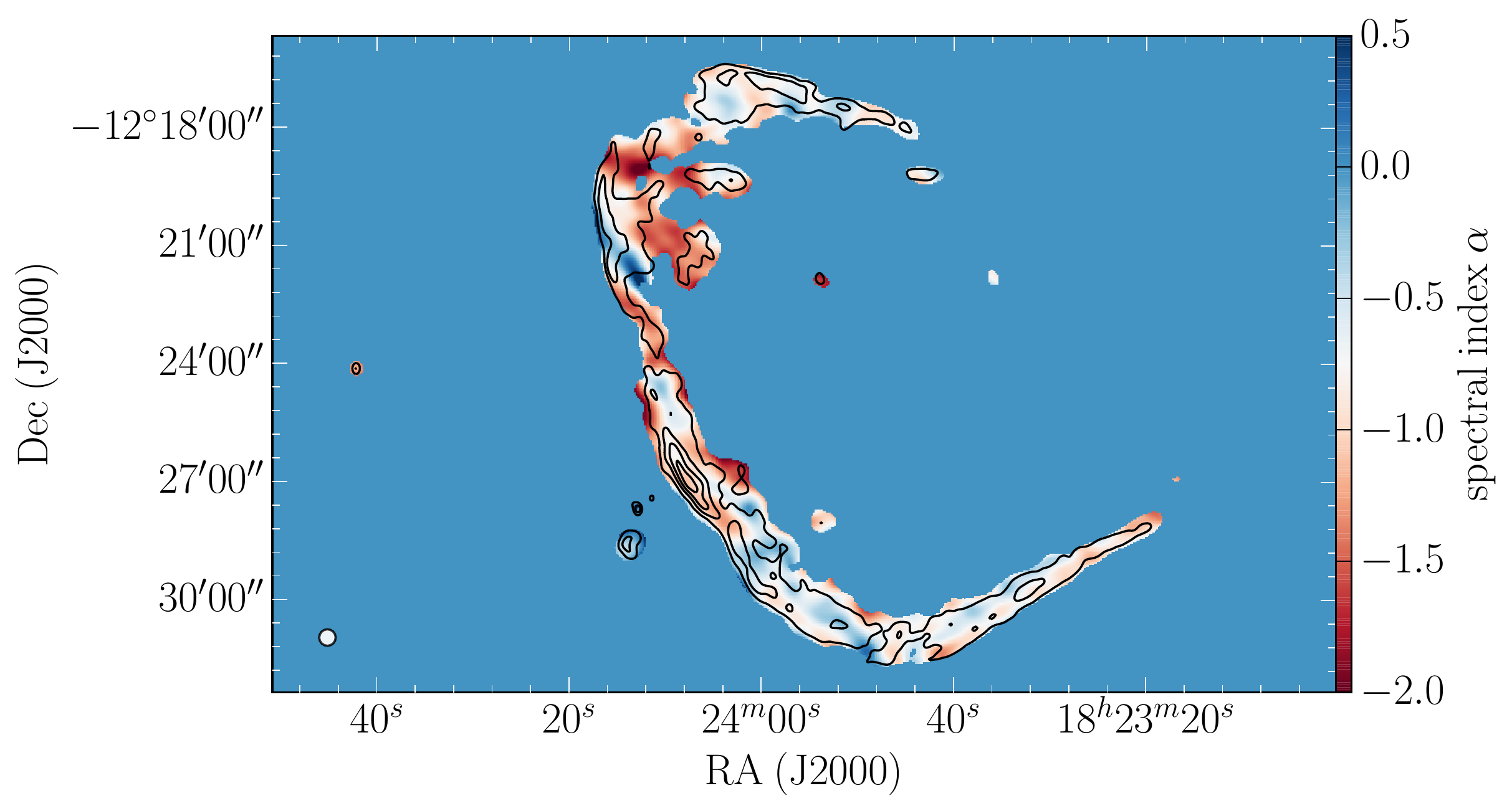}
      \caption{Spectral index $\alpha$ of the SNR G018.8+00.3 (THOR source G18.761+0.287). The black contours represent the continuum emission in steps of 10, 20, 30, and 40\,mJy\,beam$^{-1}$. }
         \label{Fig_spectral_index_SNR_18deg}
\end{figure}
\begin{figure*}
  \includegraphics[width=6cm]{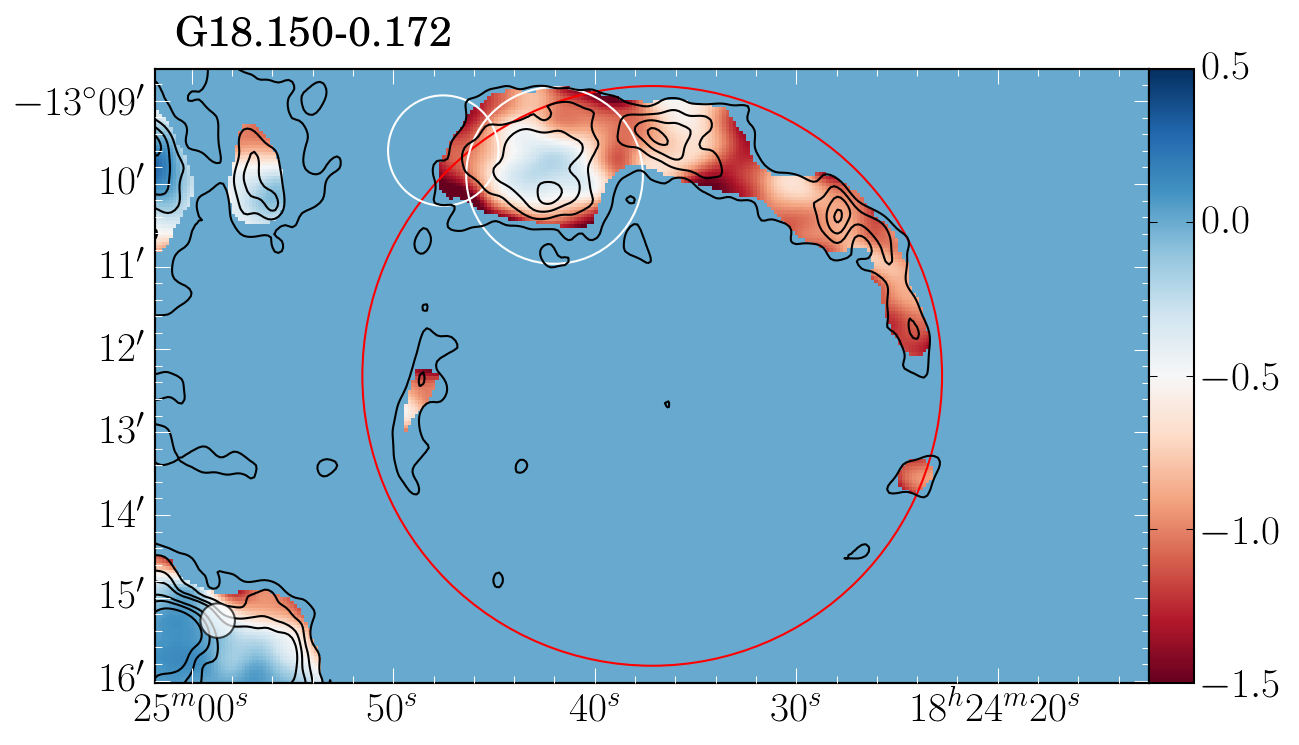}
  \includegraphics[width=6cm]{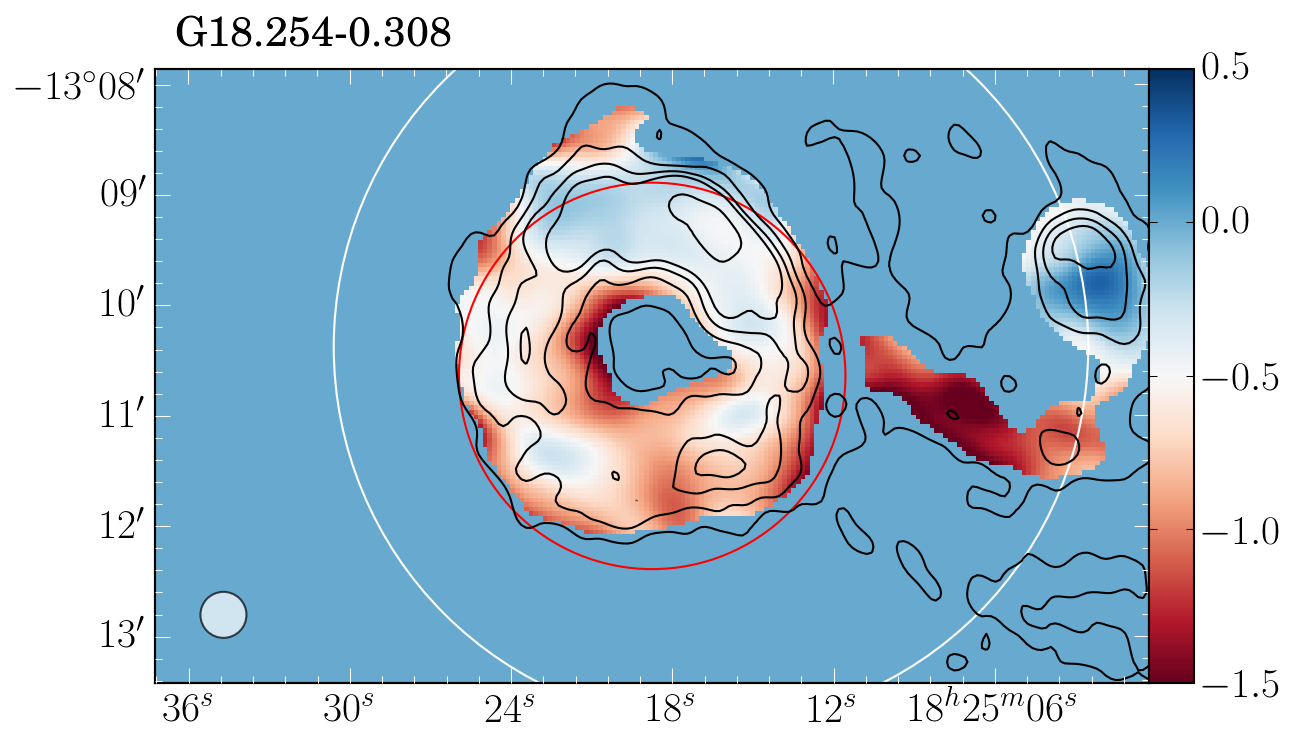}
  \includegraphics[width=6cm]{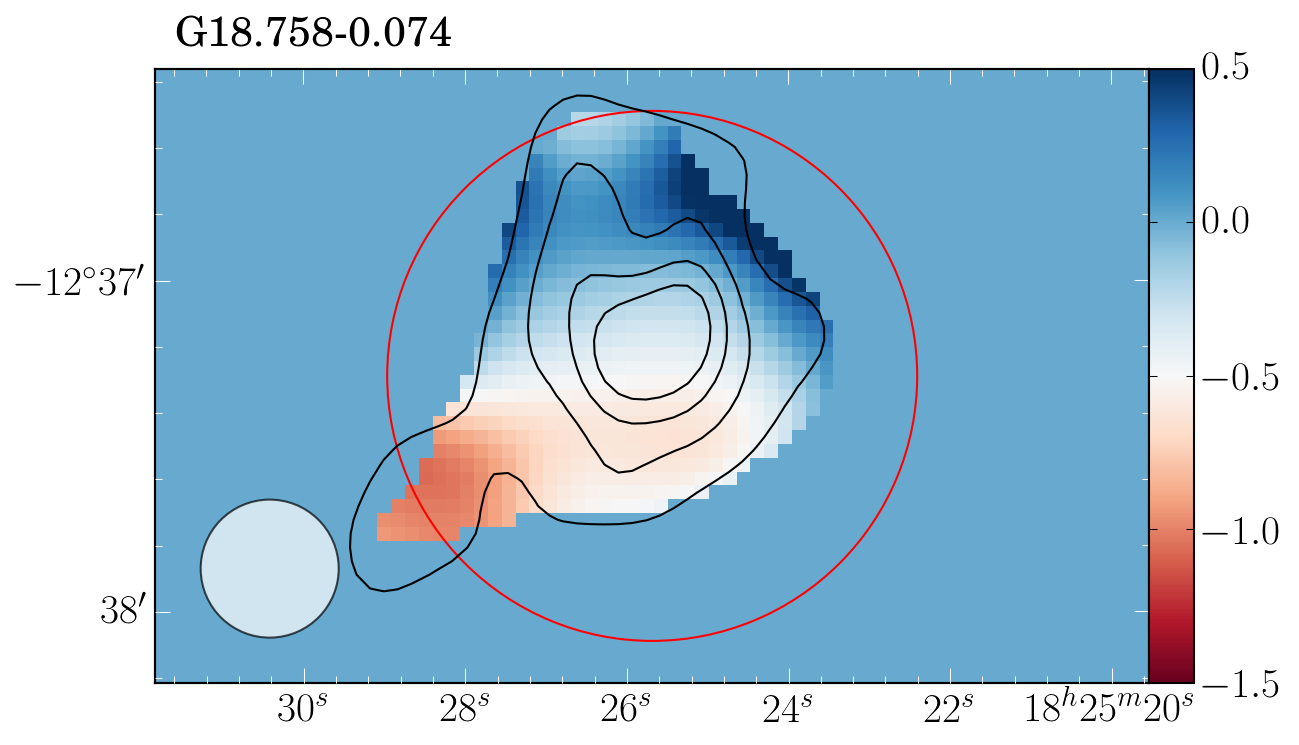} 
  \includegraphics[width=6cm]{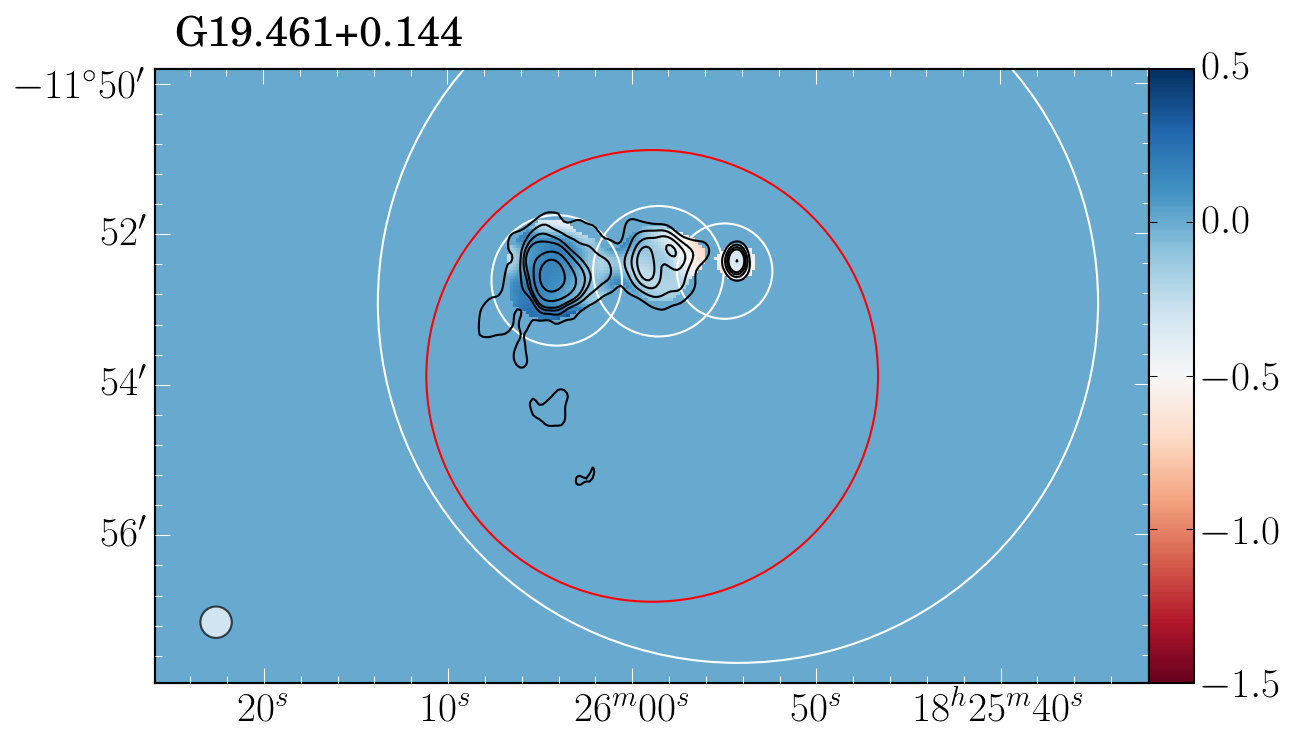}
  \includegraphics[width=6cm]{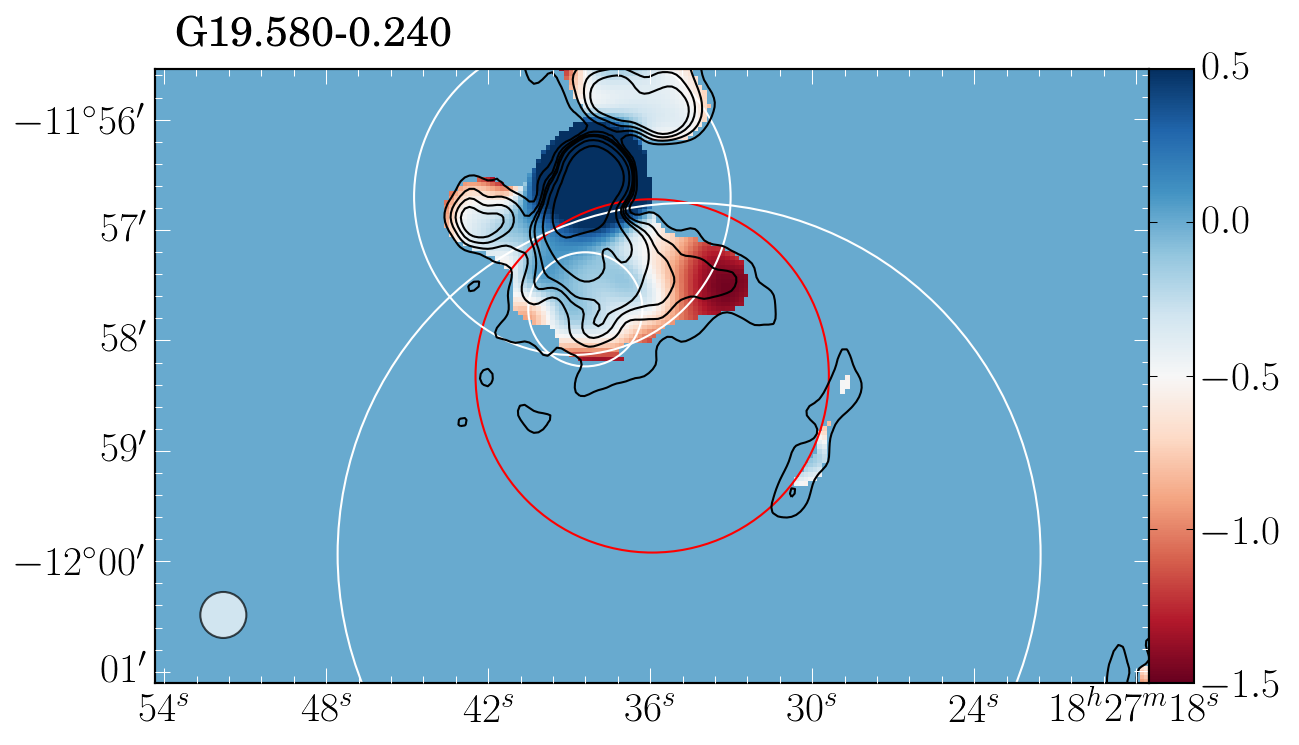}
  \includegraphics[width=6cm]{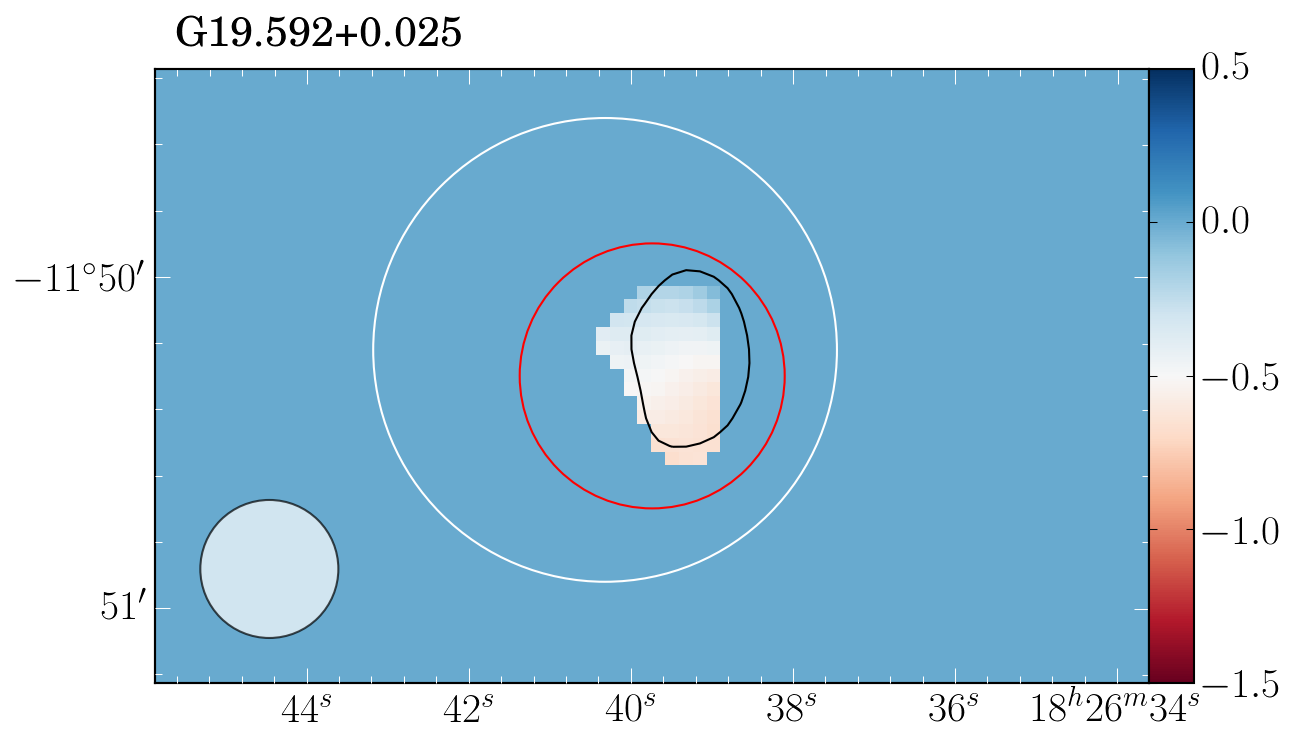} 
  \includegraphics[width=6cm]{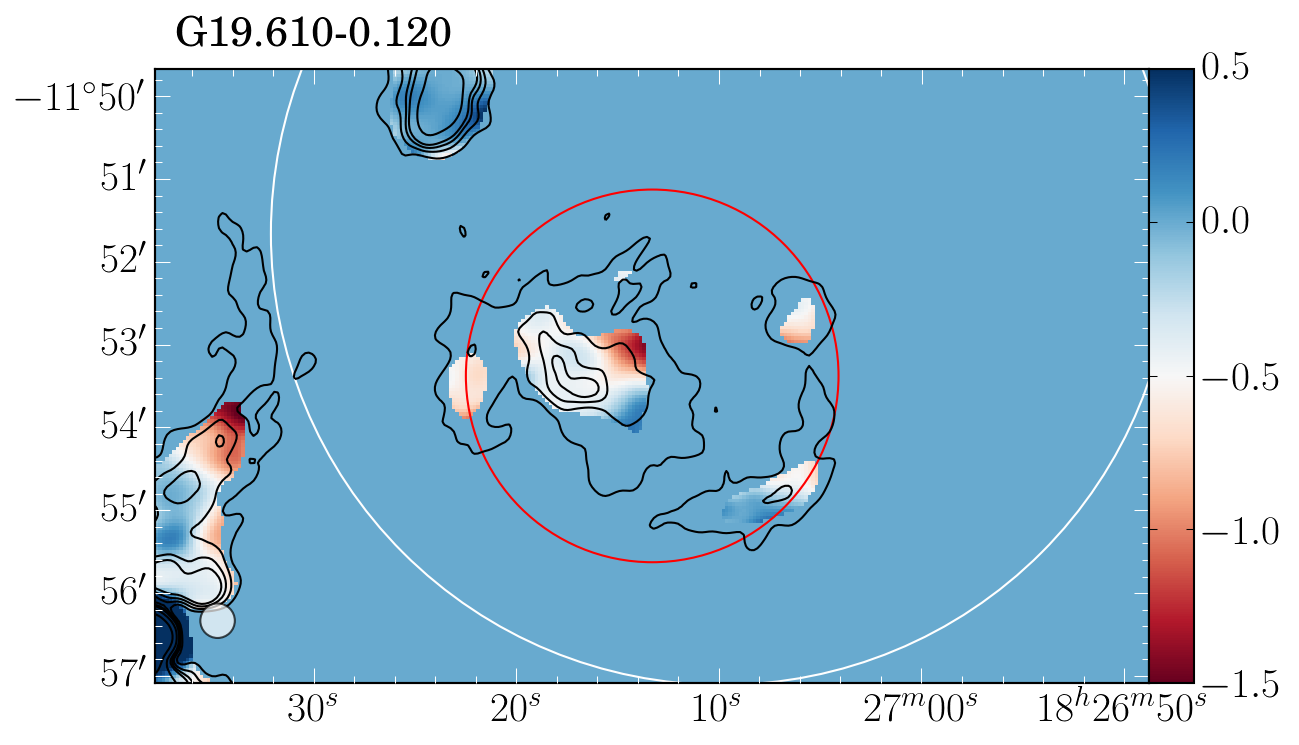}
  \includegraphics[width=6cm]{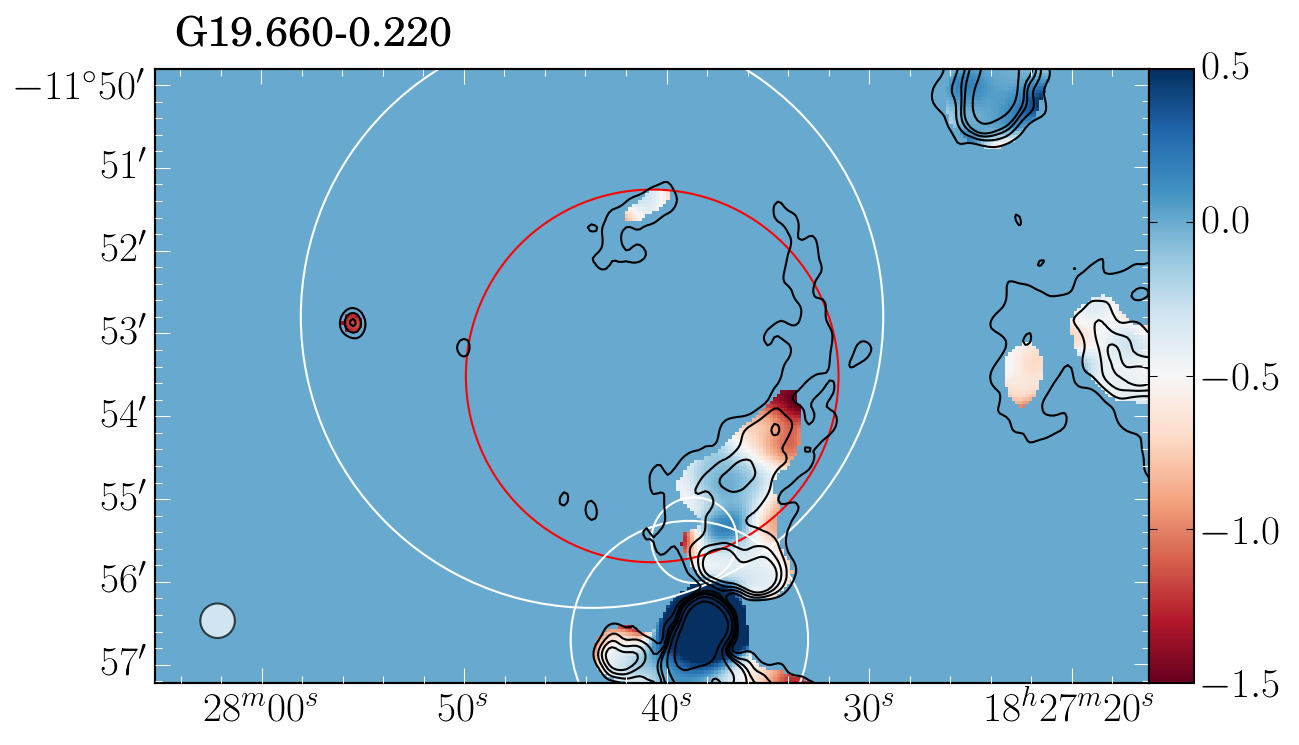}
  \includegraphics[width=6cm]{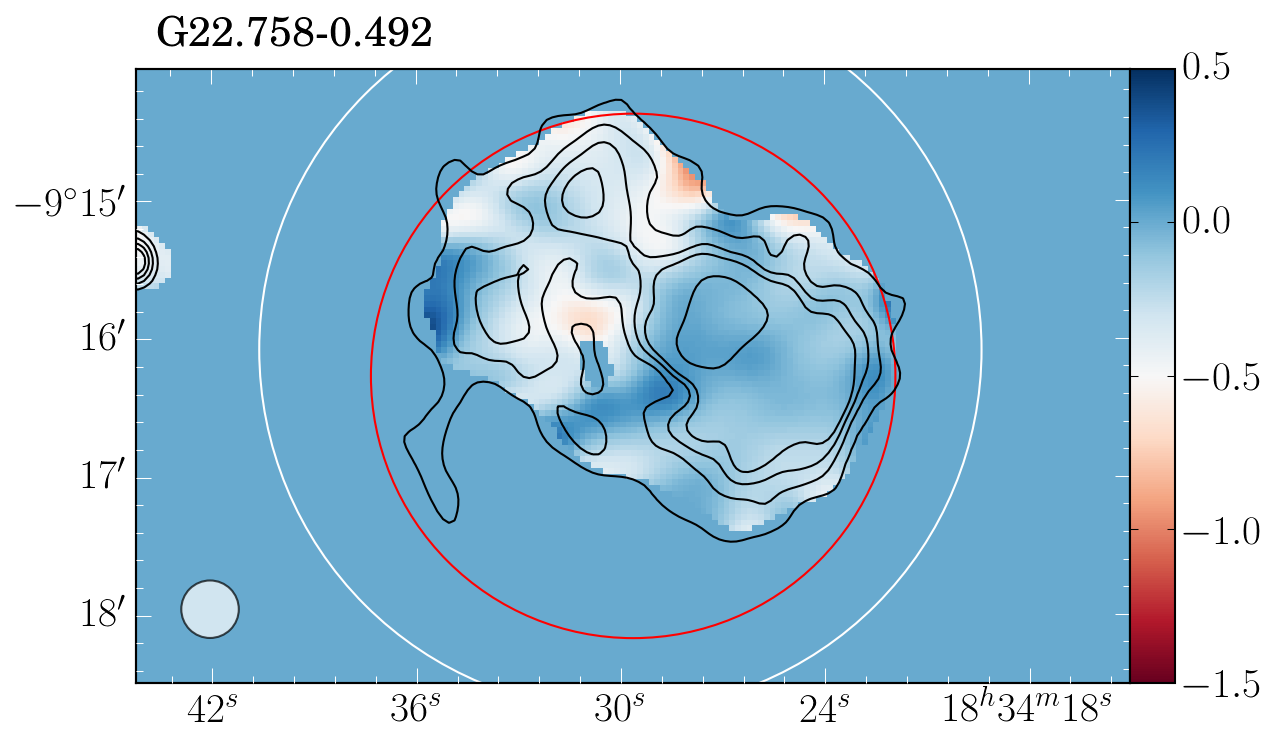}
  \includegraphics[width=6cm]{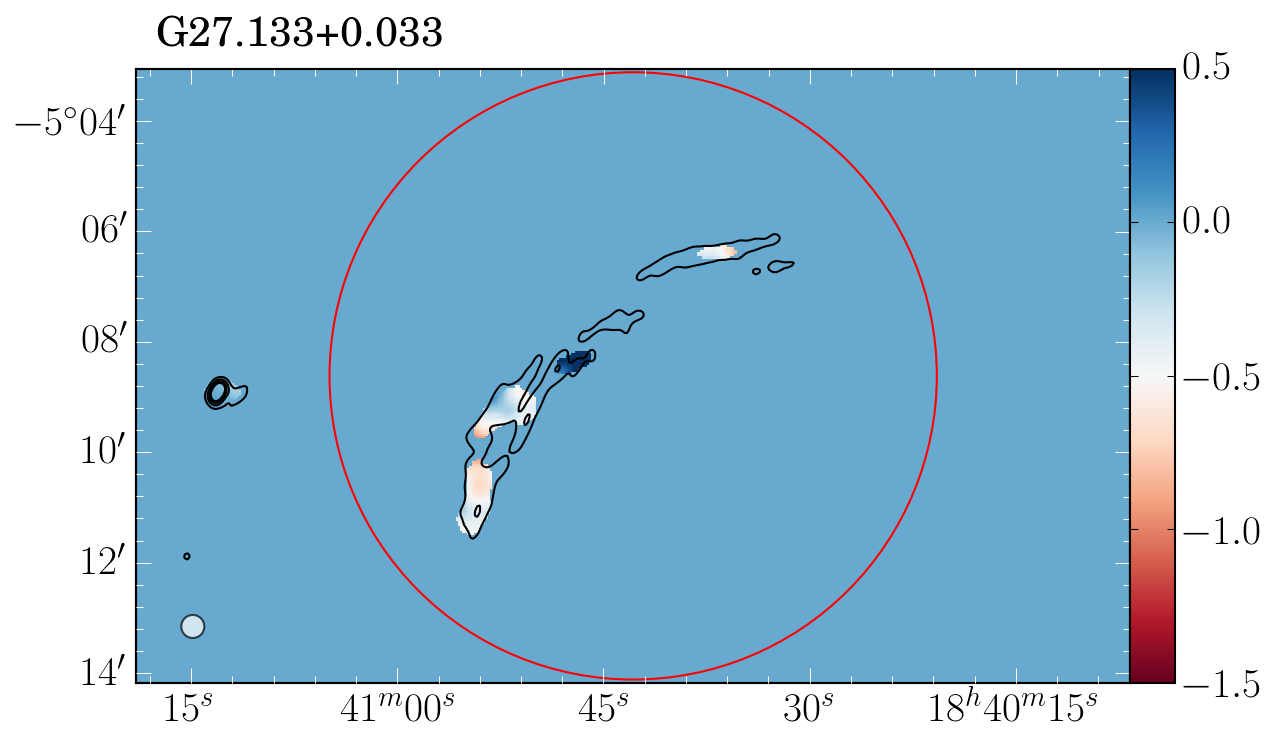}
  \includegraphics[width=6cm]{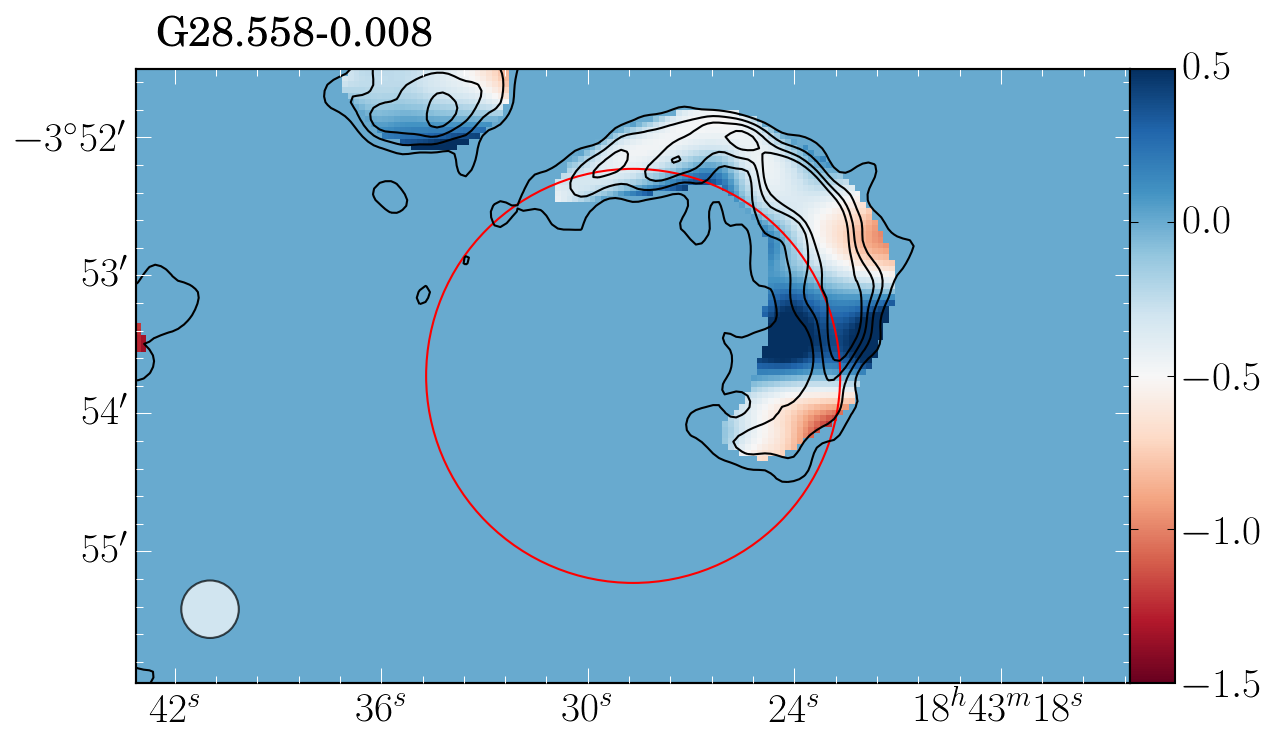}
  \includegraphics[width=6cm]{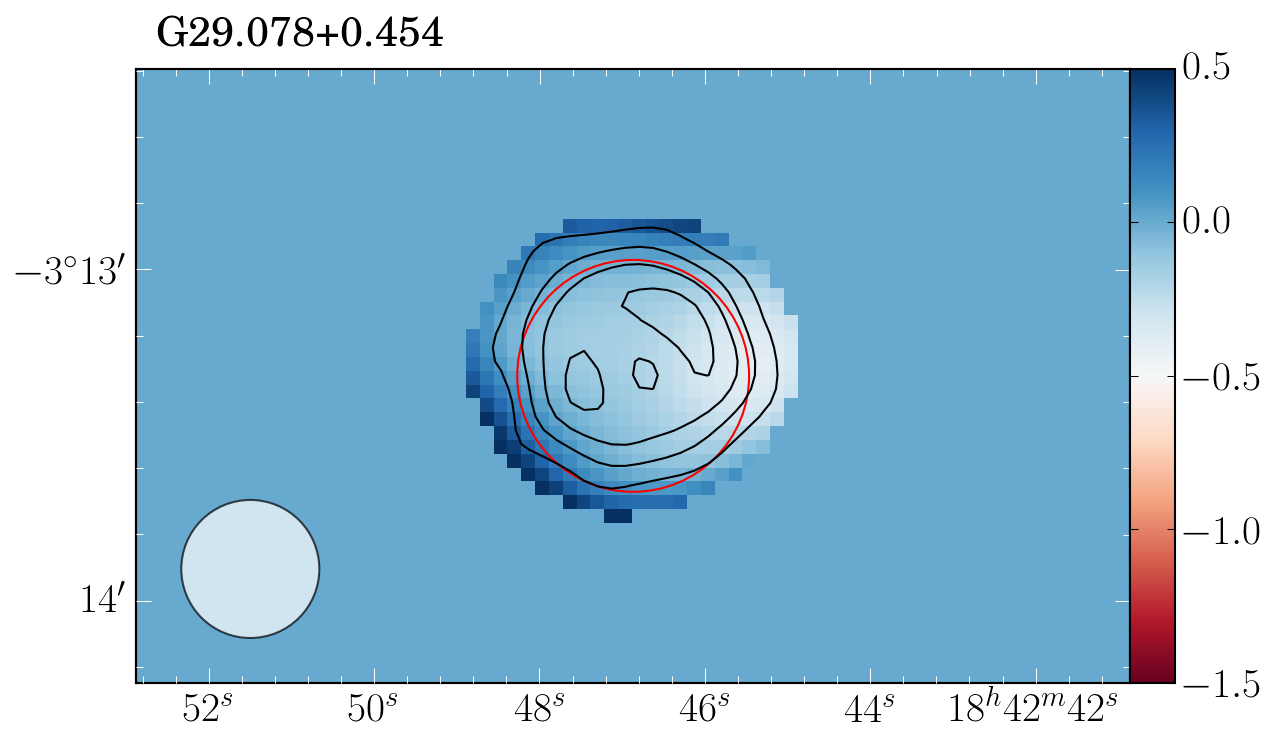}
  \includegraphics[width=6cm]{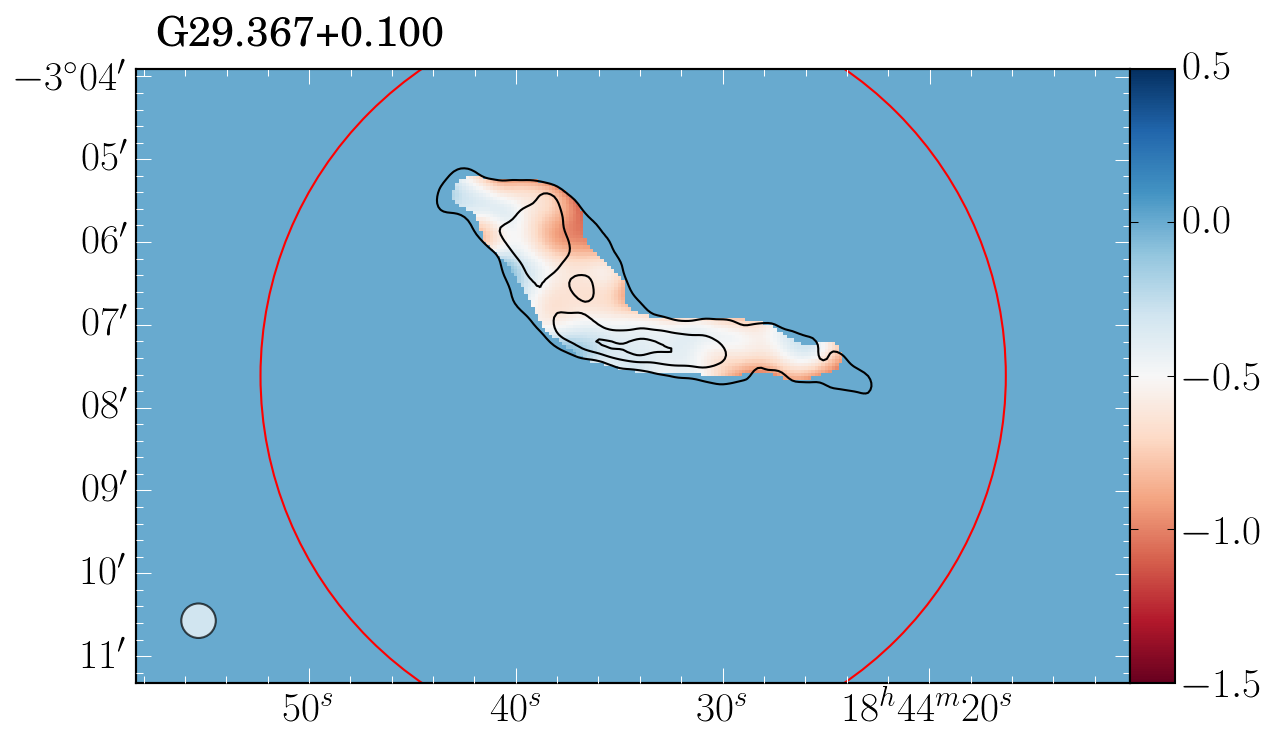}
  \includegraphics[width=6cm]{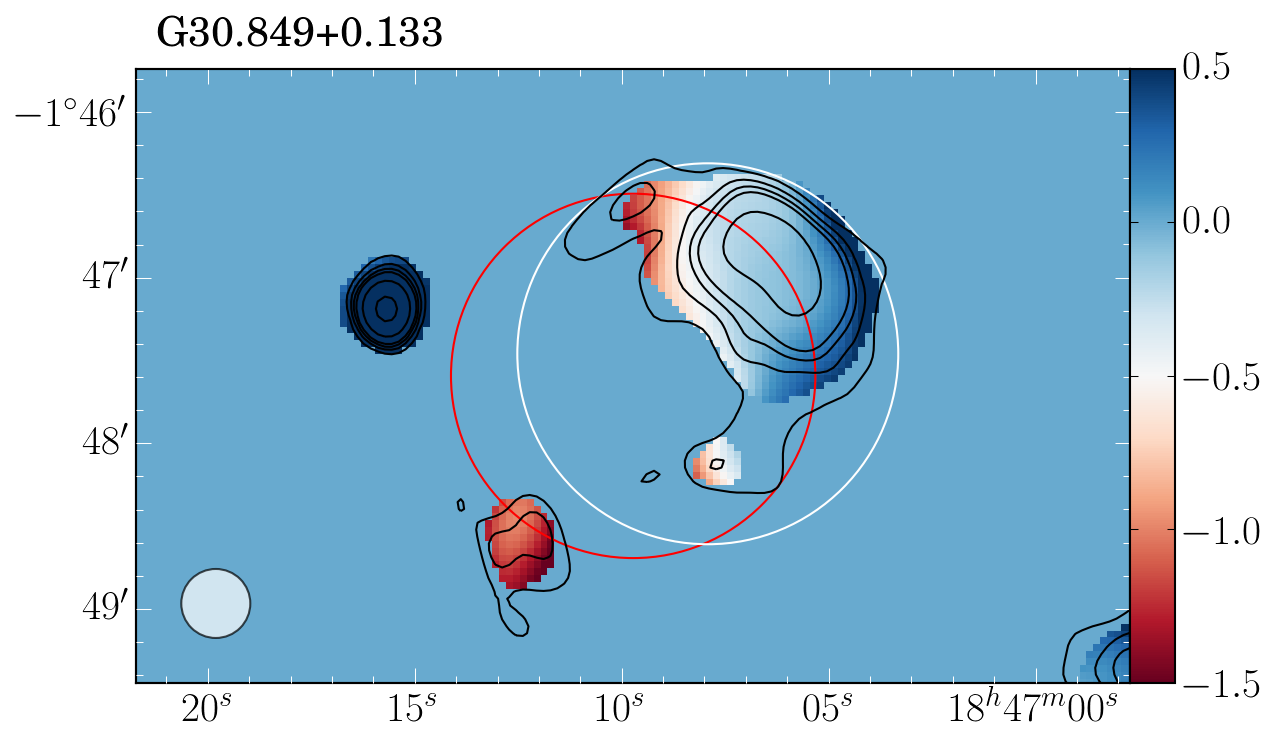}
  \includegraphics[width=6cm]{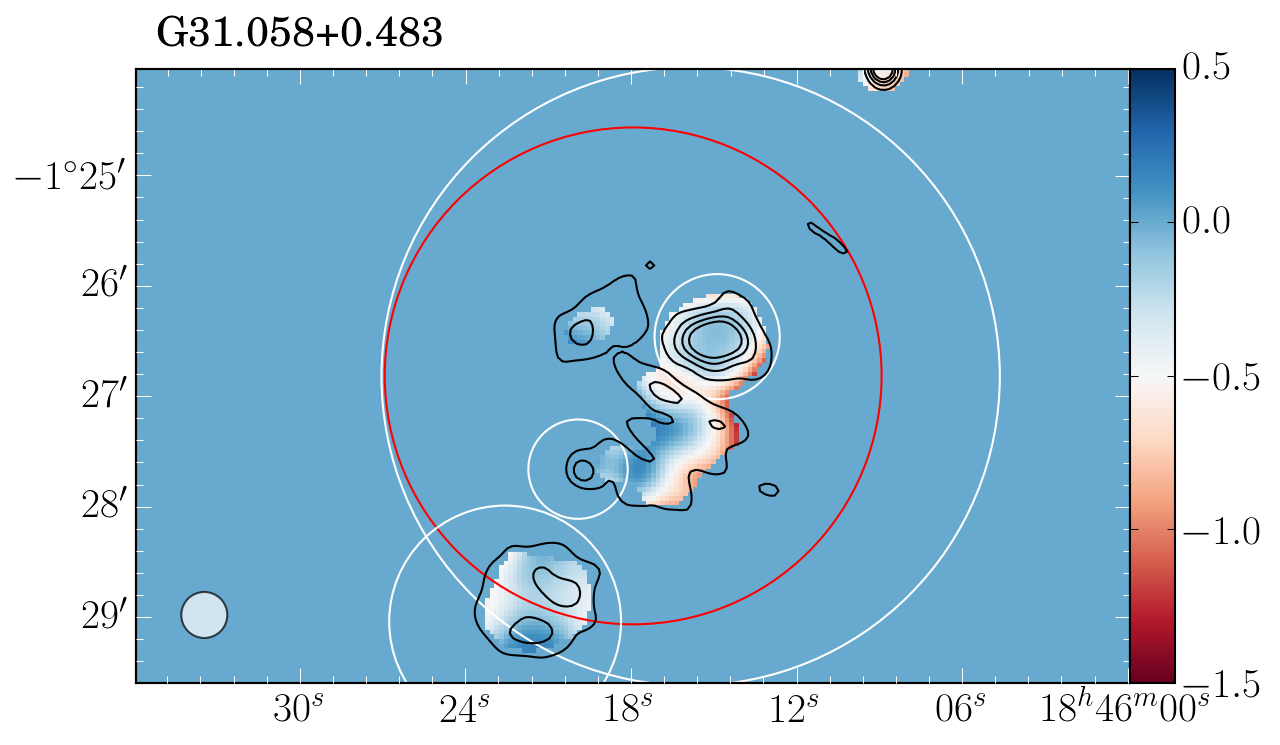}
  \includegraphics[width=6.5cm]{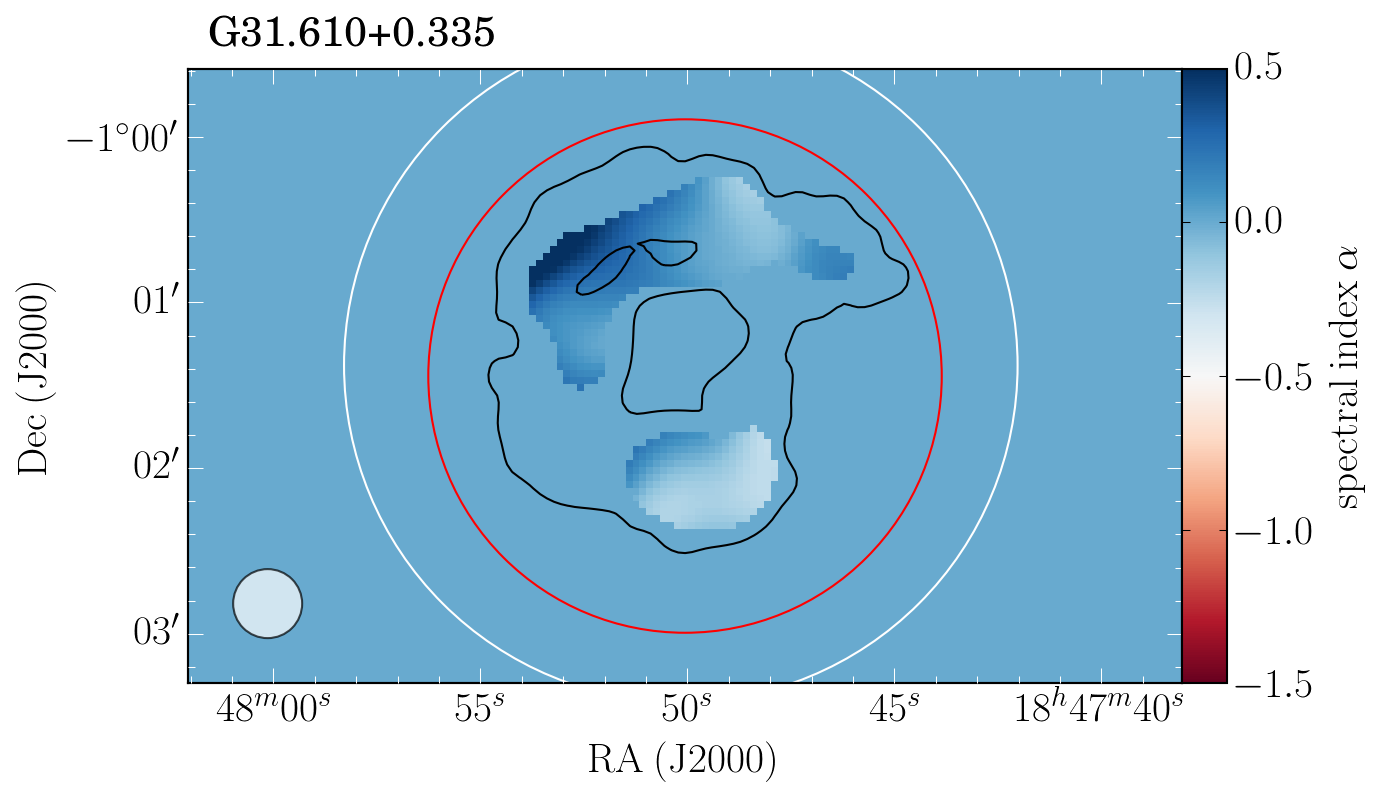}
  \caption{Spectral index maps of extended MAGPIS SNR candidates. The color scale represents the spectral index $\alpha$ from -1.5 to 0.5, the black contours show the continuum emission in steps of 5, 10, 15, 20, 40, and 100\,mJy\,beam$^{-1}$. The red and white circles indicate the sizes of the SNR candidates given in \citet{Helfand2006} and the sizes of the \ion{H}{ii} regions given in \citet{Anderson2014}, respectively. In each panel the synthesized beam is shown in the bottom left corner.}
  \label{Fig_spectral_index_map_MAGPIS_SNR}
\end{figure*}
To date the most complete catalog of Galactic SNR contains 294 sources \citep{Green2014}. Most of them are discovered or confirmed in the radio continuum. The size distribution and intensity of these SNR varies over several orders of magnitude, making them difficult to observe. Within the THOR region, the SNR catalog by \citet{Green2014} contains 43 sources. Out of these 43 sources, we identify 26 sources within our catalog. Since the SNR can be very clumpy, several sources within the THOR continuum catalog can be matched with a single SNR from the catalog by \citet{Green2014}. Table \ref{table_SNR_match} shows the sources that are matched visually. Seventeen SNR from the catalog by \citet{Green2014} are below the threshold for our THOR continuum catalog and are not extracted. These SNR are either too weak, not visible in the radio continuum, or too diffuse, and we filter them out with the VLA C-Array configuration. However, we visually inspected the missing 17 SNR and could find traces of at least nine SNR, below the used extraction threshold of 5$\sigma$. \\
The spectral index for SNR peaks around $\alpha \sim -0.5$ \citep{Green2014, Dubner2015}. The spectral index can vary spatially for the same SNR from $\alpha \sim 0$ to -1 \citep[e.g.,][]{Bhatnagar2011, Reynoso2015}. Here we highlight one example to demonstrate the capability of this data set and show the spectral index map of the well known SNR G018.8+00.3 in Fig. \ref{Fig_spectral_index_SNR_18deg}. Furthermore, we use this technique to examine several SNR candidates proposed in the MAGPIS survey.\\
MAGPIS \citep{Helfand2006} provides 49 new SNR candidates, which are only partly included in the SNR catalog by \citet{Green2014}. Their criteria for a SNR candidate is that they have: 1.) a high ratio between 20\,cm continuum and 21\,$\rm{\mu}$m flux density, 2.) a counterpart at 90\,cm continuum emission, and 3.) a distinctive SNR morphology, e.g., shell-type or filled-center \citep{Dubner2015}. Followup observations for these candidates to determine the distance using \ion{H}{i} absorption are presented in \citet{Johanson2009}. Since MAGPIS has higher spatial resolution, it is more sensitive at detecting the mentioned SNR morphology. However, the THOR survey can help to characterize the radiation and distinguish between thermal and non-thermal radiation. Within the THOR region, we find 33 MAGPIS SNR candidates, which are listed in Table \ref{table_MAGPIS_SNR_candidates}. Only five of them are included in the SNR catalog presented by \citet{Green2014}. In contrast to this, 24 of these MAGPIS SNR candidates have at least one counterpart in the WISE \ion{H}{ii} region catalog presented by \citet{Anderson2014}. However, thanks to the high density of \ion{H}{ii} regions within the Galactic plane, these can be chance alignments along the line of sight.\\ 
As explained, we can use the spectral index to distinguish between thermal and non-thermal radiation. We present spectral index maps for 16 extended MAGPIS SNR candidates in Fig. \ref{Fig_spectral_index_map_MAGPIS_SNR}. Similar to the example of the well known SNR G018.8+00.3 (Fig. \ref{Fig_spectral_index_SNR_18deg}), we find strongly varying spectral index maps. On the one hand, several sources clearly indicate a flat spectral index (e.g., G31.610+0.335), which is characteristic for an \ion{H}{ii} region. On the other hand, several sources (e.g., G18.150-0.172) show clear signatures of a negative spectral index around -0.5, which is typical of SNR. However, for some sources the classification as thermal or non-thermal radiation is difficult as the spectral index shows both positive and negative values. The source G19.580-0.240 is a good example for such a behavior. This can be explained by several different sources along the line of sight. Our goal is to use this information to classify sources as potential SNR or as \ion{H}{ii} regions. As explained in Sect. \ref{sec_extended_sources}, we have to be cautious with the spectral index for extended sources due to different filtering at different wavelengths. The spectral index maps of the MAGPIS SNR candidates G18.150-0.172, G18.758-0.074, G27.133+0.033, G28.558-0.008, G29.367+0.100 show negative values, and they are not directly correlated with any known \ion{H}{ii} region. Therefore, these five sources are excellent candidates for SNR. However, only one of them (G18.150-0.172) is listed in the SNR catalog by \citet{Green2014}. The morphology of the five proposed SNR differs widely. We find two examples of shell-type SNR (G18.150-0.172, G28.558-0.008), one small filled-center (G18.758-0.074) and two that may be part of a larger shell or a more filamentary SNR (G27.133+0.033, G29.367+0.100). \\
Beside these proposed SNR without any correlation to known \ion{H}{ii} regions, we find one source (G18.254-0.308) that is a well known \ion{H}{ii} region \citep{Anderson2014} showing the same morphology in the infrared, but the spectral index map shows mostly negative values down to $\alpha = -1$. This is an indicator of non-thermal radiation, which contradicts the expected radiation coming from an \ion{H}{ii} region. Owing to the different spatial filtering in our data (see Sect. \ref{sec_spectral_index_determination}), we cannot exclude a systematic shift of the spectral index. However, the source is strong ($\sim$30$\sigma$) and not very extended, which minimizes the filtering effects. Further analysis will be needed to reveal the nature of this source.

\begin{table*}
\caption{MAGPIS SNR candidates in comparison with THOR sources, Green SNR and WISE \ion{H}{ii} regions.}             
\label{table_MAGPIS_SNR_candidates}      
\centering          
\begin{tabular}{l l l l c c c c c }
\hline\hline       
SNR candidate & THOR source & Green SNR & WISE \ion{H}{ii} & diam. & distance & $\alpha$ & $\Delta \alpha$ \\
 &  &  & & [\arcmin] & [kpc] &  &  \\

\hline
G16.358$-$0.183 &  &  & G016.352-00.179 & 2.8 &  & -- & -- \\
G17.017-0.033 & G17.030-0.069 & G017.0-00.0 & G017.032-00.052\tablefootmark{b} & 4.0 &  & -0.19\tablefootmark{c} & 0.64\\
G17.336-0.139 & G17.335-0.139 &  & G017.336-00.146 & 1.8 &  & 0.09 & 0.29  \\
G18.150-0.172\tablefootmark{a}  & G18.193-0.174 & G018.1-00.1 & G018.195-00.171\tablefootmark{b} & 7.0 & 6.3$\pm$0.5 & -0.37 & 0.10  \\
 & G18.171-0.213 &  &  & -- &  & -0.68 & 0.25 \\
 & G18.107-0.134 &  &  & -- &  & -0.72 & 0.25 \\
G18.254-0.308\tablefootmark{a} & G18.270-0.289 &  & G018.253-00.298 & 3.5 & 4.3$\pm$0.6 & -0.46 & 0.05 \\
G18.638-0.292 & G18.610-0.316 & G018.6-00.2 &  & 4.0 & 4.6$\pm$0.6 & 0.17 & 0.22 \\
G18.758-0.074\tablefootmark{a} & G18.760-0.072 &  &  & 1.6 & 4.9$\pm$0.6 & -0.35 & 0.08 \\
G19.461+0.144\tablefootmark{a} & G19.492+0.135 &  & G019.489+00.135\tablefootmark{b} & 6.0 & 6.8-17.5 & 0.15 & 0.02  \\
 & G19.475+0.173 &  &  & -- &  & -0.30 & 0.12 \\
G19.580-0.240\tablefootmark{a} & G19.610-0.235 &  & G019.554-00.248\tablefootmark{b} & 3.2 & 6.3$\pm$0.5 & 0.95 & 0.01  \\
 & G19.555-0.230 &  &  & -- &  & -0.14 & 0.28 \\
G19.592+0.025\tablefootmark{a} & G19.592+0.028 &  & G019.594+00.024 & 0.8 &  & -0.41 & 0.19 \\
G19.610-0.120\tablefootmark{a} & G19.614-0.133 &  & G019.629-00.095 & 4.5 & 11.6$\pm$0.5 & -0.45 & 0.11  \\
G19.660-0.220\tablefootmark{a} & G19.610-0.235 &  & G019.675-00.226\tablefootmark{b} & 4.5 &  & 0.95 & 0.01 \\
 & G19.691-0.204 &  &  & -- &  & -0.41 & 0.28\\
G20.467+0.150 & G20.502+0.155 & G020.4+00.1 &  & 5.5 &  & -0.54 & 0.33 \\G21.557-0.103 &  & G021.5-00.1 & G021.560-00.108 & 4.0 &  & -- & --  \\
G21.642+0.000 & G21.632-0.007 &  & G021.634-00.003 & 2.8 &  & -0.28 & 0.49  \\
G22.383+0.100 & G22.360+0.064 &  & G022.357+00.064\tablefootmark{b} & 7.0 &  & -0.72 & 0.18  \\
G22.758-0.492\tablefootmark{a} & G22.760-0.478 &  & G022.761-00.492 & 3.8 & 5.1$\pm$0.6 & -0.04 & 0.04 \\
G22.992-0.358 & G22.980-0.370 &  & G022.988-00.360\tablefootmark{b} & 3.8 & 5.0$\pm$0.5 & -0.51 & 0.18  \\
 & G22.974-0.345 &  &  & -- &  & -0.12\tablefootmark{c} & 0.60\\
G23.567-0.033 & G23.541-0.039 &  & G023.572-00.020\tablefootmark{b} & 9.5 & 6.4$\pm$0.7 & -0.03\tablefootmark{c} & 0.14  \\
 & G23.585+0.030 &  &  & -- &  & 0.36 & 0.14\\
 & G23.645-0.028 &  &  & -- &  & 0.63\tablefootmark{c} & 0.46 \\
G24.180+0.217 & G24.200+0.192 &  & G024.185+00.211\tablefootmark{b} & 5.2 &  & 0.13\tablefootmark{c} & 0.41 \\
 & G24.197+0.243 &  &  & -- &  & -0.09\tablefootmark{c} & 0.15  \\
 & G24.166+0.251 &  &  & -- &  & -0.03 & 0.22\\
G25.222+0.292 & G25.220+0.286 &  & G025.220+00.289 & 2.0 &  & -0.00 & 0.24  \\
G27.133+0.033\tablefootmark{a} & G27.158+0.063 &  &  & 11.0 & 6.1-16.2 & -0.63 & 0.28  \\
 & G27.119-0.027 &  &  & -- &  & -0.39 & 0.21 \\
G28.375+0.203 &  &  & G028.376+00.208\tablefootmark{b} & 10.0 &  & -- & --  \\
G28.517+0.133 &  &  &  & 14.0 & 6.2-15.9 & -- & --  \\
G28.558-0.008\tablefootmark{a} & G28.569+0.020 &  &  & 3.0 & 6.5-15.9 & -0.15 & 0.09  \\
G28.767-0.425 &  &  &  & 9.5 &  & -- & --  \\
G29.067-0.675 &  &  & G029.088-00.675 & 8.0 &  & -- & --  \\
G29.078+0.454\tablefootmark{a} & G29.079+0.458 &  &  & 0.7 &  & -0.19 & 0.08 \\
G29.367+0.100\tablefootmark{a} & G29.372+0.104 &  &  & 9.0 & 5.8-15.8 & -0.40 & 0.14  \\
G30.849+0.133\tablefootmark{a} & G30.854+0.151 &  & G030.847+00.140 & 2.2 & 6.7-15.6 & -0.11 & 0.04\\
 & G30.866+0.114 &  &  & -- &  & 0.71 & 0.06\\
 & G30.839+0.117 &  &  & -- &  & -1.09 & 0.19  \\
G31.058+0.483\tablefootmark{a} & G31.057+0.497 &  & G031.054+00.491\tablefootmark{b} & 4.5 & 6.6-15.5 & -0.08 & 0.10  \\
 & G31.034+0.459 &  &  & -- &  & -0.34 & 0.19  \\
G31.610+0.335\tablefootmark{a} & G31.598+0.330 &  & G031.610+00.335 & 3.1 & 6.6-15.5 & 0.19 & 0.19 \\
G31.821-0.122 & G31.823-0.117 &  & G031.806-00.115 & 1.8 &  & -0.10 & 0.24  \\

\hline
\end{tabular}
\tablefoot{The SNR candidates are given in \citet{Helfand2006}. Since the SNR candidates can be clumpy, several THOR sources can be found within one MAGPIS SNR candidate so we list all corresponding THOR sources. The associated SNR names taken from \citet{Green2014} are given, as well as associated WISE \ion{H}{ii} regions given in \citet{Anderson2014}. The given \ion{H}{ii} regions can be associated with the SNR, but they can also be foreground or background objects. Figure \ref{Fig_spectral_index_map_MAGPIS_SNR} shows the size for each \ion{H}{ii} region. The diameter of the MAGPIS SNR candidates is taken from \citet{Helfand2006}, and the distance is determined via \ion{H}{i} absorption and taken from \citet{Johanson2009}. The spectral index $\alpha$ is measured for the peak position of the THOR source (see Sect. \ref{sec_spectral_index_determination}) and can vary significantly within the source (see Fig. \ref{Fig_spectral_index_map_MAGPIS_SNR}).
\\
\tablefoottext{a}{Spectral index map is shown in Fig. \ref{Fig_spectral_index_map_MAGPIS_SNR}.}\\
\tablefoottext{b}{Multiple \ion{H}{ii} regions can be found within the region.}\\
\tablefoottext{c}{Determination of the spectral index is uncertain, since the source is not detected in all spectral windows.}
}          
\end{table*}

\subsection{Special source G48.384+0.789}

\begin{figure}
   \centering
   \includegraphics[width=\hsize]{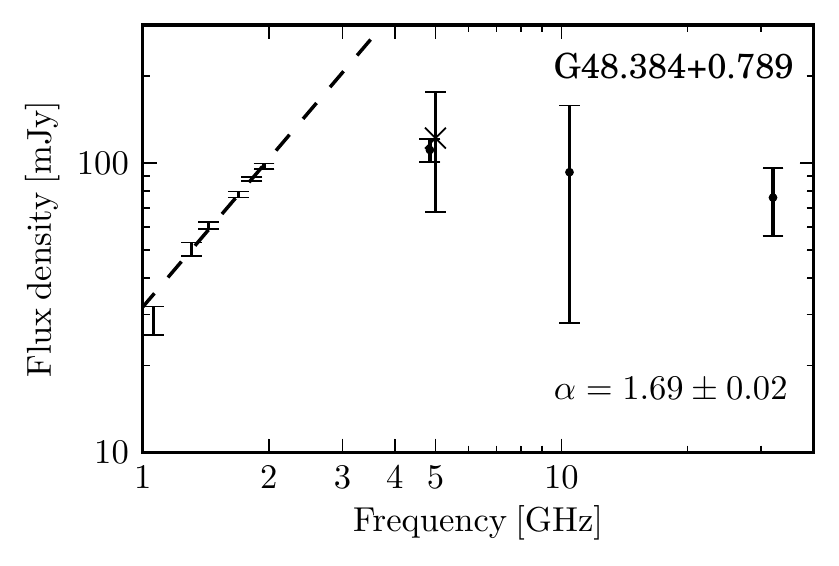}
      \caption{Spectrum of the THOR source G48.384+0.789. The data points between 1 and 2\,GHz are taken from the THOR survey, and the dashed line represents the fitted spectral index to these data points of $\alpha = 1.69\pm0.02$. The data point at 5\,GHz (x-shape) is taken from the CORNISH survey \citep{Purcell2013}, and the data points at 4.85, 10.45, and 32\,GHz (circles) are taken from \citet{Vollmer2008}. For each data point, a 5$\sigma$ uncertainty is shown.}
         \label{Fig_spectral_index_optically_thick}
\end{figure}

In this section, we introduce a special source, which has a remarkably high positive spectral index. The THOR source G48.384+0.789 is unresolved and bright (30-100\,mJy\,beam$^{-1}$) and shows a positive spectral index of $\alpha = 1.70 \pm 0.02$. Since this source is unresolved, the spectral index determination does not suffer from spatial filtering due to the VLA C-array configuration and is well constrained. Figure \ref{Fig_spectral_index_optically_thick} shows the flux density measurements for each spectral window, and the corresponding spectral index fit. This source has a counterpart in CORNISH \citep[G048.3841+00.7889,][]{Purcell2013} at 5\,GHz. The corresponding flux density measurement at 5\,GHz is also given in Fig. \ref{Fig_spectral_index_optically_thick}, but we do not consider this data point for the spectral index determination. Within CORNISH, this source is classified as "IR-quiet" and even with the high resolution of CORNISH of 1.5\arcsec, this source is unresolved. Further observations at 4.85, 10.45, and 32\,GHz using the Effelsberg 100m telescope presented by \citet{Vollmer2008} show a flat spectrum for higher frequencies (see Fig. \ref{Fig_spectral_index_optically_thick}). We do not find any counterpart in sub-mm emission (ATLASGAL) or in CO emission (GRS). However, Very Long Baseline Array (VLBA) observations presented by \citet{Immer2011} reveal a detection, and they classified this source as compact ("compactness B"). This does not translate trivially to an actual size of the object because VLBA observations suffer from severe filtering effects. But this detection shows that the object has a very compact component typical of extragalactic sources, such as AGNs. The spectral index could also be explained by an AGN as similar spectral shapes are found in the literature \citep[e.g., ][]{Brunthaler2005}.

\section{Conclusions}

The THOR survey, which is the \ion{H}{i}, OH, recombination line survey of the Milky Way is a Galactic plane survey covering a large portion of the first Galactic quadrant ($l=14-67\degr$, $|b| \leq 1.1$\degr) using the VLA in the C-Array configuration, achieving a spatial resolution of $\sim$10-25\arcsec. In this paper, we present a catalog of continuum sources within the first half of the survey ($l = 14.0-37.9\degr$ and $l = 47.1-51.2$\degr, $|b| \leq 1.1$\degr). We summarize our work and the main results below.

\begin{enumerate}
\item
With the BLOBCAT extraction software, we extracted 4422 sources. We used a spatially varying noise map, as well as automated RFI flagging methods (RFlag) to decrease the number of false detections. Furthermore, we visually inspected each source for obvious artifacts. About 75\% (3366 sources) of the extracted sources are reliable detections above 7$\sigma$. The catalog is complete up to 95\% above the 7$\sigma$ detection limit, whereas the noise is spatially varying. Half of the observed area has a noise level of $7\sigma <3$\,mJy\,beam$^{-1}$.
\item
We cross-matched the THOR catalog with the NVSS, MAGPIS, and CORNISH catalogs to validate the position and flux density. The position comparison with MAGPIS and CORNISH reveals no significant shift, and we reported a position uncertainty that depends on the strength of the source but is smaller than 2\arcsec. The flux density and peak intensity comparison with MAGPIS shows a one-to-one relation; however, we find a slight bias in comparison with the NVSS survey.
\item
Thanks to the broad bandpass between 1 and 2\,GHz, we were able to determine reliable spectral indices for 1840 sources. We extracted the peak intensity of six different spectral windows and used a linear fit to describe the spectral index $\alpha$ with the form $\rm{I(\nu) \varpropto \nu^{\alpha}}$. The spectral index distributions reveals two peaks at $\alpha=-1$ and $\alpha=0$, highlighting two groups of sources, which are dominated by thermal and non-thermal radiation, respectively. 
\item
We used the spectral index information to investigate the spectrum of \ion{H}{ii} regions. We cross-matched the THOR catalog with the WISE \ion{H}{ii} region catalog and found an overlap of 388 sources. For about 326 of these sources, we were able to determine a reliable spectral index. The distribution reveals a single peak around $\alpha = 0$, indicating thermal free-free emission.
\item
The spectral index can also be used to confirm potential SNR candidates because they exhibit a typical spectral index of $\alpha = -0.5$. We investigated the MAGPIS SNR candidate catalog and determined spectral index maps for 16 SNR candidates. Owing to potential line-of-sight contamination with \ion{H}{ii} regions, the differentiation between thermal and non-thermal radiation is difficult. However, we confirmed five SNR candidates, showing non-thermal radiation and no correlation with \ion{H}{ii} regions. Four of them are not listed in the SNR catalog presented by \citet{Green2014}.
\end{enumerate}

\begin{acknowledgements}
The authors would like to thank the referee, David J. Helfand, for insightful thoughts on the draft that improved the final paper significantly.\\The National Radio Astronomy Observatory is a facility of the National Science Foundation operated under cooperative agreement by Associated Universities, Inc.\\
We gratefully acknowledge help from the team of the Pete V. Domenici Science Operations Center (SOC) in Socorro for their help on data reduction and imaging during two extended visits in 2013 and 2015.\\
S. B. is a fellow of the International Max Planck Research School for Astronomy and Cosmic Physics (IMPRS) at the University of Heidelberg and acknowledges its support.\\
SCOG and RSK acknowledge support from the Deutsche Forschungsgemeinschaft (DFG) via the SFB 881 (subprojects B1, B2, and B8) "The Milky Way System", and the SPP (priority program) 1573, "Physics of the ISM". In addition, RSK acknowledges support from the European Research Council under the European Community's Seventh Framework Program (FP7/2007-2013) via the ERC Advanced Grant "STARLIGHT: Formation of the First Stars" (project number 339177).\\
HB acknowledges support from the European Research Council under the Horizon 2020 Framework Program via the ERC Consolidator Grant CSF-648505.

\end{acknowledgements}


\bibliographystyle{aa}
\bibliography{references.bib}

\begin{appendix}

\section{Source Examples}
\begin{figure*}
   \resizebox{\hsize}{!}
            {\includegraphics[width=18cm]{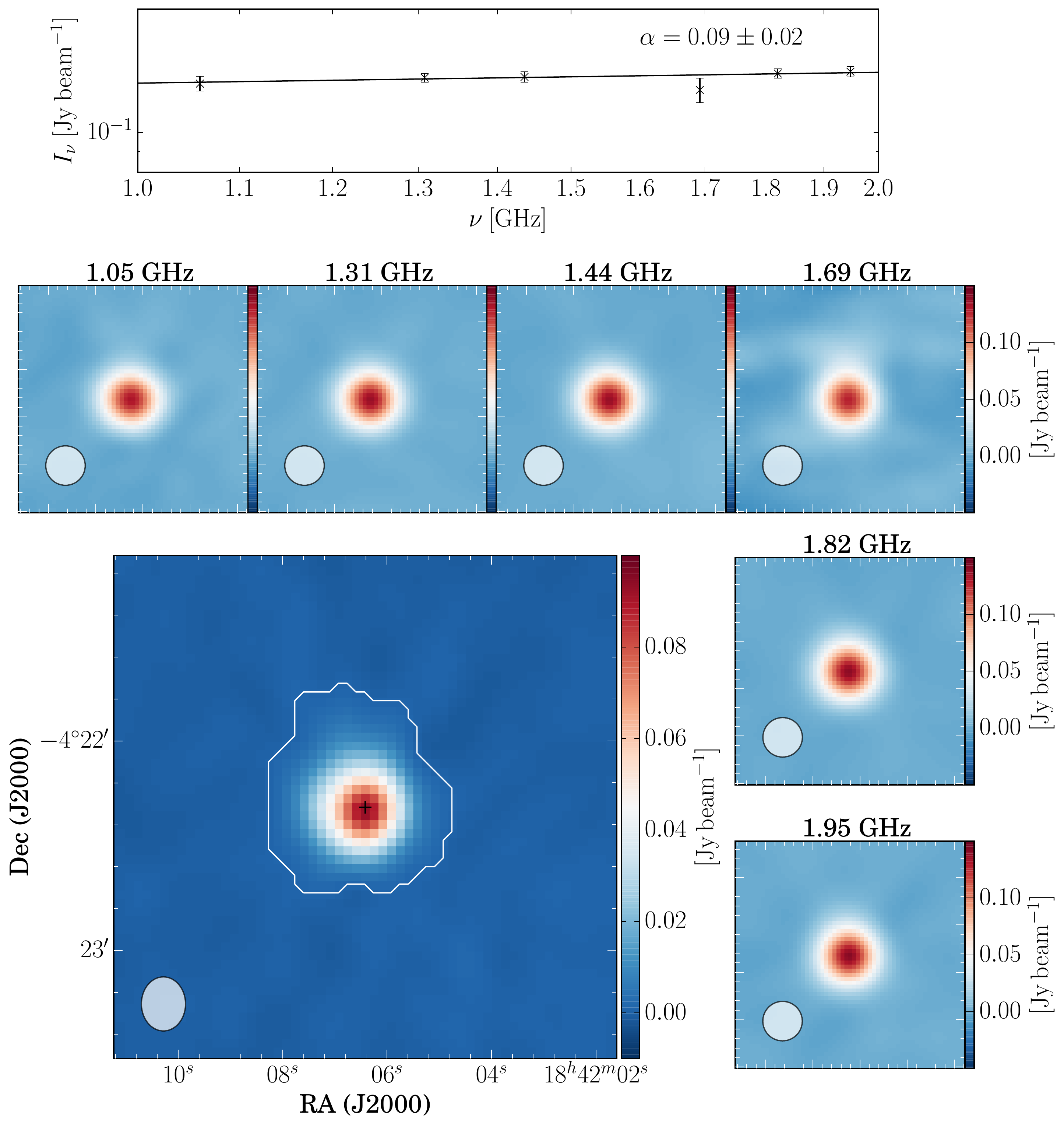}}
      \caption{Example image of the THOR source G27.978+0.078, which corresponds to the WISE \ion{H}{ii} region G027.980+00.080 \citep{Anderson2014}. The large image represents an averaged image of the two spectral windows around 1.4 and 1.8\,GHz, which we used for the source extraction (see Sects. \ref{sec_averaging_spectral_window} and \ref{sec_blobcat}). The white contours shows the extent of the source determined by the BLOBCAT algorithm. The black cross marks the peak position, which we used to determine the spectral index. The small images show each spectral window separately and the top panel presents the peak intensity for each spectral window and the corresponding spectral index fit. In each image the synthesized beam is given in the lower left corner. }
         \label{Fig_source_example_1}
\end{figure*}

\begin{figure*}
   \resizebox{\hsize}{!}
            {\includegraphics[width=18cm]{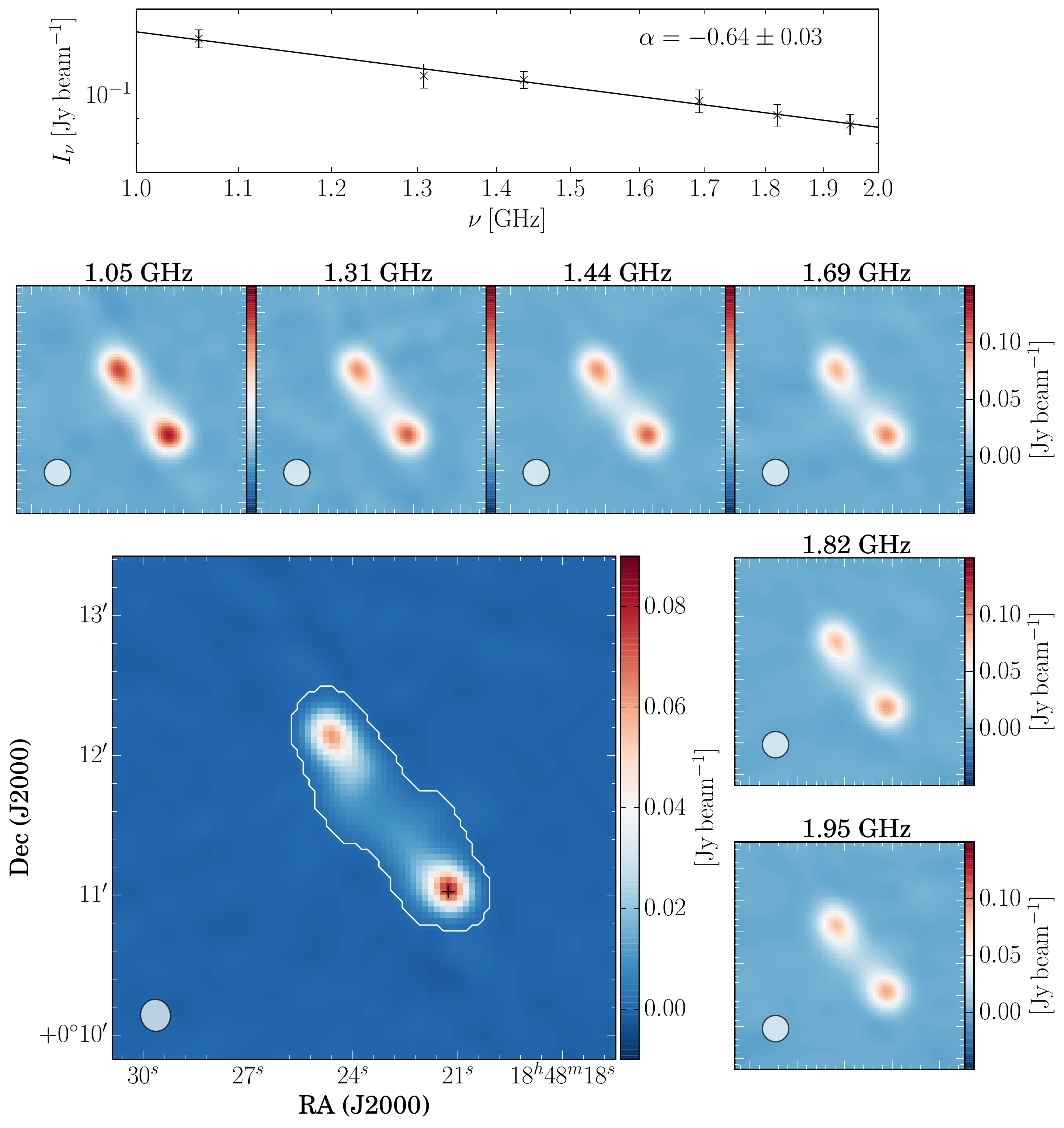}}
      \caption{Example image of the THOR source G32.744+0.770, which is most likely an extragalactic jet. The large image represents an averaged image of the two spectral windows around 1.4 and 1.8\,GHz, which we used for the source extraction (see Sects. \ref{sec_averaging_spectral_window} and \ref{sec_blobcat}). The white contours shows the extent of the source determined by the BLOBCAT algorithm. The black cross marks the peak position, which we used to determine the spectral index. The small images show each spectral window separately, and the top panel presents the peak intensity for each spectral window and the corresponding spectral index fit. In each image the synthesized beam is given in the lower left corner.  }
         \label{Fig_source_example_2}
\end{figure*}

\begin{figure*}
   \resizebox{\hsize}{!}
            {\includegraphics[width=18cm]{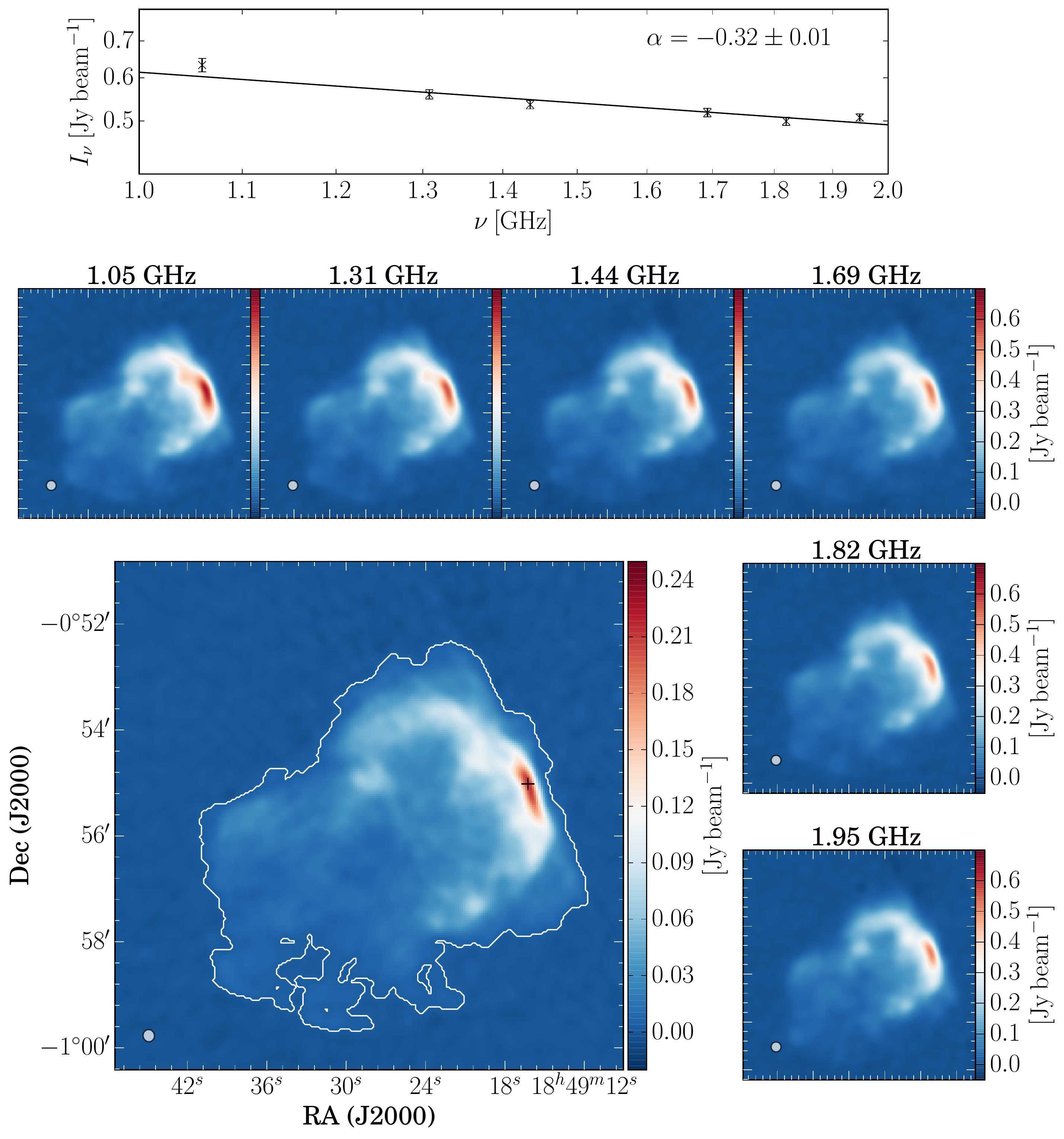}}
      \caption{Example image of the THOR source G31.869+0.064, which corresponds to the known SNR G31.9+00.0 \citep{Green2014}. The large image represents an averaged image of the two spectral windows around 1.4 and 1.8\,GHz, which we used for the source extraction (see Sects. \ref{sec_averaging_spectral_window} and \ref{sec_blobcat}). The white contours show the extent of the source determined by the BLOBCAT algorithm. The black cross marks the peak position, which we used to determine the spectral index. The small images show each spectral window separately, and the top panel presents the peak intensity for each spectral window and the corresponding spectral index fit. In each image the synthesized beam is given in the lower left corner. }
         \label{Fig_source_example_3}
\end{figure*}

\section{Completeness maps}

\begin{figure*}
   \resizebox{\hsize}{!}
            {\includegraphics[width=18cm]{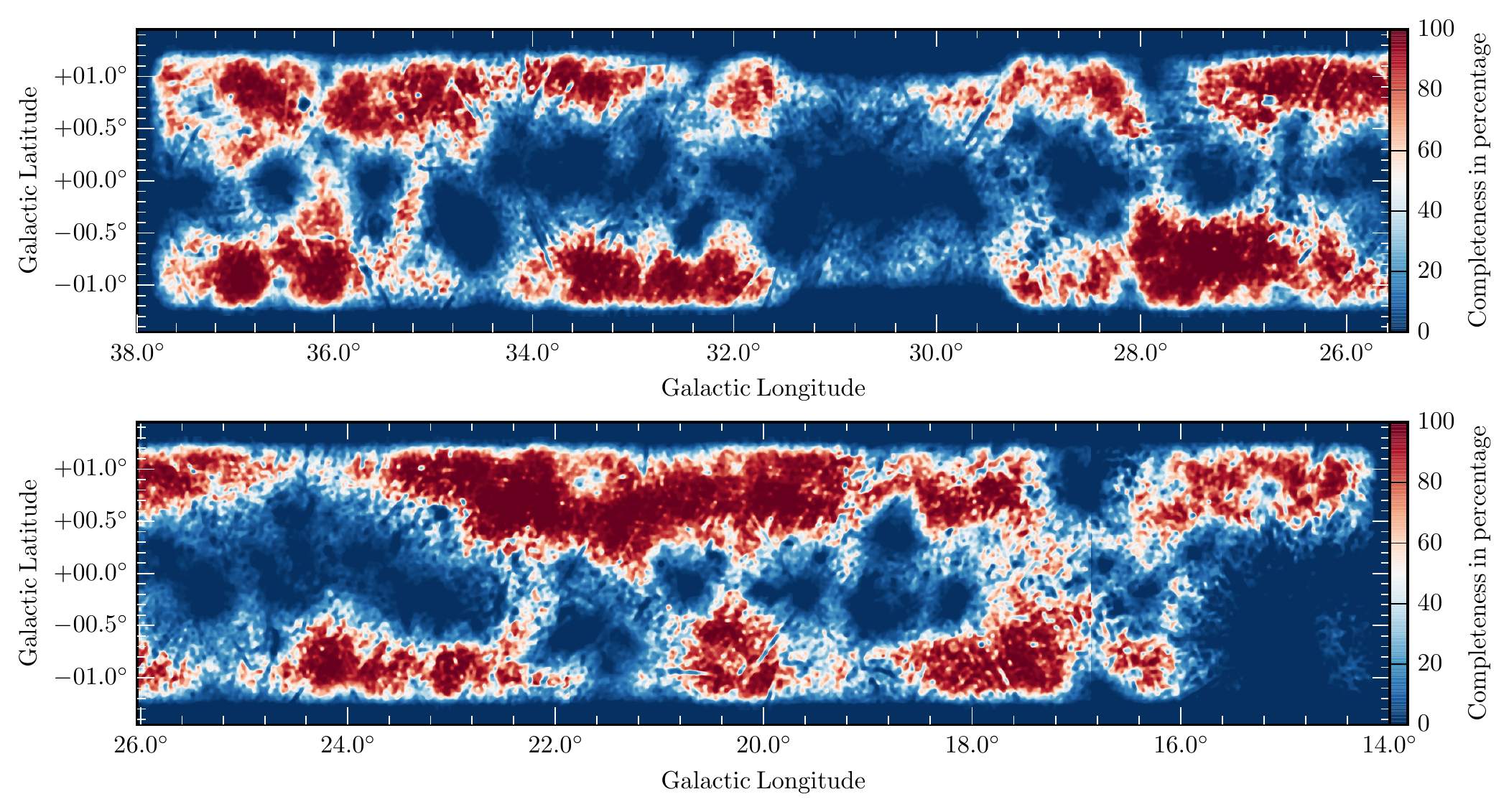}}
      \caption{Completeness map in percentage for sources with a peak intensity of 2\,mJy\,beam$^{-1}$. }
         \label{Completeness_map_large_2mJy}
\end{figure*}

\begin{figure*}
   \resizebox{\hsize}{!}
            {\includegraphics[width=18cm]{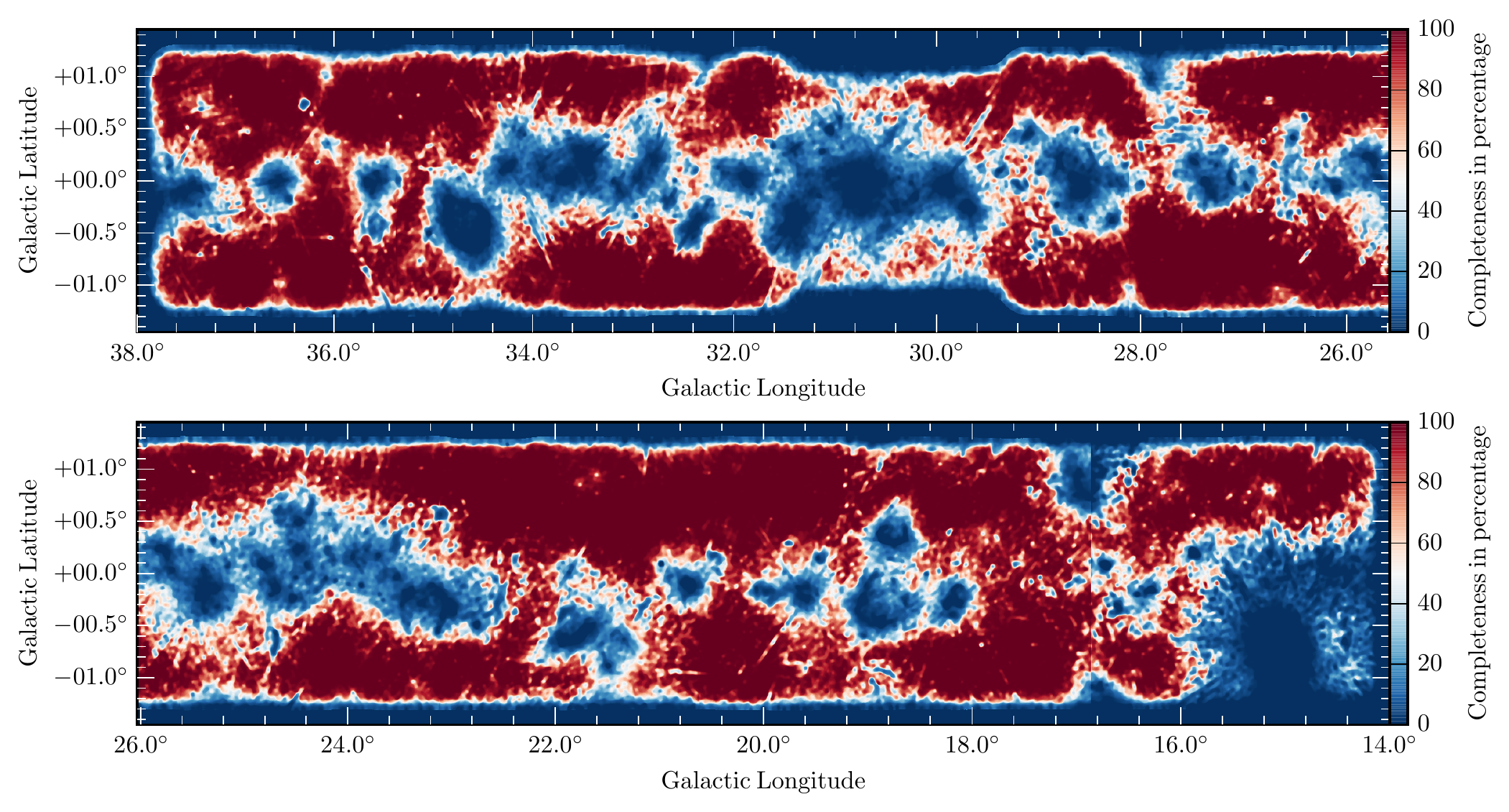}}
      \caption{ Completeness map in percentage for sources with a peak intensity of 3\,mJy\,beam$^{-1}$.}
         \label{Completeness_map_large_3mJy}
\end{figure*}

\begin{figure*}
   \resizebox{\hsize}{!}
            {\includegraphics[width=18cm]{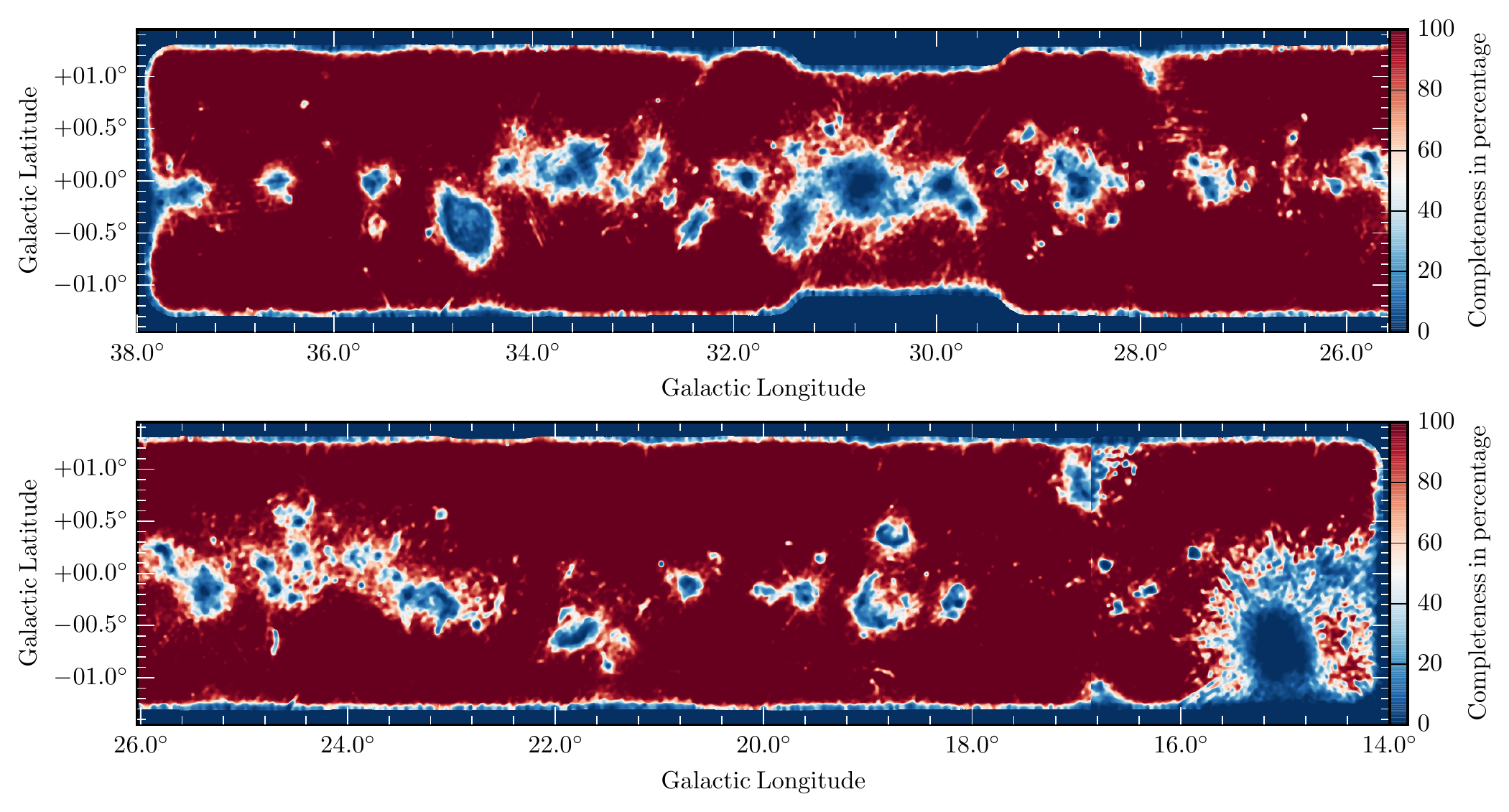}}
      \caption{Completeness map in percentage for sources with a peak intensity of 5\,mJy\,beam$^{-1}$. }
         \label{Completeness_map_large_5mJy}
\end{figure*}

\begin{figure*}
   \resizebox{\hsize}{!}
            {\includegraphics[width=18cm]{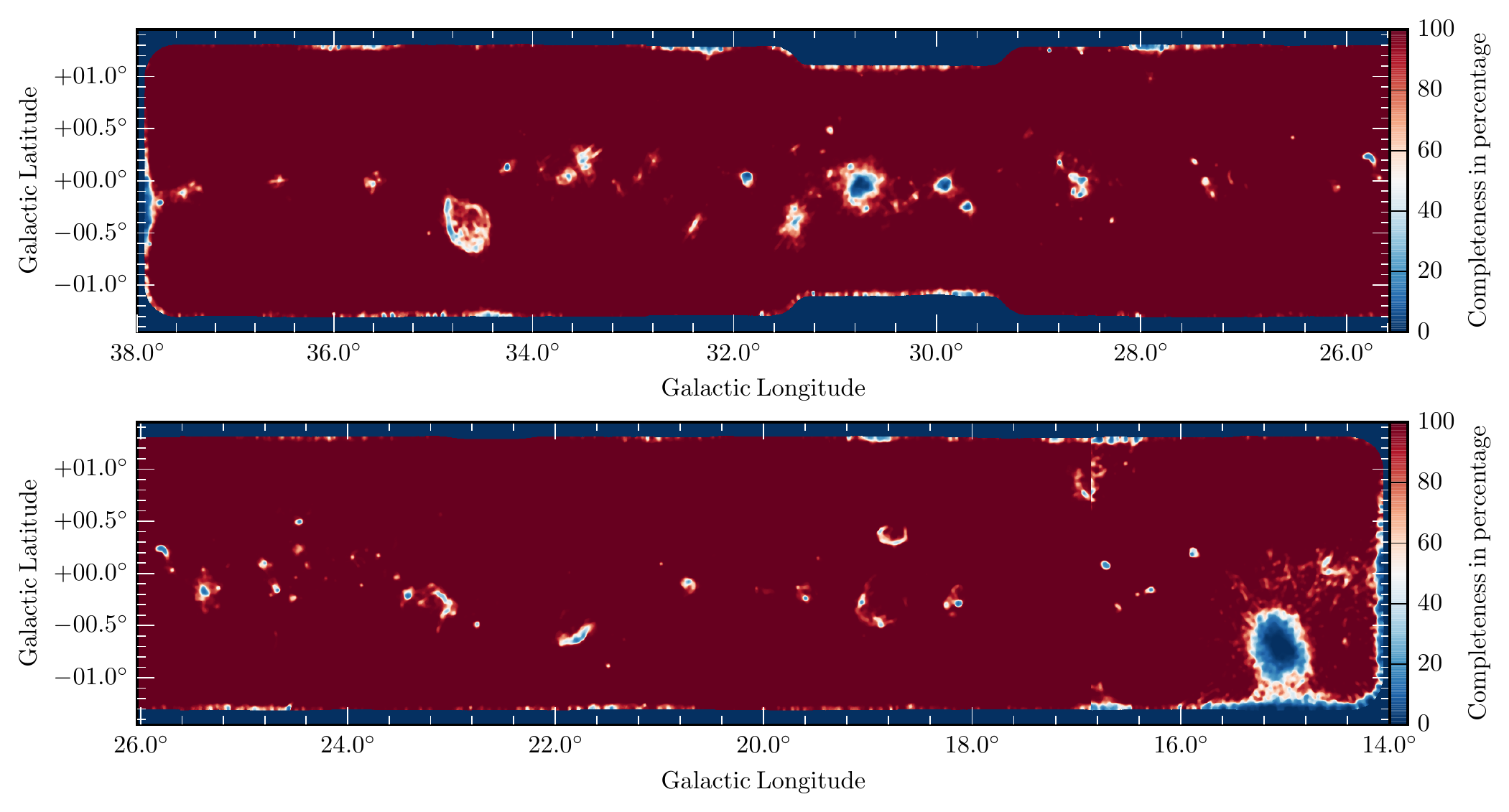}}
      \caption{ Completeness map in percentage for sources with a peak intensity of 10\,mJy\,beam$^{-1}$.}
         \label{Completeness_map_large_10mJy}
\end{figure*}

\begin{figure}
   \centering
   \includegraphics[width=\hsize]{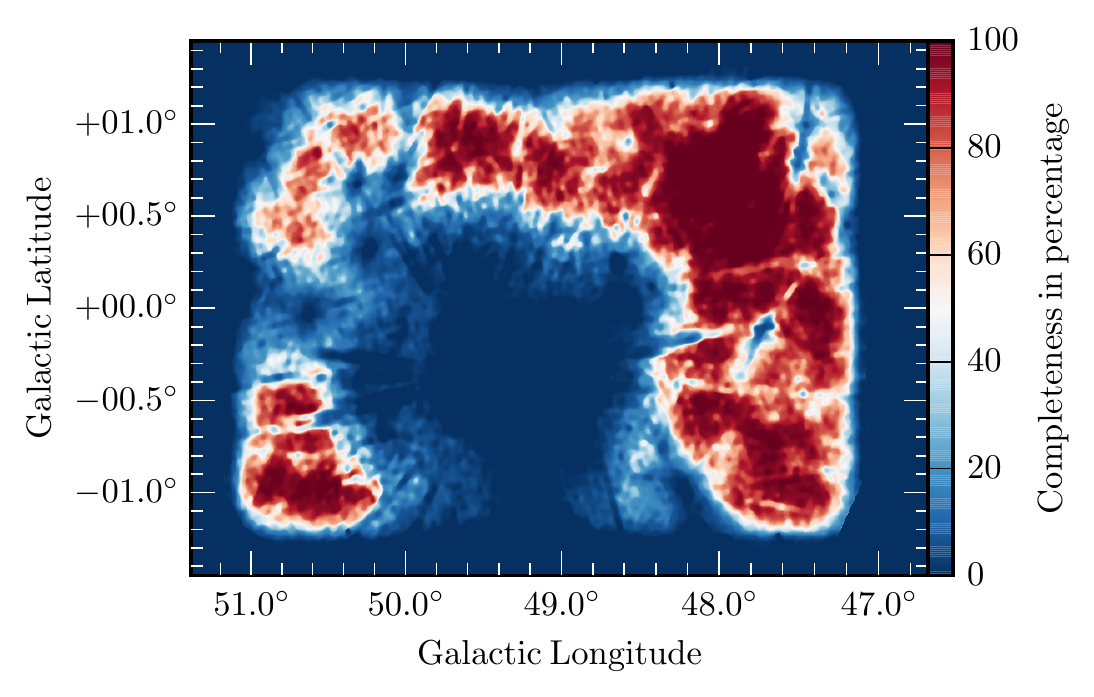}
      \caption{Completeness map in percentage for sources with a peak intensity of 2\,mJy\,beam$^{-1}$.}
         \label{Completeness_map_49deg_2mJy}
\end{figure}

\begin{figure}
   \centering
   \includegraphics[width=\hsize]{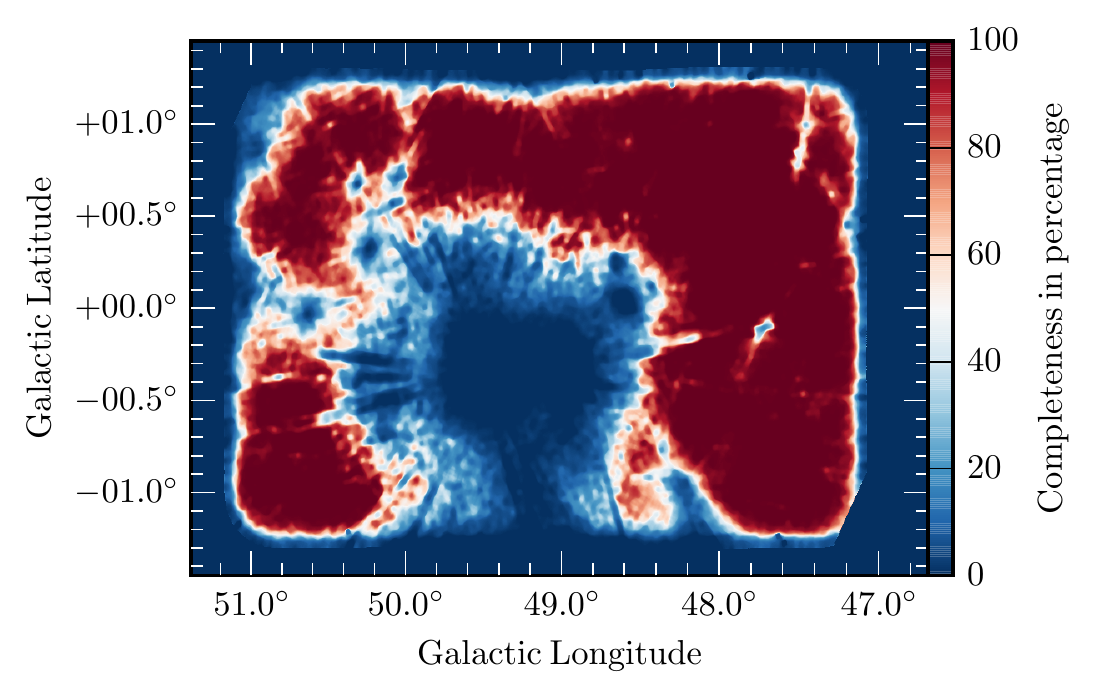}
      \caption{Completeness map in percentage for sources with a peak intensity of 3\,mJy\,beam$^{-1}$.}
         \label{Completeness_map_49deg_3mJy}
\end{figure}

\begin{figure}
   \centering
   \includegraphics[width=\hsize]{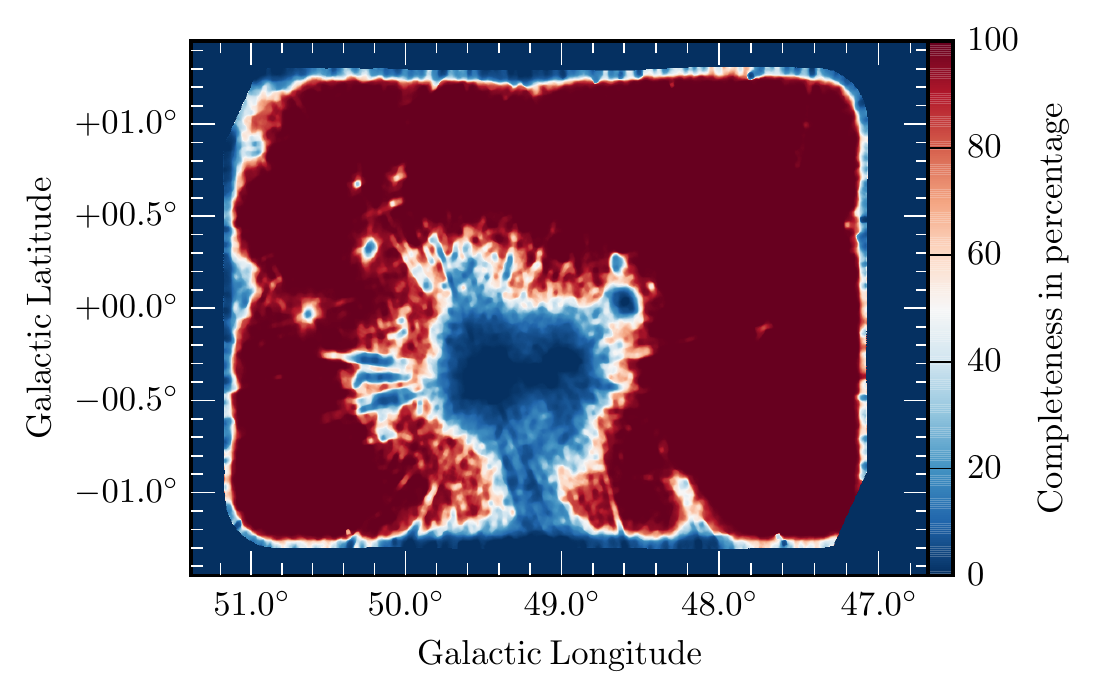}
      \caption{Completeness map in percentage for sources with a peak intensity of 5\,mJy\,beam$^{-1}$.}
         \label{Completeness_map_49deg_5mJy}
\end{figure}

\begin{figure}
   \centering
   \includegraphics[width=\hsize]{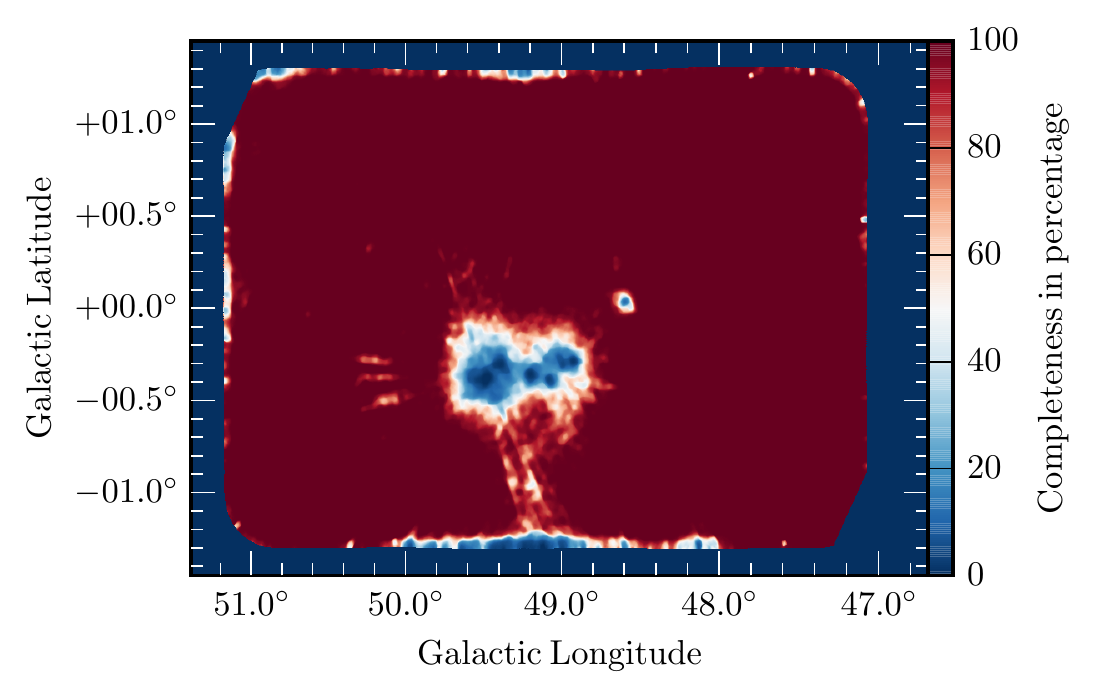}
      \caption{Completeness map in percentage for sources with a peak intensity of 10\,mJy\,beam$^{-1}$.}
         \label{Completeness_map_49deg_10mJy}
\end{figure}

\section{Example Table}
\begin{sidewaystable*}
\scriptsize
\caption{Example sources of the published catalog.}
\label{table_catalog_example}
\centering
\begin{tabular}{cccccccccccccccccccccc}
\hline\hline    
Galactic\_ID & RA & Dec & S\_p & SNR & S\_int & BMAJ & BMIN & BPA & n\_pix & ext. & art. & S\_p & S\_p &  S\_p &  S\_p & S\_p & S\_p & alpha & delta & reliable \\
&  &  &  &  &  &  &  &  &  &  &  & 1.06\,GHz & 1.31\,GHz &  1.44\,GHz & 1.69\.GHz & 1.82\,GHz & 1.95\,GHz & alpha & \_alpha & \_alpha \\
&  &  & [mJy\,b$^{-1}$]  &  &[mJy]  & [$\arcsec$]  & [$\arcsec$] & [$\degr$] &  &  &  & [mJy\,b$^{-1}$]  & [mJy\,b$^{-1}$]  &  [mJy\,b$^{-1}$]  &  [mJy\,b$^{-1}$]  & [mJy\,b$^{-1}$]  & [mJy\,b$^{-1}$]  &  &  &  \\
\hline
G16.008+1.022 & 274.0295 & -14.5304 & 4.00 & 13.48 & 3.81 & 18.1 & 11.1 & -176.7 & 62 & 0 & 0 & 8.61 & 6.07 & 5.44 & 5.27 & 2.73 & 3.08 & -1.88 & 0.37 & 1 \\
G16.026-0.804 & 275.7037 & -15.3780 & 12.56 & 27.21 & 10.12 & 18.1 & 11.1 & -176.7 & 64 & 0 & 0 & 19.46 & 12.56 & 12.70 & 10.06 & 8.88 & 6.22 & -1.58 & 0.17 & 1 \\
G16.029-0.663 & 275.5768 & -15.3096 & 57.97 & 101.00 & 56.76 & 18.1 & 11.1 & -176.7 & 121 & 0 & 0 & 78.56 & 63.65 & 63.81 & 56.13 & 53.21 & 50.58 & -0.71 & 0.05 & 1 \\
G16.055+0.828 & 274.2288 & -14.5818 & 9.92 & 24.01 & 7.64 & 18.1 & 11.1 & -176.7 & 59 & 0 & 0 & 6.43 & 9.15 & 8.75 & 7.56 & 8.69 & 6.69 & -0.05 & 0.26 & 1 \\
G16.059+0.875 & 274.1880 & -14.5560 & 5.97 & 15.21 & 6.97 & 18.1 & 11.1 & -176.7 & 98 & 0 & 0 & 9.28 & 5.04 & 6.29 & 4.47 & 4.57 & 4.02 & -1.30 & 0.37 & 1 \\
G16.070+0.346 & 274.6755 & -14.7975 & 2.36 & 7.05 & 2.23 & 18.1 & 11.1 & -176.7 & 39 & 0 & 1 & 3.16 & 2.21 & 3.68 & -0.14 & 1.30 & 1.32 & 0.00 & 0.00 & 0 \\
G16.079+0.812 & 274.2554 & -14.5680 & 3.42 & 9.09 & 2.61 & 18.1 & 11.1 & -176.7 & 37 & 0 & 0 & 3.46 & 0.82 & 3.24 & 1.75 & 2.88 & 1.71 & -0.50 & 1.61 & 0 \\
G16.084+0.042 & 274.9591 & -14.9290 & 11.56 & 17.24 & 16.22 & 18.1 & 11.1 & -176.7 & 113 & 1 & 0 & 23.84 & 15.58 & 16.39 & 12.55 & 10.79 & 10.41 & -1.37 & 0.24 & 1 \\
G16.086-0.546 & 275.4974 & -15.2042 & 2.77 & 8.36 & 4.34 & 18.1 & 11.1 & -176.7 & 74 & 1 & 1 & 9.11 & -0.64 & 2.37 & 3.30 & 2.97 & 3.46 & -1.78 & 0.68 & 0 \\
G16.091+0.333 & 274.6970 & -14.7844 & 3.58 & 9.75 & 3.89 & 18.1 & 11.1 & -176.7 & 57 & 0 & 0 & 4.77 & 5.34 & 4.78 & 3.38 & 3.42 & 2.33 & -1.46 & 1.70 & 0 \\
G16.101+0.472 & 274.5751 & -14.7105 & 2.27 & 8.53 & 1.94 & 18.1 & 11.1 & -176.7 & 41 & 0 & 0 & 6.50 & 2.66 & 2.43 & 0.50 & 1.84 & 2.82 & -1.29 & 0.63 & 0 \\
G16.106-0.984 & 275.9083 & -15.3919 & 13.10 & 30.19 & 14.38 & 18.1 & 11.1 & -176.7 & 95 & 0 & 0 & 21.56 & 14.68 & 14.93 & 12.35 & 14.24 & 13.22 & -0.56 & 0.23 & 1 \\
G16.112+0.702 & 274.3722 & -14.5913 & 7.47 & 20.13 & 5.72 & 18.1 & 11.1 & -176.7 & 55 & 0 & 0 & 10.07 & 6.94 & 7.45 & 5.96 & 5.31 & 3.50 & -1.31 & 0.27 & 1 \\
G16.125+1.227 & 273.9014 & -14.3305 & 5.73 & 9.45 & 5.66 & 18.1 & 11.1 & -176.7 & 50 & 0 & 0 & 11.28 & 8.18 & 6.72 & 5.17 & 5.63 & 6.13 & -1.44 & 0.65 & 0 \\
G16.134+0.399 & 274.6576 & -14.7155 & 4.16 & 14.27 & 2.94 & 18.1 & 11.1 & -176.7 & 43 & 0 & 0 & 6.60 & 5.71 & 4.63 & 4.00 & 1.85 & 2.79 & -1.28 & 0.46 & 0 \\
G16.134+0.491 & 274.5744 & -14.6717 & 4.91 & 17.16 & 3.98 & 18.1 & 11.1 & -176.7 & 57 & 0 & 0 & 7.03 & 5.70 & 5.44 & 3.01 & 3.19 & 2.10 & -1.79 & 0.34 & 1 \\
G16.140-0.126 & 275.1396 & -14.9589 & 2.20 & 5.49 & 1.62 & 18.1 & 11.1 & -176.7 & 23 & 0 & 1 & 4.08 & 1.87 & 2.67 & 2.21 & 0.60 & -0.04 & 0.00 & 0.00 & 0 \\
G16.143+0.041 & 274.9893 & -14.8769 & 7.18 & 14.44 & 5.12 & 18.1 & 11.1 & -176.7 & 45 & 0 & 0 & 12.19 & 6.49 & 6.01 & 4.40 & 4.57 & 3.83 & -2.09 & 0.59 & 0 \\
G16.145+0.009 & 275.0188 & -14.8901 & 9.50 & 17.45 & 8.14 & 18.1 & 11.1 & -176.7 & 62 & 0 & 0 & 6.98 & 4.07 & 7.31 & 9.86 & 11.06 & 7.90 & 0.48 & 0.66 & 0 \\
G16.155-0.432 & 275.4266 & -15.0891 & 1.87 & 6.24 & 1.67 & 18.1 & 11.1 & -176.7 & 32 & 0 & 1 & 2.73 & -0.66 & 2.75 & 1.08 & 1.22 & 2.66 & -0.11 & 1.34 & 0 \\
G16.170+0.953 & 274.1720 & -14.4219 & 3.97 & 11.65 & 3.05 & 18.1 & 11.1 & -176.7 & 43 & 0 & 0 & 6.62 & 7.24 & 4.11 & 4.32 & 2.89 & 2.74 & -1.49 & 0.44 & 1 \\
G16.174+0.411 & 274.6663 & -14.6746 & 1.61 & 5.84 & 1.43 & 18.1 & 11.1 & -176.7 & 31 & 0 & 1 & 2.56 & 1.88 & 1.81 & 1.33 & 1.19 & -0.89 & 0.00 & 0.00 & 0 \\
G16.178-0.711 & 275.6924 & -15.2003 & 2.39 & 8.01 & 1.74 & 18.1 & 11.1 & -176.7 & 33 & 0 & 0 & -1.97 & 6.91 & 2.87 & 1.34 & 1.13 & 0.75 & 0.00 & 0.00 & 0 \\
G16.182+0.950 & 274.1799 & -14.4122 & 3.16 & 8.97 & 2.45 & 18.1 & 11.1 & -176.7 & 39 & 0 & 0 & 5.25 & 5.15 & 3.36 & 4.76 & 2.27 & 2.96 & -1.09 & 0.59 & 0 \\
G16.183+0.768 & 274.3460 & -14.4975 & 36.51 & 72.15 & 35.82 & 18.1 & 11.1 & -176.7 & 113 & 0 & 0 & 57.19 & 44.52 & 41.68 & 32.96 & 32.53 & 28.64 & -1.10 & 0.06 & 1 \\
G33.118-1.244 & 284.0517 & -0.4017 & 4.61 & 5.86 & 7.68 & 13.5 & 13.3 & 3.3 & 51 & 1 & 1 & 9.64 & 7.78 & 7.09 & 6.05 & 4.63 & 4.70 & -1.01 & 0.64 & 0 \\
G33.120-0.894 & 283.7413 & -0.2406 & 23.31 & 93.90 & 23.35 & 13.5 & 13.3 & 3.3 & 134 & 0 & 0 & 23.04 & 23.44 & 23.27 & 23.87 & 22.79 & 22.14 & -0.05 & 0.06 & 1 \\
G33.128+0.364 & 282.6253 & 0.3406 & 8.34 & 20.43 & 7.06 & 13.5 & 13.3 & 3.3 & 60 & 0 & 0 & 8.67 & 9.58 & 6.72 & 7.72 & 7.60 & 5.90 & -0.39 & 0.32 & 1 \\
G33.133-0.093 & 283.0336 & 0.1365 & 187.07 & 176.50 & 315.49 & 13.5 & 13.3 & 3.3 & 392 & 1 & 0 & 130.33 & 165.62 & 179.39 & 206.92 & 219.04 & 228.78 & 0.86 & 0.03 & 1 \\
G33.141-0.673 & 283.5538 & -0.1212 & 2.45 & 12.09 & 2.56 & 13.5 & 13.3 & 3.3 & 63 & 0 & 0 & 4.45 & 4.41 & 2.60 & 2.04 & 2.08 & 2.64 & -1.66 & 0.60 & 0 \\
G33.143-0.066 & 283.0149 & 0.1580 & 122.27 & 111.80 & 230.62 & 13.5 & 13.3 & 3.3 & 252 & 1 & 0 & 239.64 & 190.95 & 180.47 & 157.71 & 144.76 & 128.60 & -0.96 & 0.03 & 1 \\
G33.144-0.041 & 282.9927 & 0.1698 & 5.22 & 5.02 & 29.34 & 13.5 & 13.3 & 3.3 & 135 & 1 & 1 & 11.20 & 9.81 & 12.98 & 15.20 & 12.06 & 7.48 & -0.54 & 0.62 & 0 \\
G33.147+1.044 & 282.0287 & 0.6677 & 26.93 & 104.80 & 26.50 & 13.5 & 13.3 & 3.3 & 114 & 0 & 0 & 30.77 & 28.74 & 27.58 & 26.61 & 25.98 & 24.19 & -0.36 & 0.06 & 1 \\
G33.149-0.557 & 283.4545 & -0.0607 & 1.61 & 5.32 & 1.13 & 13.5 & 13.3 & 3.3 & 19 & 0 & 1 & 1.82 & 0.98 & 1.24 & 0.08 & 0.61 & 0.29 & 0.00 & 0.00 & 0 \\
G33.166-1.165 & 284.0031 & -0.3233 & 6.74 & 23.71 & 6.30 & 13.5 & 13.3 & 3.3 & 69 & 0 & 0 & 8.94 & 9.12 & 6.65 & 6.63 & 5.77 & 6.75 & -0.73 & 0.23 & 1 \\
G33.169-0.016 & 282.9816 & 0.2038 & 7.30 & 7.41 & 339.77 & 13.5 & 13.3 & 3.3 & 2011 & 1 & 0 & 22.89 & 22.11 & 23.00 & 20.30 & 20.61 & 19.51 & -0.30 & 0.28 & 1 \\
G33.176+0.424 & 282.5933 & 0.4101 & 2.34 & 5.56 & 4.74 & 13.5 & 13.3 & 3.3 & 56 & 1 & 1 & 5.19 & 4.40 & 4.72 & 2.68 & 2.62 & 1.80 & -0.31 & 1.13 & 0 \\
G33.176-0.196 & 283.1454 & 0.1275 & 6.55 & 9.84 & 7.09 & 13.5 & 13.3 & 3.3 & 50 & 0 & 0 & 11.93 & 8.06 & 8.66 & 7.00 & 5.04 & 5.57 & -1.18 & 0.47 & 0 \\
G33.181-0.187 & 283.1392 & 0.1365 & 4.22 & 6.45 & 5.61 & 13.5 & 13.3 & 3.3 & 44 & 1 & 1 & 6.54 & 3.79 & 5.27 & 3.46 & 4.14 & 2.85 & 0.00 & 0.00 & 0 \\

\hline
\end{tabular}
\end{sidewaystable*}

\section{SNR Green and THOR comparison}

\begin{table*}
\caption{Matching of SNR between the THOR catalog and the SNR catalog presented by \citet{Green2014}.}             
\label{table_SNR_match}      
\centering          
\begin{tabular}{l c l c c c l l }
\hline\hline       
Galactic\_ID & art. & npix & res. & $\alpha$ & $\Delta \alpha$ & SNR name\tablefootmark{a} & SNR $\alpha\tablefootmark{b}$ \\
\hline

G15.913+0.183 & 0 & 2753 & 1 & -0.78 & 0.07 & G015.9+00.2 & -0.63 \\
G15.907+0.233 & 0 & 1303 & 1 & -1.12 & 0.20 & G015.9+00.2 & -0.63 \\
G16.742+0.088 & 0 & 9207 & 1 & -0.17 & 0.10 & G016.7+00.1 & -0.60 \\
G17.030-0.069 & 1 & 620 & 1 & -0.19 & 0.64 & G017.0-00.0 & -0.50 \\
G17.448-0.063 & 0 & 815 & 1 & 0.19 & 0.29 & G017.4-00.1 & -0.70 \\
G18.107-0.134 & 0 & 760 & 1 & -0.72 & 0.25 & G018.1-00.1 & -0.50 \\
G18.193-0.174 & 0 & 6624 & 1 & -0.37 & 0.10 & G018.1-00.1 & -0.50 \\
G18.171-0.213 & 0 & 952 & 1 & -0.68 & 0.25 & G018.1-00.1 & -0.50 \\
G18.128-0.218 & 1 & 35 & 0 & -0.36 & 1.37 & G018.1-00.1 & -0.50 \\
G18.610-0.316 & 0 & 2784 & 1 & 0.17 & 0.22 & G018.6-00.2 & -0.40 \\
G18.761+0.287 & 0 & 31818 & 1 & -1.07 & 0.06 & G018.8+00.3 & -0.46 \\
G18.908-0.922 & 1 & 1385 & 1 & -0.85 & 0.27 & G018.9-01.1 & -0.39 \\
G19.954-0.250 & 0 & 162 & 1 & -1.12 & 0.34 & G020.0-00.2 & -0.10 \\
G19.952-0.169 & 0 & 11499 & 1 & -0.32 & 0.19 & G020.0-00.2 & -0.10 \\
G20.075-0.181 & 0 & 913 & 1 & -1.30 & 0.44 & G020.0-00.2 & -0.10 \\
G20.502+0.155 & 0 & 1411 & 1 & -0.54 & 0.33 & G020.4+00.1 & -0.10 \\
G21.503-0.884 & 0 & 1713 & 1 & -0.02 & 0.00 & G021.6-00.8 & -0.50 \\
G21.765-0.631 & 0 & 49153 & 1 & -0.77 & 0.04 & G021.8-00.6 & -0.56 \\
G21.948-0.416 & 0 & 11333 & 1 & -0.31 & 0.12 & G021.8-00.6 & -0.56 \\
G23.124-0.199 & 0 & 6108 & 1 & -1.13 & 0.11 & G023.3-00.3 & -0.50 \\
G23.015-0.288 & 0 & 3619 & 1 & -1.27 & 0.10 & G023.3-00.3 & -0.50 \\
G23.105-0.411 & 0 & 10590 & 1 & 0.04 & 0.18 & G023.3-00.3 & -0.50 \\
G23.062-0.376 & 0 & 935 & 1 & -1.71 & 0.18 & G023.3-00.3 & -0.50 \\
G23.539+0.268 & 0 & 9419 & 1 & -0.26 & 0.39 & G023.6+00.3 & -0.30 \\
G24.664+0.620 & 0 & 2798 & 1 & -0.66 & 0.22 & G024.7+00.6 & -0.20 \\
G24.689-0.589 & 0 & 7553 & 1 & -0.73 & 0.17 & G024.7-00.6 & -0.50 \\
G27.365+0.014 & 0 & 8048 & 1 & -0.49 & 0.04 & G027.4+00.0 & -0.68 \\
G28.610-0.142 & 0 & 2919 & 1 & -0.79 & 0.04 & G028.6-00.1 & -- \\
G28.672-0.108 & 0 & 2015 & 1 & -0.64 & 0.04 & G028.6-00.1 & -- \\
G29.567+0.094 & 0 & 468 & 1 & -0.36 & 0.54 & G029.6+00.1 & -0.50 \\
G29.689-0.242 & 0 & 5059 & 1 & -0.64 & 0.01 & G029.7-00.3 & -0.63 \\
G31.869+0.064 & 0 & 18727 & 1 & -0.32 & 0.01 & G031.9+00.0 & -- \\
G32.423+0.079 & 0 & 60 & 0 & -0.74 & 0.44 & G032.4+00.1 & -- \\
G32.415+0.076 & 1 & 37 & 1 & 0.00 & 0.00 & G032.4+00.1 & -- \\
G32.929+0.021 & 0 & 5006 & 1 & -0.74 & 0.24 & G032.8-00.1 & -0.20 \\
G33.748+0.025 & 0 & 3496 & 1 & -1.04 & 0.13 & G033.6+00.1 & -0.51 \\
G33.651+0.051 & 0 & 6623 & 1 & -0.82 & 0.10 & G033.6+00.1 & -0.51 \\
G33.667+0.100 & 0 & 3002 & 1 & -1.31 & 0.15 & G033.6+00.1 & -0.51 \\
G33.607+0.089 & 0 & 2519 & 1 & -0.55 & 0.19 & G033.6+00.1 & -0.51 \\
G34.588-0.238 & 0 & 8050 & 1 & -0.34 & 0.06 & G034.7-00.4 & -0.37 \\
G34.568-0.630 & 0 & 11697 & 1 & -1.01 & 0.07 & G034.7-00.4 & -0.37 \\
G34.834-0.439 & 0 & 45702 & 1 & -0.80 & 0.06 & G034.7-00.4 & -0.37 \\
G34.681-0.635 & 0 & 3854 & 1 & -1.06 & 0.09 & G034.7-00.4 & -0.37 \\
G35.583-0.448 & 0 & 37 & 0 & -3.09 & 1.18 & G035.6-00.4 & -0.50 \\
G35.602-0.548 & 1 & 226 & 1 & 0.55 & 0.94 & G035.6-00.4 & -0.50 \\
G49.016-0.731 & 0 & 4875 & 1 & -1.05 & 0.12 & G049.2-00.7 & -0.30 \\
G49.059-0.777 & 1 & 473 & 1 & 0.42 & 0.38 & G049.2-00.7 & -0.30 \\
G49.190-0.801 & 0 & 5343 & 1 & -1.00 & 0.21 & G049.2-00.7 & -0.30 \\

\hline
\end{tabular}
\tablefoot{Visually matched sources between the THOR catalog and the SNR catalog by \citet{Green2014}. The first six columns are taken from the THOR continuum catalog, whereas the last two columns are presented in \citet{Green2014}. As the SNR are very clumpy, we find several THOR continuum sources, which overlap with the same SNR.\\
\tablefoottext{a}{Following the naming in \citet{Green2014}.}\\
\tablefoottext{b}{Taken from \citet{Green2014}. The spectral index in \citet{Green2014} is negatively defined, and we adapt the values according to our definition of the spectral index.}
}
\end{table*}
\end{appendix}
\end{document}